\documentclass[aps,rmp,nofootinbib,letterpaper,showkeys,%
superscriptaddress,amsmath,amssymb,showpacs,floatfix]{revtex4}
\usepackage{bm}
\usepackage{graphics,graphicx}
\usepackage{amssymb}
\begin{document}
\newcommand{\be}{\begin{equation}}
\newcommand{\ee}{\end{equation}}
\newcommand{\anu}{\ensuremath{\bar{\nu}}}
\newcommand{\anue}{\ensuremath{\bar{\nu}_e}}
\newcommand{\rE}{\ensuremath{R_{\oplus}}}
\newcommand{\nucleus}[2][]{\ensuremath{{}^{#1}\mathrm{#2}}}
\newcommand{\U}[1][]{\nucleus[#1]{U}}
\newcommand{\Th}[1][]{\nucleus[#1]{Th}}
\newcommand{\K}[1][]{\nucleus[#1]{K}}
\newcommand{\Rb}{\nucleus[87]{Rb}}
\newcommand{\eg}{\emph{e.g.},\ }
\newcommand{\ie}{\emph{i.e.\ }}
\newcommand{\et}{\emph{et al.\ }}
\title{Geo-neutrinos and Earth's interior}
\author{Gianni Fiorentini}
\email{fiorenti@fe.infn.it} \affiliation{Dipartimento di Fisica,
Universit\`a di Ferrara, I-44100 Ferrara, Italy}
\affiliation{Istituto Nazionale di Fisica Nucleare, Sezione di
Ferrara, I-44100 Ferrara, Italy}
\author{Marcello Lissia}
\email{marcello.lissia@ca.infn.it} \affiliation{Istituto Nazionale
di Fisica Nucleare, Sezione di Cagliari,
             I-09042 Monserrato, Italy}
\affiliation{Dipartimento di Fisica, Universit\`a di Cagliari,
             I-09042 Monserrato, Italy}
\author{Fabio Mantovani}
 \email{fabio.mantovani@unisi.it}
 \affiliation{Dipartimento di Scienze
della Terra, Universit\`a di Siena, I-53100 Siena, Italy}
\affiliation{Centro di GeoTecnologie CGT,I-52027 San Giovanni
Valdarno, Italy} \affiliation{Istituto Nazionale di Fisica
Nucleare, Sezione di Ferrara, I-44100 Ferrara, Italy}
%

\date{June 11th, 2007}
\begin{abstract}
  The deepest hole that has ever been dug is about 12 km deep.
Geochemists analyze samples from the
Earth's crust and from the top of the mantle. Seismology can
reconstruct the density profile throughout all Earth, but not its
composition. In this respect, our planet is mainly unexplored.  
Geo-neutrinos, the antineutrinos from the progenies of  \U, \Th\ and
\K[40] decays in the Earth, bring to the surface information from
the whole  planet, concerning its content of natural radioactive
elements.    Their detection can shed light on the sources of the
terrestrial heat flow, on the present composition, and on the
origins of the Earth.   Geo-neutrinos represent a   new probe of our
planet, which can be exploited  as a consequence of  two fundamental
advances that occurred in the  last few years: the  development of
extremely low background neutrino detectors and the  progress on
understanding neutrino propagation.      We review the status and
the prospects of the field.
\end{abstract}

\maketitle

\tableofcontents

\section{\label{sec:intro}Introduction}

The deepest hole that has ever been dug is about 12 km deep, a mere dent
in planetary terms. Geochemists analyze samples from the Earth's crust
and from the top of the mantle. Seismology can reconstruct the density
profile throughout all Earth, but not its composition. In this respect,
our planet is mainly unexplored.

Geo-neutrinos, antineutrinos from the progenies of \U, \Th, and \K\ decays
in the Earth, bring to Earth's surface information coming from the whole
planet. Differently form other emissions of the planet (\eg heat, noble gases…),
they are unique in that they can escape freely and instantaneously from Earth's interior.

Detection of geo-neutrinos is becoming practical as a consequence of
two fundamental advances that occurred in the last few years:
a) development of extremely low background neutrino detectors and
b) progress on understanding neutrino propagation. In fact, KamLAND
has reported in 2005~\cite{Araki:2005qa} evidence of a signal
originating from geo-neutrinos,
showing that the technique for geo-neutrino detection is now available.

Geo-neutrinos look thus a promising new probe for the study of
global properties of Earth and one has to examine their potential.
Let us enumerate a few items which, at least in principle, can be
addressed by means of geo-neutrinos\footnote{Additional goals for
geo-neutrinos (\eg the distribution of radio-elements in the core,
discrimination among models of mantle circulation, and the
possibility of detecting plumes in the
mantle~\cite{Fiorentini:2005mr}) appear presently too ambitious for
the available technology.}.

\begin{description}
  \item \emph{What is the radiogenic contribution to terrestrial heat production?}

There are large uncertainties on Earth's energetics, both on the
value of the heat flow (estimated between 30 and 45~TW) and on the
separate contributions to Earth's energy supply (radiogenic,
gravitational, chemical \ldots). Estimates of radioactivity in the
Earth's crust, based on observational data, account for at least
some 8~TW. The canonical Bulk Silicate Earth (BSE) model provides
about 20~TW of radiogenic heat. However, on the grounds of available
geochemical and/or geophysical data, one cannot exclude that
radioactivity in the present Earth is enough to account for even the
highest estimate of terrestrial heat flow.

An unambiguous and observationally based determination of the
radiogenic heat production would provide an important contribution
for understanding Earth's energetics. It requires determining how
much uranium, thorium and potassium are inside the Earth, quantities
which are strictly related to the anti-neutrino luminosities from
these elements.

  \item \emph{Test of the Bulk Silicate Earth model.}

The BSE model presents a chemical composition of the Earth similar
to that of CI chondritic meteorites see, \eg
\cite{McDonough:2003,Palme:2003}. The consistency between their
composition and that of the solar photosphere points towards
considering CI representatives of the material available in the
pre-solar nebula and the basic material from which our planet has
been formed. Some authors, however, have argued for a genetic
relationship of our planet with other chondrites, such as enstatite
chondrites,  which are richer in long lived radioactive elements
\cite{Javoy:1995}.

We remind that BSE is a basic geochemical paradigm consistent with
most observational data, which however regard mostly the crust and
an undetermined portion of the mantle. The global abundance of no
element in the Earth can be estimated on the basis of observational
data only. Geo-neutrinos could provide the first direct test of BSE
(and/or its variants) by measuring the global abundances of natural
heat radiogenic elements.

  \item \emph{Heat generating elements in the crust: a test of the estimated abundances.}

  The amount of radioactivity in the Earth's crust is reasonably well
constrained by observational data, with the exception of the lowest
portion. Most of the uncertainty on the amount of radioactivity in
the crust arises from the different estimates about the lower crust.
In this respect, a detector located well in the middle of a
continent, being most sensitive to geo-neutrinos from the crust,
might provide a significant check of the estimates on the crustal
content of heat generating elements.

\item\emph{A measurement of heat generating elements in the mantle.}

The estimated content in the mantle is based on cosmochemical
arguments and implies that abundances in deep layers have to be much
larger than those measured in samples originating from the uppermost
layer~\cite{Jochum:1983,Zartman:1988}. Uncertainties on the heat
generating elements content of the Earth essentially reflect the
lack of observational data on the bulk of the mantle. A geo-neutrino
detector located far from continents would be mainly sensitive to
heat radiogenic elements in the whole mantle, as the oceanic crust
is thin and poor in these elements.

\item \emph{What can be said about the core?}

Geochemical arguments are against the presence of radioactive
elements in the core, although alternative hypothesis have been
advanced see, \eg \cite{Herndon:1996,RamaMurthy:2003}.

Present non directional detectors can say little about the core;
however some extreme hypothesis can already be tested. If a natural
fission reactor were present in the Earth's core, as advocated by
Herndon in a series of paper
\cite{Herndon:1998,Herndon:2001,Herndon:2003}, it would produce
antineutrinos with a spectrum similar to that of man-made reactors.
An excess of "reactor like" antineutrinos events could be detected.
A detailed analysis already excludes a natural reactor producing
more than about 20~TW~\cite{Dye:2006,Fogli:2005qa}.
\end{description}

On the other hand, ``non c'\`e rosa senza spine''\footnote{There is
no such thing as a rose without a thorn.}. We list here the main
difficulties and limitations encountered when detecting
geo-neutrinos:
\begin{itemize}
  \item
First of all, even huge detectors cannot provide more than some hundreds of
geo-neutrino events per year.
  \item
Geo-neutrino events are to be disentangled from reactor neutrino events,
which provide a severe background at many locations.
  \item
Some 80\% of the geo-neutrino events are expected to arise from uranium
decay chain and only 20\% from thorium chain.
Due to the low yield, it will be hard to extract information on thorium
abundance from the difference in the spectra.
  \item
Geo-neutrinos from \K\ cannot be observed by means of inverse beta
on free protons, the classical reaction for antineutrinos detection.
  \item
Present detectors cannot provide directional information.
\end{itemize}
In the next section we shall outline the main properties (sources,
spectra and cross sections) of geo-neutrinos and in
section~\ref{sec:histo} we present how the field has evolved.
Available information on the radioactivity content of the Earth is
summarized in Sec.~\ref{sec:radio} and the debated issue of the
sources and flow of terrestrial heat is examined in
Sec.~\ref{sec:heat}. Section~\ref{sec:refmod} presents a reference
model for geo-neutrino production, \ie a calculation of geo-neutrino
fluxes based upon the best available information on Earth's
interior. This model is refined in section~\ref{sec:refine} for a
specific location (the Kamioka mine, Japan) with a detailed
calculation of the flux generated in the region.
Section~\ref{sec:beyond} provides a strategy for determining Earth's
radioactivity from geo-neutrino measurements. This approach is
developed in detail for KamLAND, the results of this experiment
being presented and interpreted in section~\ref{sec:kam}. The role
of reactor neutrinos, which are generally a significant background
for geo-neutrino detection, is discussed in
section~\ref{sec:reactor}. The prospects of the field are summarized
in the final section.

As a rule, when a section is divided into subsections, the first
one contains an overview of the main points, so that the reader
can decide whether the more detailed information presented in the
foregoing subsections is of interest to him/her.

\section{\label{sec:geonu}Geo-neutrino properties}
\subsection{Overview}
The natural radioactivity of present Earth arises mainly from the
decay (chains) of nuclear isotopes with half-lives comparable to or
longer than Earth's age\footnote{Isotopes in the list have
abundances and decay rates sufficiently large to give contributions
of order 1\% or more to the estimated radiogenic heat production:
other radioactive elements such as \nucleus[176]{Lu}, \nucleus[147]{Sm},
\nucleus[187]{Rn}, give contributions of order $10^{-4}$ or less.}:
\U[238], \Th[232], \K[40], \U[235], and \Rb.

Properties\footnote{In the Table and in the rest of the paper,
unless differently specified, nuclear data are taken
from~\cite{Firestone:1996}.} of these isotopes and of the
(anti)neutrinos produced from their decay (chains) are summarized in
Table~\ref{tab:propertiesIsotopes}. Actually neutrinos are produced
only in electron capture of \K[40]. In contrast to the Sun, Earth
shines essentially in antineutrinos.


\begin{table}[htb] \caption[aaa]{Properties of \U[238], \Th[232], \K[40], \U[235], and \Rb\ and of their (anti)neutrinos.
For each parent nucleus the table presents the natural isotopic mass
abundance, half-life, antineutrino maximal energy (or neutrino
energy), $Q$ value, $Q_{\mathrm{eff}}= Q-\langle
E_{(\nu,\anu)}\rangle$, antineutrino and heat production rates for
unit mass of the isotope ($\varepsilon_{\anu}$, $\varepsilon_{H}$),
and for unit mass at natural isotopic composition
($\varepsilon'_{\anu}$, $\varepsilon'_{H}$). Note that antineutrinos
with energy above threshold for inverse beta decay on free proton
($E_{th} = 1.806$~MeV) are produced only in the firsts two decay
chains. \label{tab:propertiesIsotopes} }
\begin{tabular}{lccccccccc}
 \hline\hline
      & Natural & &&&&&&& \\
Decay & isotopic  & $T_{1/2}$           & $E_\mathrm{max}$ & $Q$   & $Q_\mathrm{eff}$ & $\varepsilon_{\anu}$ & $\varepsilon_{H}$ &$\varepsilon'_{\anu}$ & $\varepsilon'_{H}$ \\
      & abundance & [$10^{9}$~yr] & [MeV]            & [MeV] & [MeV]            & [kg$^{-1}$s$^{-1}$]  & [W/kg]            & [kg$^{-1}$s$^{-1}$]  & [W/kg] \\
 \hline
$ \U[238] \to \nucleus[206]{Pb}+8\; \nucleus[4]{He}+6 e + 6\anu$
  & 0.9927 & 4.47  &  3.26 &   51.7  &  47.7 &   $7.46\times 10^7$ & $0.95\times 10^{-4}$ &  $7.41\times 10^7$ & $0.94\times 10^{-4}$ \\
$ \Th[232] \to \nucleus[208]{Pb}+6\; \nucleus[4]{He}+4 e + 4\anu$
  & 1.0000 & 14.0  &  2.25 &   42.7  &  40.4 &   $1.62\times 10^7$ & $0.27\times 10^{-4}$ &  $1.62\times 10^7$ & $0.27\times 10^{-4}$ \\
\hline
 $\K[40] \to \nucleus[40]{Ca}+ e + \anu$ (89\%)
    &$1.17\times 10^{-4}$ & 1.28 & 1.311 & 1.311 & 0.590 & $2.32\times
10^{8}$ & $0.22\times 10^{-4}$ & $2.71\times 10^{4}$ & $2.55\times
  10^{-9}$\\
  $\K[40]+ e \to \nucleus[40]{Ar} + \nu$ (11\%)
  &$1.17\times 10^{-4}$ & 1.28 & 0.044 & 1.505 & 1.461 & = & $0.65\times
10^{-5}$ & = & $0.78\times
  10^{-9}$\\
$ \U[235] \to \nucleus[207]{Pb}+7\; \nucleus[4]{He}+4 e + 4\anu$
&0.0072 & 0.704 & 1.23 & 46.4 & 44 & $3.19\times 10^8$ & $0.56\times
10^{-3}$ & $2.30\times 10^6$ & $0.40\times
  10^{-5}$\\
  $\Rb \to \nucleus[87]{Sr}+ e + \anu$
  &0.2783 & 47.5 & 0.283 & 0.283 & 0.122 & $3.20\times 10^6$ & $0.61\times 10^{-7}$ & $8.91\times 10^5$ & $0.17\times
  10^{-7}$\\
   \hline \hline
\end{tabular}\\[2pt]
\end{table}

The energy of \Rb\ neutrinos is so low that it is very unlikely that
its flux could be measured. Also heat production from \Rb\ is at the
level of 1\% of the total\footnote{This estimate is obtained
assuming an abundance of \Rb\ about 50 times the one of uranium.}.
For these reasons, from now on we shall consider only \U, \Th, and
\K[40] and refer to these three elements as the Heat Generating
Elements (HGEs) and to the antineutrinos from their decay (chains)
as geo-neutrinos.

\begin{figure}[hp]
\includegraphics[width=0.6\textwidth,angle=0]{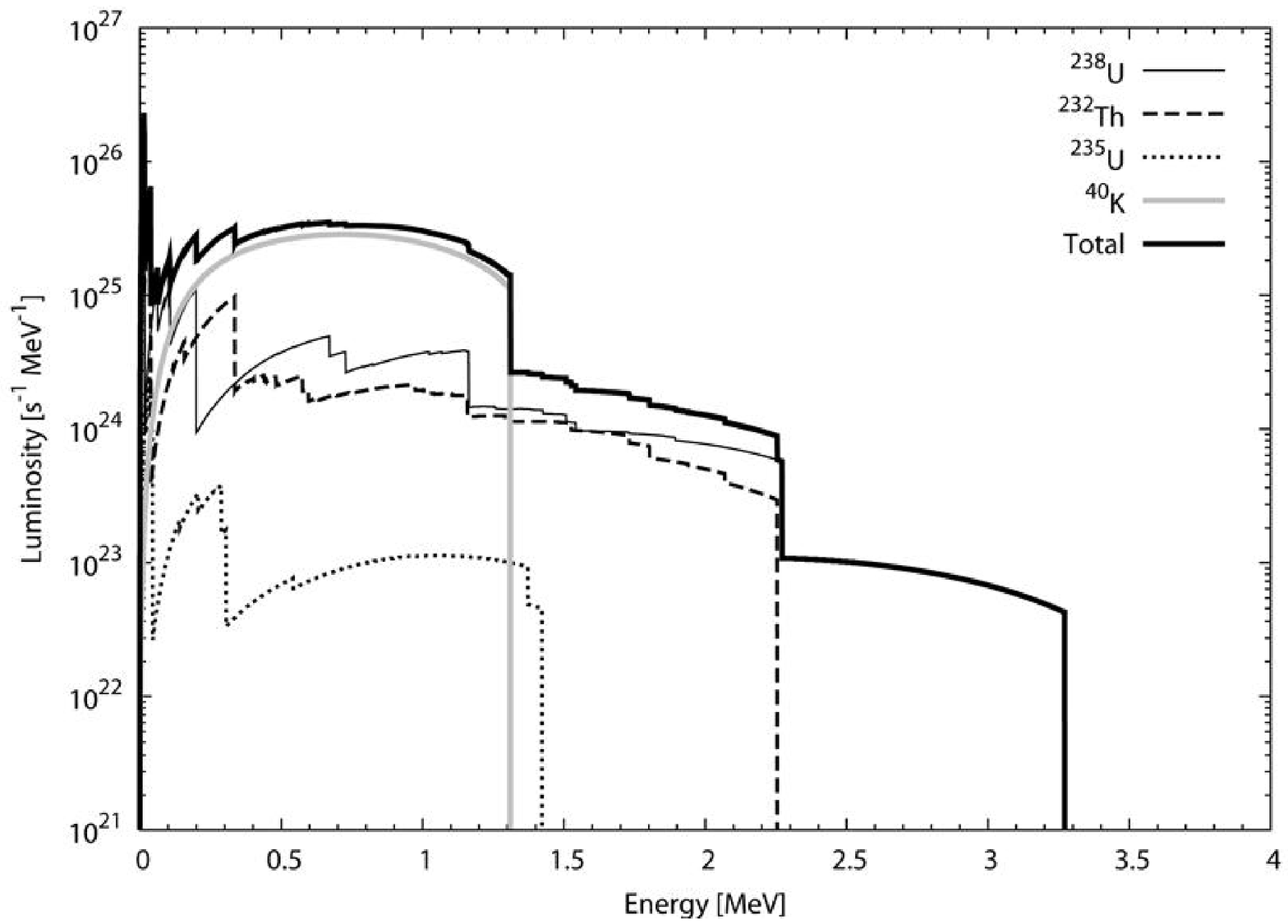}
 \caption[bbb]{Differential geo-neutrino luminosity, from~\cite{Enomoto:2005}~\protect\footnote{Data
 are from Enomoto's web
 page: \texttt{http://www.awa.tohoku.ac.jp/\~{}sanshiro/geoneutrino/spectrum/index.html}}.
 One assumes the following global
 abundances: $a(\U[238]) = 15$~ppb, $a(\U[235]) = 0.1$~ppb, $a(\Th[232]) = 55$~ppb,
 $a(\K[40]) = 160$~ppm~\cite{McDonough:1999}.
\label{fig:geoNuLumiAll}
}
\end{figure}

For each isotope there is a strict connection between the
geo-neutrino luminosity $L$ (anti-neutrinos produced in the Earth
per unit time), the radiogenic heat production rate $H_R$ and the
mass $m$ of that isotope  in the Earth:
\begin{equation}\label{eq:lumIsoto}
  L = 7.46 \times m(\U[238]) + 31.94 \times m(\U[235]) + 1.62 \times m(\Th[232])
    + 23.16 \times m(\K[40])
\end{equation}
\begin{equation}\label{eq:heatIsoto}
  H_R = 9.52 \times m(\U[238]) + 55.53 \times m(\U[235]) + 2.67 \times m(\Th[232])
    + 2.85 \times m(\K[40])
\end{equation}
where units are $10^{24}$~s$^{-1}$, $10^{12}$~W and $10^{17}$~kg,
respectively. By using the natural isotopic abundances  in
Table~\ref{tab:propertiesIsotopes} these equations can be written in
terms of the masses of the three elements\footnote{The coefficients
are slightly different from those quoted
in~\cite{Fiorentini:2002bp,Fiorentini:2005zc}, which did not include
\U[235] contribution.}:

\begin{equation}\label{eq:lumNatural}
  L = 7.64 \times m(\U) + 1.62 \times m(\Th)+ 27.10\times 10^{-4} \times m(\K)
\end{equation}
\begin{equation}\label{eq:heatNatural}
  H_R = 9.85 \times m(\U) + 2.67 \times m(\Th) + 3.33 \times 10^{-4}\times m(\K)
  \quad .
\end{equation}
The geo-neutrino spectrum depends on the shapes and rates of the
individual decays, and on the abundances and spatial distribution of
the terrestrial elements. It is shown in Fig.~\ref{fig:geoNuLumiAll}
for a specific model.

The complete geo-neutrino spectrum depends on a large number of beta
transitions in the uranium and thorium decay chains and it is
essentially a result of theoretical calculations. These should be
checked by measurements of the corresponding beta spectra, at least
for the most important decays which contribute to the geo-neutrino
signal: those of $^{214}$Bi and $^{234}$Pa$_m$ in the uranium chain,
$^{212}$Bi and $^{228}$Ac in the thorium chain.

Geo-neutrinos originating from different elements can be
distinguished -- at least in principle -- due to their different
energy spectra, \eg geo-neutrinos with $E > 2.25$ MeV are produced
only in the uranium chain.

Geo-neutrinos from \U[238] and \Th[232] (not those from \U[235] and
\K[40]) are above threshold for the classical anti-neutrino detection
reaction, the inverse beta on free protons:

\begin{equation}\label{eq:inverseBeta}
    \anue + p \to e^+ + n - 1.806 \mathrm{\ MeV}\quad .
\end{equation}
Note that anti-neutrinos from the Earth are not obscured by solar
neutrinos, which cannot yield reaction (\ref{eq:inverseBeta}). On
the other hand, antineutrinos from nuclear power plants are a
significant source of background, as first observed in
\cite{Lagage:1985tq} and discussed in more detail in
section~\ref{sec:reactor}.

An order of magnitude estimate of the geo-neutrino luminosity can be
obtained by assuming that a large fraction of the heat released from
Earth, $H \approx 40$~TW, arises from the decay chains of uranium
and thorium. Table~\ref{tab:propertiesIsotopes} shows that each of
the $N$ geo-neutrinos from each chain is associated with energy
release  $\Delta E \approx Q/N \approx 10$~MeV, so that:

\begin{equation}
  L(\U + \Th) \approx H / \Delta E \approx 2.5\times 10^{25}
       \mathrm{\ s}^{-1}\quad .
\end{equation}

The order of magnitude of the produced flux is
$\Phi^{\mathrm{(pro)}}(\U+\Th) \approx L/(4 \pi \rE^2)$, where \rE\
is the Earth's radius. The flux arriving at detectors will be
smaller than that produced due to neutrino oscillations,
$\Phi^{\mathrm{(arr)}}(\U+\Th)= \langle
P_{ee}\rangle\Phi^{\mathrm{(pro)}}(\U+\Th)$, where $ \langle
P_{ee}\rangle\approx 0.6$ is the average survival probability. All
this gives:
\begin{equation}
\Phi^{\mathrm{(arr)}}(\U+\Th) \approx 2\times 10^{6}\mathrm{\ cm}^{-2}\mathrm{\
 s}^{-1}\quad .
\end{equation}
This is a flux comparable to that of solar neutrinos from
\nucleus[8]{B} decay~\cite{Castellani:1996cm}, however the detection
of geo-neutrinos is a much more difficult task: their smaller energy
implies that the signal is smaller and is in an energy region where
background is larger.

For an order of magnitude estimate of the signal rate in a one-kton
detector (containing some $10^{32}$ free protons), we observe that
the cross section for inverse beta decay at few MeV is $\sigma\sim
10^{-43}$~cm$^2$ and the fraction of antineutrinos above threshold
is $f\approx0.05$. This gives a signal $S(\U+\Th) \approx \sigma f
\Phi^{\mathrm{(arr)}}(\U+\Th) N_p \approx 30 \mathrm{\ yr}^{-1}$.

More precisely,  the signal rates $S(\U)$ and $S(\Th)$ in a detector
containing $N_p$ free protons are
\begin{eqnarray}
\label{eq:signU}
  S(\U)  &=& 13\times
  \frac{\Phi^{\mathrm{(arr)}}(\U)}{10^6\mathrm{cm}^{-2}\mathrm{s}^{-1}}
                  \times\frac{N_p}{10^{32}}\mathrm{\ yr}^{-1}\\
                  \label{eq:signTh}
  S(\Th) &=& 4.0\times
  \frac{\Phi^{\mathrm{(arr)}}(\Th)}{10^6\mathrm{cm}^{-2}\mathrm{s}^{-1}}
                  \times\frac{N_p}{10^{32}}\mathrm{\ yr}^{-1} \quad ,
\end{eqnarray}
where  $\Phi^{\mathrm{(arr)}}(\U)$ and  $\Phi^{\mathrm{(arr)}}(\Th)$
are the fluxes of antineutrinos from \U[238] and \Th\ arriving at
the detector.

Events rates are conveniently expressed in terms of a Terrestrial
Neutrino Unit (TNU), defined as one event per $10^{32}$ target
nuclei per year, or $3.17\times 10^{-40}$~s$^{-1}$ per target
nucleus. This unit, which is analogous to the solar neutrino unit
(SNU)~\cite{Bahcall:1989}, is practical since one kton of liquid
scintillator contains about $10^{32}$ free protons (the precise
value depending on the chemical composition) and the exposure times
are of order of a few years.

Concerning the relative contributions of thorium and uranium to
geo-neutrino events, Eqs.~(\ref{eq:signU}) and (\ref{eq:signTh}),
together with Eq.~(\ref{eq:lumIsoto}) give:
\begin{equation}\label{eq:stimaThSuU}
    \frac{S(\Th)}{S(\U)}=0.32\times\frac{\Phi^{\mathrm{(arr)}}(\Th[232])}{\Phi^{\mathrm{(arr)}}(\U[238])}\approx
    0.32\times\frac{L(\Th[232])}{L(\U[238])}\approx \frac{1}{16}\times\frac{m(\Th[232])}{m(\U[238])}
    \quad .
\end{equation}
Since one estimates that in our planet $m(\Th)/m(\U) \approx 4$,
one expects $S(\Th) / S(\U) \approx 1/4$. Note that, although the
global thorium mass is four times than that of uranium, it contributes
just 1/5 of the total signal $S(\U+\Th)$.

\subsection{Decay chains and geo-neutrino spectra from uranium and thorium}
One needs antineutrino spectra for two main reasons: the calculation of the
specific elemental heat production and of the signal in the detector.

Heat production rate is calculated by subtracting from the $Q$ value the
energy $\langle E\rangle$ of antineutrinos averaged over the whole spectrum.
In the case of \U[238] and \Th[232]
chains\footnote{Concerning  \U[235], its contribution to the heat production
is just a few per cent, so that the
energy subtracted by antineutrinos is not relevant.}
the average antineutrino
energy is about 8\% and 6\% of the total available energy: an error of
10\% on the calculation of $\langle E\rangle$ is sufficient to determine
the elemental heat production to better the 1\%. For this reason
in the literature individual determinations of beta spectra have not
been used to determine neutrino energy loss. Instead, the approximate
relationship that, on average, neutrinos carry 2/3 of the decay
energy for beta decay has been applied~\cite{vanSchmus:1995}.
This approximation can be checked or improved if the complete spectrum is known.

For calculating the signal in a detector we need to integrate the
spectrum times the cross section: only the spectrum above the
detection threshold is needed for this aim.

On these grounds we shall concentrate on the antineutrino energy
spectra from \U[238] and \Th[232] decay chains. In general, the chain
involves many different  $\beta$ decays and the total antineutrino
spectrum results from the sum of the individual spectra.

For each decay chain, if the sample of material contains $n_i$ nuclei
of type $i$, the number of alpha and beta decays $i\to j$  per unit time is:
\begin{equation}\label{eq:rate}
    r_{i,j}= n_i \lambda_i b_{i,j} \quad ,
\end{equation}
where  $\lambda_i$, is the inverse of the mean-life and $b_{i,j}$ is the branching ratio,
$\sum_j b_{i,j}=1$. The probability of each decay in the chain is:
\begin{equation}\label{eq:probRij}
    R_{i,j}=\frac{n_i \lambda_i b_{i,j} }{\sum_j r_{h,j}} \quad ,
\end{equation}
where $h$ indicates the decay-chain head.
The $R_{i,j}$  form a network, with an isotope at each node.
Generally the network has the following properties:
\begin{itemize}
  \item $ \frac{R_{i,j}}{R_{i,k}}=\frac{b_{i,j}}{b_{i,k}}$ (by definition),
  \item  $\sum_j R_{h,j} = 1$ (normalization);
\end{itemize}
assuming that the chain is in secular equilibrium, one has:
\begin{itemize}
  \item   $\sum_k R_{k,i}=\sum_j R_{i,j}$, at each node $i$   (equilibrium).
\end{itemize}
These three conditions fully determine the
network\footnote{It can be seen as a circuit where
$R_{i,j}$  are the currents and $b_{i,j}$ the inverse
of the resistance, and where it flows a unit of current.}.

In general the beta decay $i\to j$  involves transitions to
different nuclear states which yield spectra with different
endpoints: we call $I_{i,j;k}$ the percentage intensity of
the $k$-th beta
transition\footnote{Our notation corresponds to the
normalization  $\sum_k I_{i,j;k} = 1$. A different
normalization, $\sum_k I_{i,j;k} = b_{i,j}$ is used in~\cite{Firestone:1996}.}
and $f_{i,j;k}(E)$
the corresponding antineutrino energy spectrum normalized to 1 (see below).

Then the antineutrino spectrum generated from the sample is:
\begin{equation}\label{eq:totalSpectrum}
f(E)= \sum_{ij}R_{i,j}\sum_k I_{i,j;k}  f_{i,j;k}(E) \quad .
\end{equation}
Lifetimes $1/\lambda_i$, branching ratios $b_{i,j}$
and intensities $I_{i,j;k}$, can be found in~\cite{Firestone:1996}.

A somehow delicate point is the expression to be used for the
antineutrino spectra $ f_{i,j;k}(E) $ of the  $\beta$ decay of
nucleus $i$ to the nucleus $j$ into the state $k$. It can be derived
from that for electron energy spectrum $ \phi_{i,j;k}(W)$  by using
energy conservation:
\begin{equation}
  f_{i,j;k}(E) = \left. \phi_{i,j;k}(W) \right|_{W=W_{max}-E} \quad,
\end{equation}
where $W$ is the total electron energy and $W_{max} = m_{e}c^2 +
E_{max}$ with $E_{max}$ being the maximal neutrino energy for the
transition and $m_e$ the electron mass.

For allowed decays the electron energy spectrum has the
well-known universal shape:
\begin{equation}\label{eq:electrSpectrum}
 \phi_{i,j;k}(W) = \frac{1}{N} W (W_{max}-W)^2 (W^2-m_e^2 c^4)^{\gamma-1/2}
    \, e^{\pi y} \left| \Gamma(\gamma + i y)\right|^2 \quad,
\end{equation}
where:
\begin{equation}\label{eq:gammaY}
    \gamma =\sqrt{1-(\alpha Z)^2},\quad\quad y=\alpha Z \frac{W}{\sqrt{W^2-m_e^2 c^4}}\quad,
\end{equation}
with $Z$ denoting the nuclear charge of the daughter nucleus and
$\alpha$ the fine structure constant. $N$ is a normalization constant such that:
\begin{equation}
\int_{m_e c^2}^{W_{max}} dW\,  \phi_{i,j;k}(W)  = 1 \quad .
\end{equation}
Equation~(\ref{eq:electrSpectrum}) is generally used to estimate
geo-neutrino spectra and this requires a few comments.
\begin{enumerate}
  \item[1)]
  Equation~(\ref{eq:electrSpectrum}) considers the effect
  of the bare Coulomb field through the relativistic Fermi function.
  Electron screening and finite nuclear size effects are not considered.
  These provide corrections to the spectrum shape of order of few per cent,
  a quantity which is not significant in comparison with the uncertainties mentioned below.
  \item[2)]
  All important contributions actually arise from (first) parity forbidden decays.
  In this case the spectrum does not need to have a universal shape, since it
  involves also momentum-dependent nuclear matrix elements. However experimental data show that many
  forbidden decays of high-$Z$ nuclei have spectra \emph{close} to the allowed one:
  the theoretical explanation is that these decays are dominated by momentum-independent
  matrix elements or matrix elements whose relevant momentum is the electron momentum near the nucleus
  $pR \approx Z\alpha $, which is weakly dependent on the emerging momentum ($\xi$  approximation).
  This provides a partial justification for using equation~(\ref{eq:electrSpectrum}).
  The resemblance with the allowed spectrum depends on the nucleus and it is difficult to study
  at low electron energy; in few cases one finds significant
  differences\footnote{Spectra of high-$Z$ nuclei, that do not follow the allowed spectra,
  are explained theoretically by cancelations of dominant terms:
  a detailed knowledge of the relative weights and signs of the
  nuclear matrix elements becomes necessary.},
  \eg  \nucleus[210]{Bi} (for an experimental review
  see, \eg~\cite{Daniel:1968}).
  \item[3)]
  Measurements of electron spectra would be very useful -- in particular at low energy --
  in order to check the predictions for geo-neutrino spectra, which are mostly theoretical.
  In this respect an experimental study of the beta decay of \nucleus[214]{Bi} would be most significant.
\end{enumerate}

Regarding the intensities $I_{i,j;k}$, the experimental errors on some of them
should be reduced: at the moment they imply a few percent uncertainty on the total geo-neutrino
signal (see Tables~\ref{tab:uBetaTransitions} and \ref{tab:thBetaTransitions} and relative comments).

\subsubsection{The \U[238] decay chain}
\begin{figure}[p]
\includegraphics[width=0.5\textwidth,angle=0]{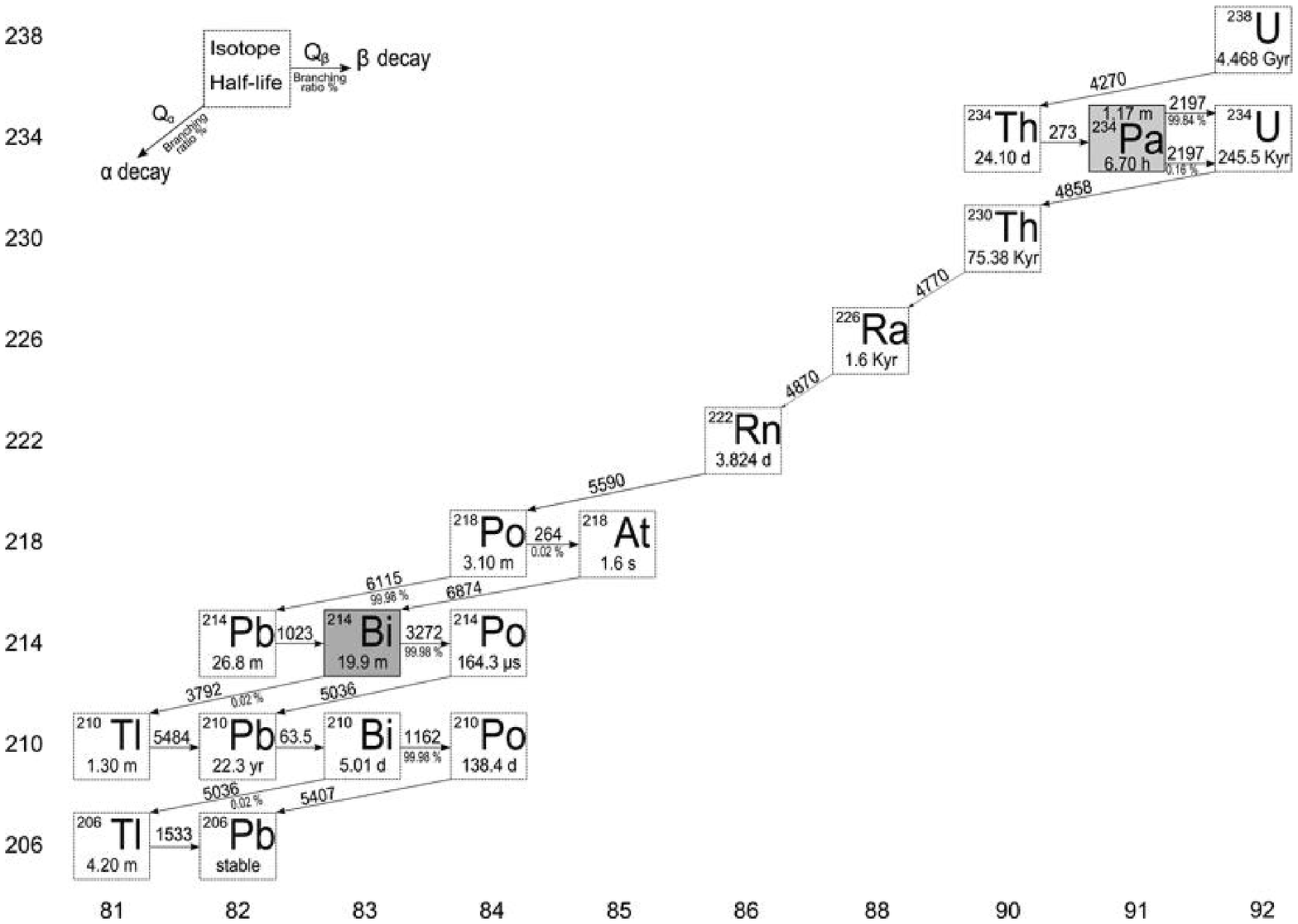}
 \caption[ccc]{The \U[238] decay chain. The two nuclides inside the grey boxes
(\nucleus[234]{Pa} and \nucleus[214]{Bi}) are the main sources of
geo-neutrinos.} \label{fig:uChain}
\vspace{2cm}
\includegraphics[width=0.45\textwidth,angle=-90]{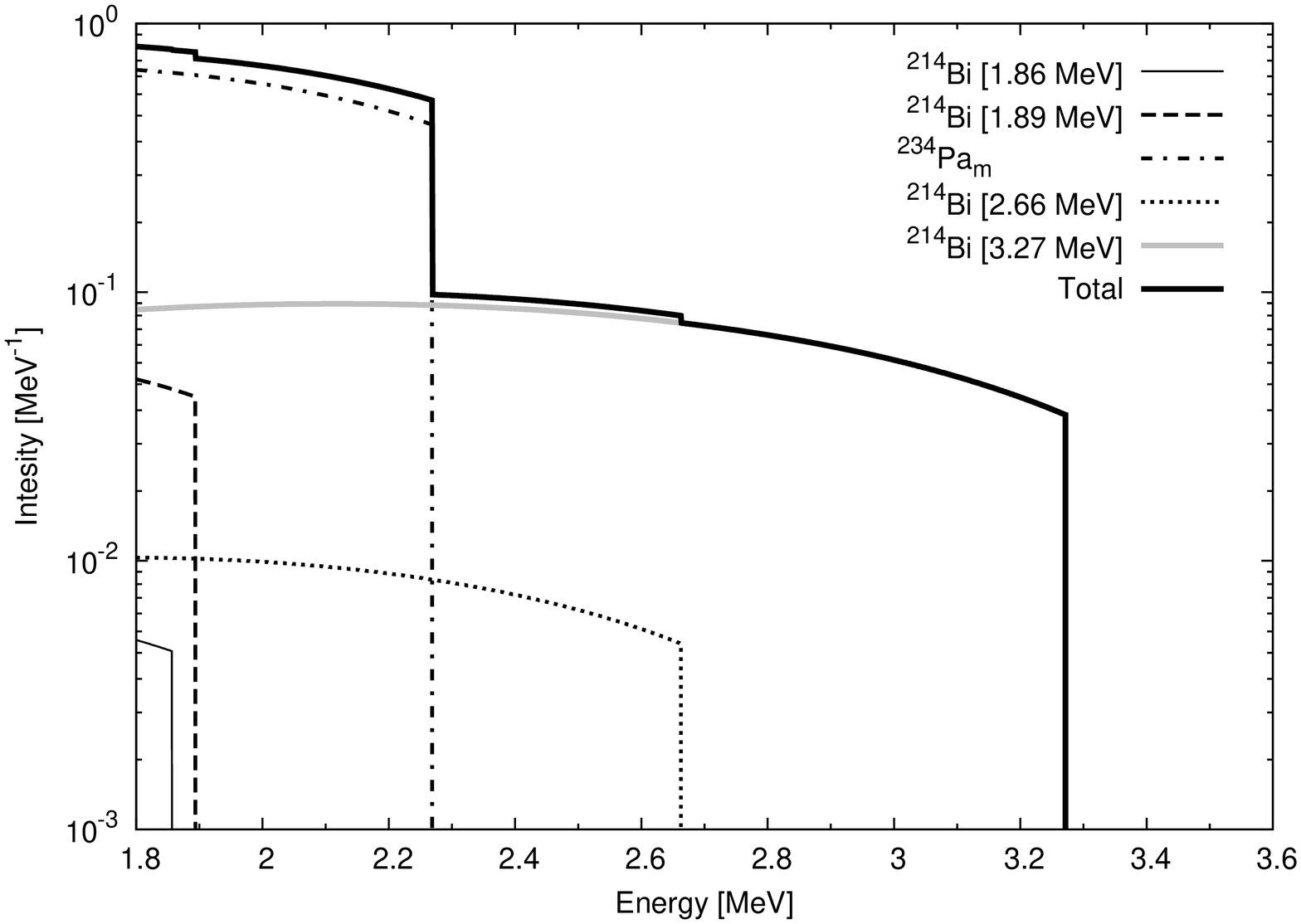}
 \caption[fff]{Geo-neutrino spectra from the five main $\beta$ decays of the \U[238]
 chain.
 All spectra are normalized to one decay of the head element of the chain.
 Since the \U[238] chain contains six $\beta$ decays, the integral from zero to the
 end point of the total spectrum is 6. Note that only 0.38 neutrinos per chain are above thresholds.}
\label{fig:uSpectrum}
\end{figure}

\U[238] decays into \nucleus[206]{Pb} through a chain of eight
$\alpha$ decays and six  $\beta$ decays\footnote{If we call
$N_{\alpha}$ the number of $\alpha$ decays and $N_{\beta}$  the
number of $\beta$ decays, $A$ and $Z$ ($A'$ and $Z'$) the atomic
number and charge of the initial (final) nucleus, then $N_{\alpha} =
(A-A')/4$ and $N_{\beta}  = Z'-Z + (A-A')/2$}. In secular
equilibrium the complete network (see Fig.~\ref{fig:uChain})
includes nine  $\beta$-decaying nuclei\footnote{This accounts for
all branches with probability $> 10^{-5}$.} summarized in
Table~\ref{tab:uBetaNuclei}.

\begin{table}[htb] \caption[ddd]{
Beta decays in the \U[238] chain. For each decay we present the
probability, the maximal antineutrino energy and the number of
effective transitions, defined as those producing antineutrinos with
$E > 1806$~keV. \label{tab:uBetaNuclei} }
\renewcommand{\tabcolsep}{2pc} 
\begin{tabular}{cccc}
 \hline\hline
    $i\to j $  & $R_{i,j}$ & $E_{max}$  & Effective \\
            &    &      [keV] & transitions \\
 \hline
 $ \nucleus[234]{Th}\to  \nucleus[234]{Pa} $ & 1.0000 & 199.08  & 0 \\
 $ \nucleus[234]{Pa}_m\to  \nucleus[234]{U} $ & 0.9984 & 2268.92 & 1 \\
  $ \nucleus[214]{Pb}\to  \nucleus[214]{Bi}  $ & 0.9998 & 1024    & 0 \\
   $ \nucleus[214]{Bi}\to  \nucleus[214]{Po}  $ & 0.9998 & 3272    & 4 \\
    $ \nucleus[210]{Pb}\to  \nucleus[210]{Bi}  $ & 1.0000 & 63.5    & 0 \\
     $ \nucleus[210]{Bi}\to  \nucleus[210]{Po}  $ & 0.9999 & 1162.1  & 0 \\
      $ \nucleus[234]{Pa}\to  \nucleus[234]{U}   $ & 0.0016 & 1247.15 & 0 \\
       $ \nucleus[218]{Po}\to  \nucleus[218]{At}  $ & 0.0002 & $< 265$ & 0 \\
        $ \nucleus[206]{Tl}\to  \nucleus[206]{Pb}  $ & 0.0001 & 1533.5  & 0 \\
         $ \nucleus[210]{Tl}\to  \nucleus[210]{Pb}  $ & 0.0002 & 4391.3  & 5 \\
   \hline \hline
\end{tabular}\\[2pt]
\end{table}

Only three nuclides (\nucleus[234]{Pa}, \nucleus[214]{Bi},
\nucleus[210]{Tl}) yield antineutrinos with energy larger than
1.806~MeV and contribute to the geo-neutrino signal. The
contribution from \nucleus[210]{Tl} is negligible, due to its small
occurrence probability and the uranium contribution to the
geo-neutrino signal comes from five  $\beta$ decays: one from
\nucleus[234]{Pa} and four from \nucleus[214]{Bi} (see
Table~\ref{tab:uBetaTransitions} and Fig.~\ref{fig:uSpectrum}). In
fact, 98\% of the uranium signal arises from the first two
transitions in Table~\ref{tab:uBetaTransitions} and an accuracy
better than 1\% is achieved by adding the third one.

\begin{table}[htb] \caption[eee]{Effective transitions in the \U[238] chain. In addition to
quantities defined in Table~\ref{tab:uBetaNuclei} we present the
intensity $I_k$, its error $\Delta I_k$, type and percentage contributions to the uranium
geo-neutrino signal,  and to the  (\U + \Th) geo-neutrino signal.
For this last column we assume the chondritic ratio for the masses
($\Th/\U= 3.9$), which implies that 79\% of the geo-neutrino signal
comes from uranium. \label{tab:uBetaTransitions} }
\newcommand{\dg}{\hphantom{$0$}}
\renewcommand{\tabcolsep}{1pc} 
\begin{tabular}{cccccccc}
 \hline\hline
    $i\to j $  & $R_{i,j}$ & $E_{max}$  & $I_k$ & $\Delta I_k$ & Type & $S_{\U}$ & $S_{\mathrm{tot}}$ \\
            &    &      [keV] & & & [\%] & [\%]\\
 \hline
 $ \nucleus[234]{Pa}_m\to  \nucleus[234]{U} $ & 0.9984 & 2268.92 & 0.9836 & 0.002\dg & \dg 1st forbidden $(0^-) \to 0^+$  & 39.62 & 31.21 \\
\hline
   $ \nucleus[214]{Bi}\to  \nucleus[214]{Po}  $ & 0.9998 & 3272.00    & 0.182\dg & 0.006\dg & 1st forbidden $1^- \to 0^+$ &   58.21 &   45.84 \\
                                                &        & 2662.68    & 0.017\dg & 0.006\dg & 1st forbidden $1^- \to 2^+$ & \dg1.98 & \dg1.55 \\
                                                &        & 1894.32    & 0.0743   & 0.0011   & 1st forbidden $1^- \to 2^+$ & \dg0.18 & \dg0.14 \\
                                                &        & 1856.51    & 0.0081   & 0.0007   & 1st forbidden $1^- \to 0^+$ & \dg0.01 & \dg0.01 \\
   \hline \hline
\end{tabular}\\[2pt]
\end{table}

In the last column of Table~\ref{tab:uBetaTransitions} we show the
contribution of each decay to the total ($\U + \Th$) geo-neutrino
signal: this is calculated using a ratio of \Th\ to \U\ signal
$S_{\Th}/S_{\U}=0.270$, that comes from the ratio between the
average cross sections
$\langle\sigma\rangle_{\Th[232]}/\langle\sigma\rangle_{\U[238]}=0.127/0.404
= 0.314$ (see subsection~\ref{subsec:XsctRates}) and an assumed
chondritic ratio\footnote{The corresponding ratio of fluxes is
$\Phi_{\Th}/\Phi_{\U}=(4/6)\times (m(\Th)/m(\U)) \times (238/232)
\times (\tau_{\U}/\tau_{\Th})\times (1/0.9927) = 0.8579$.}
 for the masses $m(\Th)/m(\U)= 3.9$.
Present errors on the intensities of the second and third decay
of Table~\ref{tab:uBetaTransitions} imply
corresponding errors to the total signal of 1.5\% and 0.5\%,
respectively.

\subsubsection{The \Th[232] decay chain}

\begin{figure}[hp]
\includegraphics[width=0.60\textwidth,angle=0]{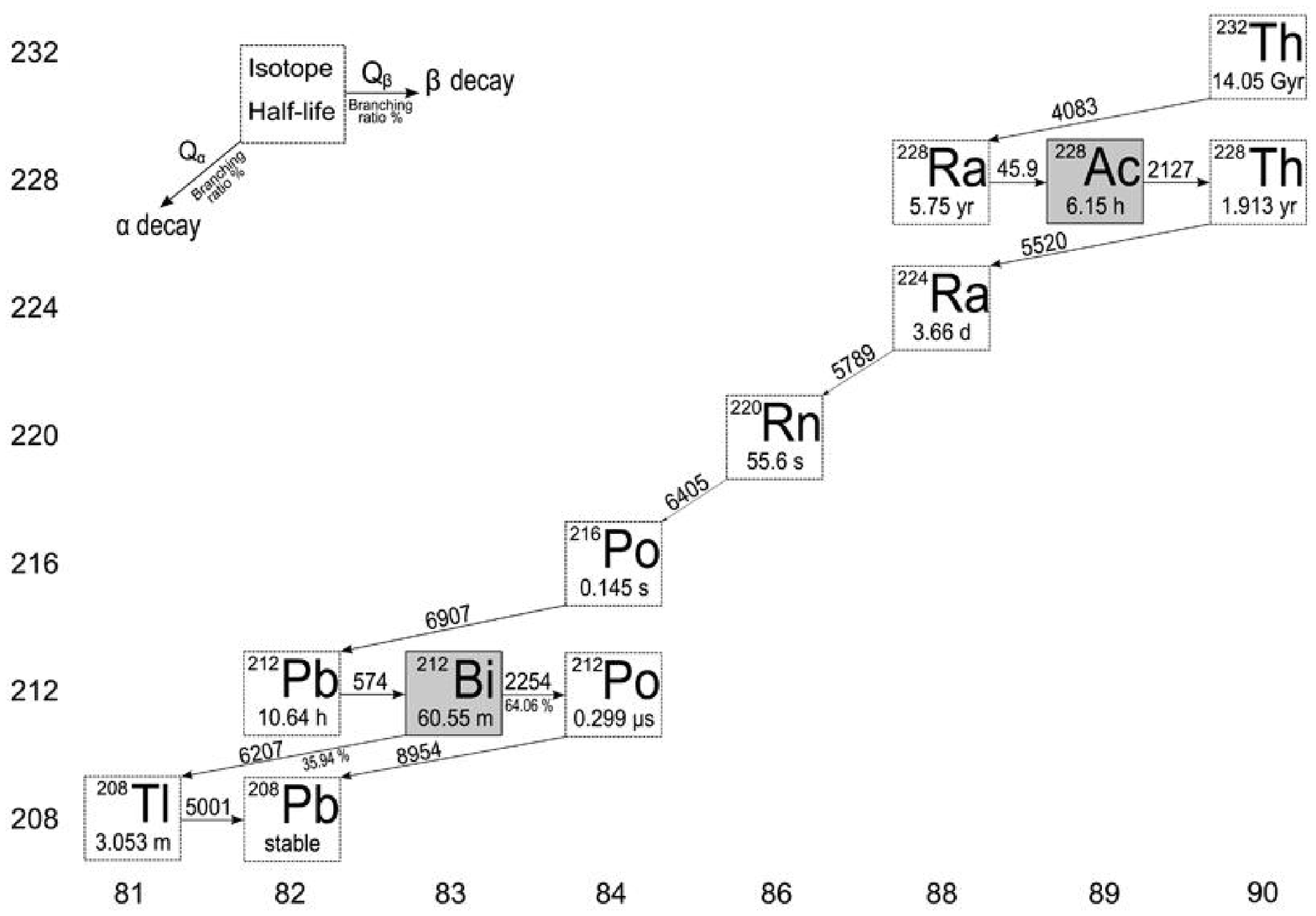}
 \caption[hhh]{The \Th[232] decay chain. The two nuclides inside the grey boxes
(\nucleus[228]{Ac} and \nucleus[212]{Bi}) are the main sources of
geo-neutrinos.} \label{fig:thChain}
\vspace{0.5cm}
\includegraphics[width=0.45\textwidth,angle=-90]{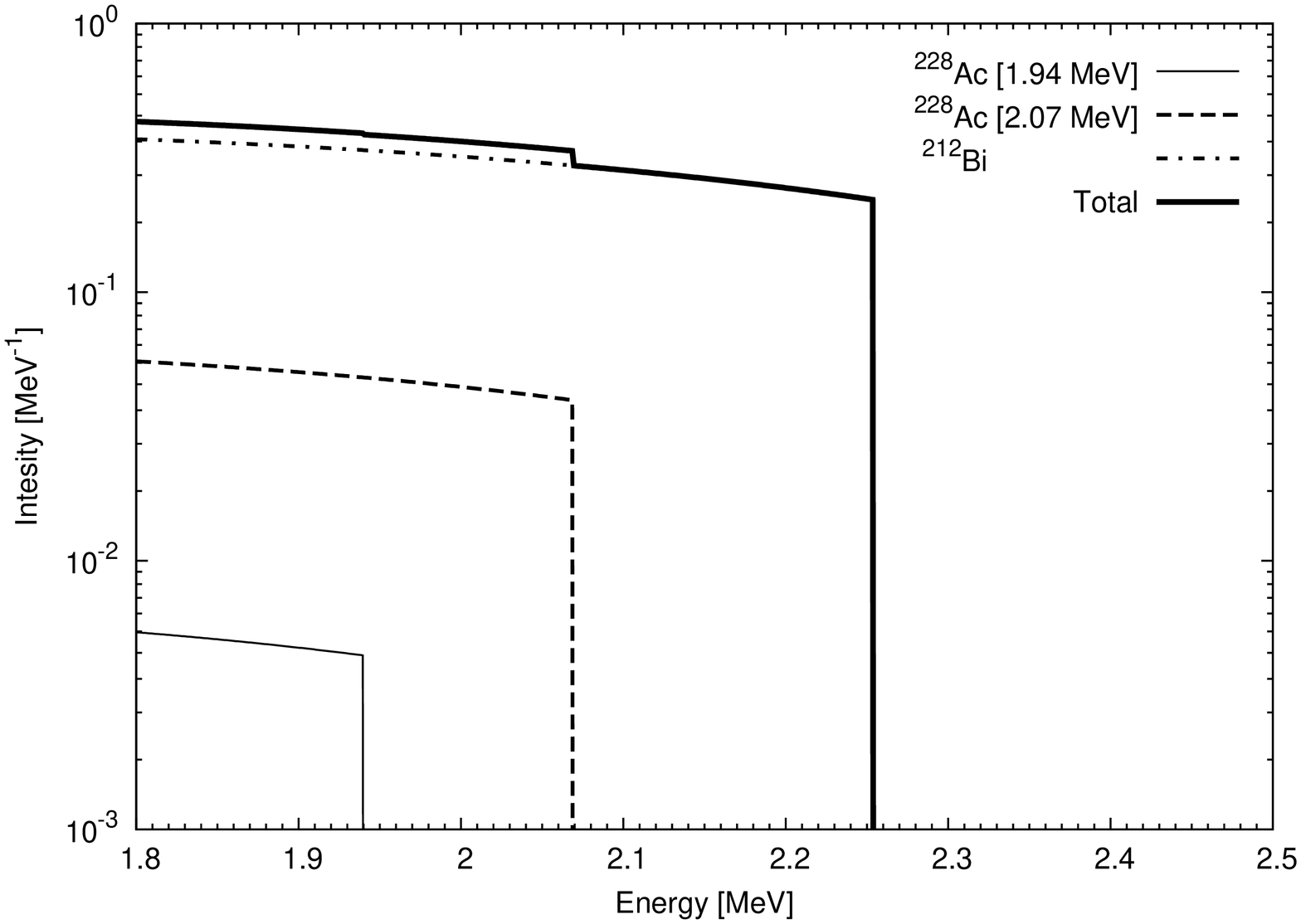}
 \caption[kkk]{Geo-neutrino spectra from the three main $\beta$ decays of the \Th[232]
 chain.
 All spectra are normalized to one decay of the head element of the chain.
 Since the \Th[232] chain contains four $\beta$ decays, the integral from zero to the
 end point of the total spectrum is 4. Note that only 0.15 neutrinos per chain are above thresholds.}
\label{fig:thSpectrum}
\end{figure}

\Th[232] decays into \nucleus[208]{Pb} through a chain of six
$\alpha$ decays and four $\beta$ decays. In secular equilibrium the
complete network (see Fig.~\ref{fig:thChain}) includes five
$\beta$-decaying nuclei\footnote{This accounts for all the branches
with probability $> 10^{-5}$.} summarized in
Table~\ref{tab:thBetaNuclei}.

\begin{table}[htbp] \caption[iii]{Beta decays in the \Th[232] chain.
For each decay we present the probability, the maximal antineutrino energy
and the number of effective transitions, defined as those producing antineutrinos with $E > 1806$~keV.
\label{tab:thBetaNuclei}}
\renewcommand{\tabcolsep}{2pc} 
\begin{tabular}{cccc}
 \hline\hline
    $i\to j $  & $R_{i,j}$ & $E_{max}$  & Effective \\
            &    &      [keV] & transitions \\
 \hline
 $ \nucleus[228]{Ra}\to  \nucleus[228]{Ac} $ & 1.0000 & 39.62  & 0 \\
  $ \nucleus[228]{Ac}\to  \nucleus[228]{Th} $ & 1.0000 & 2069.24 & 2 \\
   $ \nucleus[212]{Pb}\to  \nucleus[214]{Bi}  $ & 1.0000 & 573.8    & 0 \\
    $ \nucleus[212]{Bi}\to  \nucleus[212]{Po}  $ & 0.6406 & 2254    & 1 \\
     $ \nucleus[208]{Tl}\to  \nucleus[208]{Pb}  $ & 0.3594 & 1803.26  & 0 \\
   \hline \hline
\end{tabular}\\[2pt]
\end{table}

Only two nuclides (\nucleus[228]{Ac} and \nucleus[212]{Bi}) yield
antineutrinos with energy larger than 1.806~MeV. The thorium
contribution to the geo-neutrino signal comes from three  $\beta$
decays: one from \nucleus[212]{Bi} and two from \nucleus[228]{Ac}
(see Table~\ref{tab:thBetaTransitions} and
Fig.~\ref{fig:thSpectrum}). In fact, 99.8\% of the signal arises
from the first two transitions in Table~\ref{tab:thBetaTransitions}.
The present error on the intensity of the second decay of
Table~\ref{tab:thBetaTransitions} implies
a corresponding error to the total signal of 0.9\%.

\newpage

\begin{table}[htb] \caption[jjj]{Effective transitions in the \Th[232] chain. In addition to
quantities defined in Table~\ref{tab:thBetaNuclei} we present the
intensity $I_k$, its error $\Delta I_k$, type and percentage contributions to the thorium
geo-neutrino signal, and to the total (\U + \Th) geo-neutrino
signal. For this last column we assume the chondritic ratio for the
masses ($\Th/\U= 3.9$), which implies that 21\% of the geo-neutrino
signal comes from thorium. \label{tab:thBetaTransitions} }
\newcommand{\dg}{\hphantom{$0$}}
\renewcommand{\tabcolsep}{1pc} 
\begin{tabular}{cccccccc}
 \hline\hline
    $i\to j $  & $R_{i,j}$ & $E_{max}$  & $I_k$ & $\Delta I_k$ & Type & $S_{\Th}$ & $S_{\mathrm{tot}}$ \\
            &    &      [keV] & & & [\%] & [\%]\\
 \hline
 $ \nucleus[212]{Bi}\to  \nucleus[212]{Po} $    & 0.6406 & 2254\dg\dg & 0.8658 & 0.0016 & 1st forbidden  $1^{(-)} \to 0^+$  & 94.15 & 20.00 \\
\hline
   $ \nucleus[228]{Ac}\to  \nucleus[228]{Th}  $ & 1.0000 & 2069.24    & 0.08\dg\dg & 0.06\dg\dg& Allowed $3^+ \to 2^+$ & \dg5.66 & \dg1.21 \\
                                                &        & 1940.18    & 0.008\dg & 0.006\dg & Allowed $3^+ \to 4^+$ & \dg0.19 & \dg0.04 \\
   \hline \hline
\end{tabular}\\[2pt]
\end{table}

\subsection{Geo-neutrinos from \K[40]}

\begin{figure}[p]
\includegraphics[width=0.4\textwidth,angle=0]{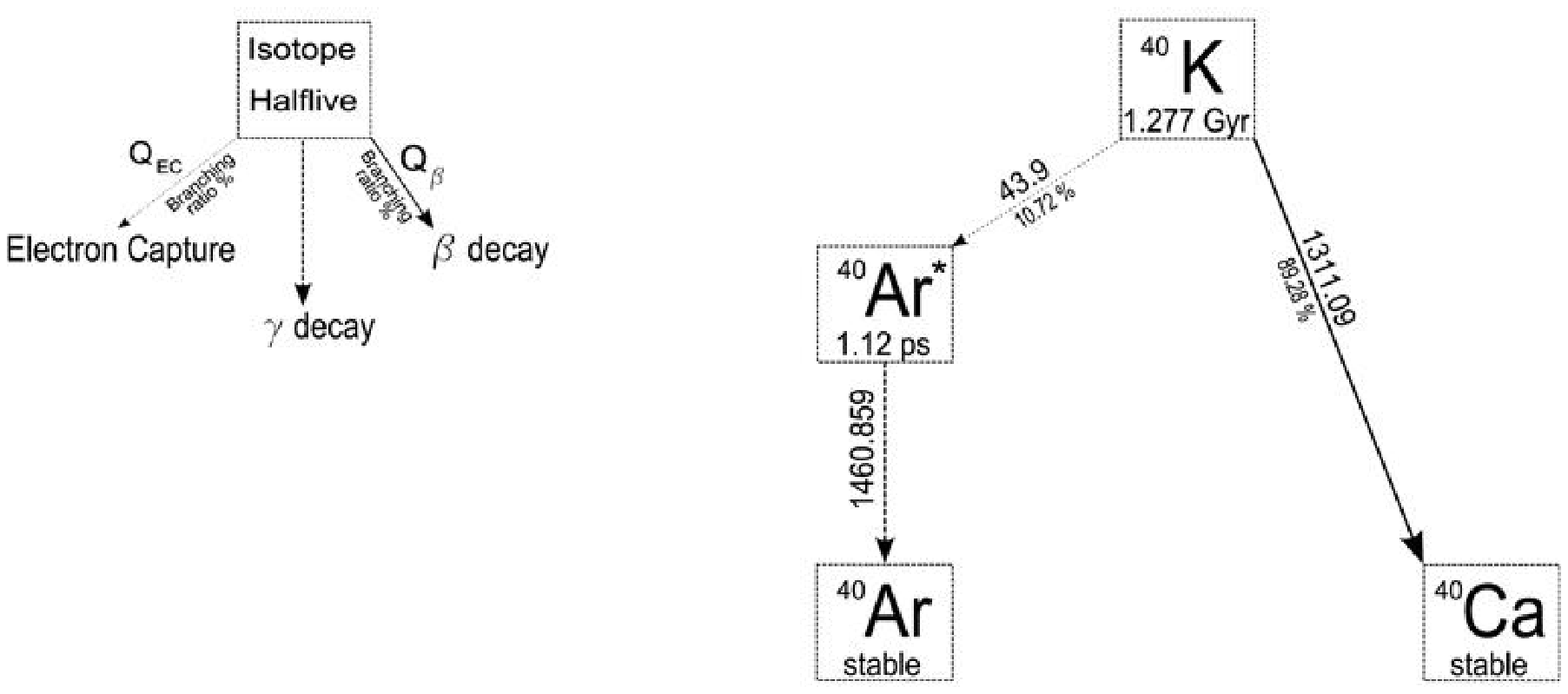}
 \caption[mmm]{Simplified decay scheme for \K[40].}
\label{fig:KdecayScheme}
\vspace{4cm}
\includegraphics[width=0.5\textwidth,angle=0]{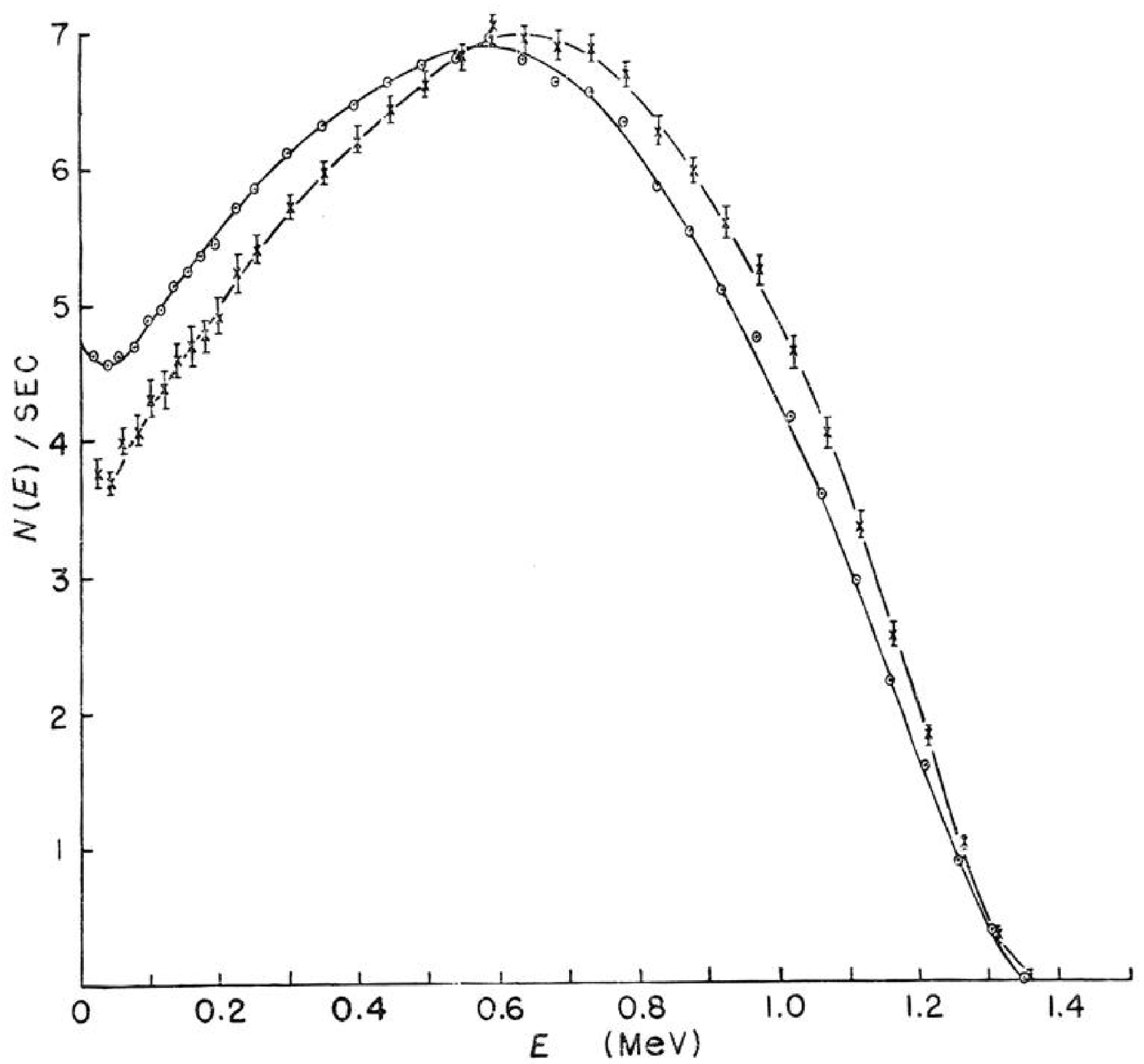}
 \caption[nnn]{Experimental spectrum of electron kinetic energy for the decay of \K[40] into \nucleus[40]{Ca},
 from~\cite{Kelley:1959}. The circles show the measured spectrum including background, 1.46~MeV gamma
 and finite resolution corrections. The x-s show the spectrum after the electron escape corrections.
 The flags represent total estimated error at each point, due mostly to the uncertainty in the
 electron escape correction.
\label{fig:Kspectrum} }
\end{figure}

\K[40] undergoes branching decay to \nucleus[40]{Ca} (via  $\beta$
decay) and \nucleus[40]{Ar} (via electron capture), both of which
are stable: the simplified decay scheme of \K[40] is shown in
Fig.~\ref{fig:KdecayScheme}.  The half life is $1.277\times
10^9$~yr, with a 10.7 percent probability of decaying
  to \nucleus[40]{Ar} and an 89.3 percent probability of
decaying to \nucleus[40]{Ca}. All decays to \nucleus[40]{Ca}
 proceed directly to the ground state, but most of the decays to \nucleus[40]{Ar}
 reach an excited state, see Table~\ref{tab:Kdecays}.

\begin{table}[htb] \caption[mmm]{Decays of \K[40].
For each decay we show the maximal antineutrino/neutrino energy, the
intensity and the type of transition. \label{tab:Kdecays} }
\newcommand{\dg}{\hphantom{$0$}}
\renewcommand{\tabcolsep}{1pc} 
\begin{tabular}{cccc}
 \hline\hline
    $i\to j $  &  $E_{max}$  & $I_k$ & Type  \\
               &      [keV]  &       &       \\
 \hline
    $ \K[40]\to  \nucleus[40]{Ca} + e^- + \anu $ &  1311.09 & 0.8928  & 3rd forbidden $4^- \to 0^+$ \\\hline
   $ e^- + \K[40]\to  \nucleus[40]{Ar}^* + \nu  $ & \dg\dg44.04   & 0.1067\dg  & 1st forbidden $4^- \to 2^+$  \\
  $ e^- + \K[40]\to  \nucleus[40]{Ar} + \nu  $  & \dg\dg1504.9 & 0.00047   & 3rd forbidden  $4^- \to 0^+$  \\
 $  \K[40]\to  \nucleus[40]{Ar} + e^+ + \nu  $   & \dg\dg482.9 & 0.00001   & 3rd forbidden  $4^- \to 0^+$  \\
   \hline \hline
\end{tabular}\\[2pt]
\end{table}

Kelley \et~\cite{Kelley:1959} determined the
 spectrum of  $\beta$ particles emitted in the decay to  \nucleus[40]{Ca} (Fig.~\ref{fig:Kspectrum}).
 From these data Van Schmus~\cite{vanSchmus:1995} obtained a mean
 $\beta$ energy of 0.598~MeV,
 or about 45\% of the total;
 the remainder, 0.722 MeV (55\%), is carried away by the
  antineutrino\footnote{We checked that by using Eq.~(\ref{eq:electrSpectrum})
  times the non-relativistic correction factor appropriate for a
  3rd forbidden decay,
  $ S(p_e,p_{\nu})\sim p^6_{\anu} + p^6_{e} + 7  p^2_{\anu}  p^2_{e} (p^2_{\anu} + p^2_{e})$,
  one finds the same average energy as Van Schmus. Note that the value of the maximal energy
  used by Van Schmus,  $W_{\mathrm{max}}=1.32$, should be replaced with the more resent
  value: $W_{\mathrm{max}}=1.31109$. In this case the average $\beta$ energy becomes 0.588~MeV.}.

 We remind that  the antineutrinos from \K[40] ($E_{max} = 1.311$~MeV)
 are below the threshold for inverse beta on free protons.
 Note also that the monochromatic neutrinos from \K[40] have a very small energy (44~keV).

\subsection{From cross sections to event rates}
\label{subsec:XsctRates}
\begin{figure}[hp]
\includegraphics[width=0.4\textwidth,angle=-90]{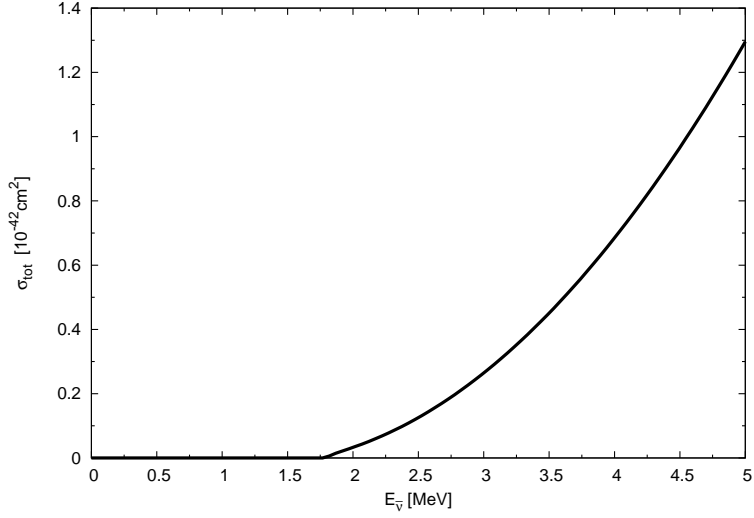}
 \caption[ooo]{Total cross section for $\anu_e+p \to e^+ + n$  as a function of the antineutrino energy, Eq.~(\ref{eq:crossSection}).}
\label{fig:Xsection}
\end{figure}

As already mentioned, the classical process for detection of low energy
antineutrinos is the inverse beta decay on free protons:
\begin{equation}\label{eq:inverseBetaSenzaE}
    \anu_e+p \to e^+ + n \quad .
\end{equation}
The threshold of the reaction is:
\begin{equation}\label{eq:threshold}
    E^{thr}_{\nu}=\frac{(M_n+m_e)^2-M_p^2}{2M_p}c^2=1.806 \mathrm{\ MeV}\quad .
\end{equation}
The total cross section, neglecting terms of order $E_e/M_p$, is
given by the standard formula:
\begin{equation}\label{eq:crossSection}
    \sigma = 0.0952\times \left( \frac{E_e p_e c}{\mathrm{MeV}^2}\right)\times 10^{-42}\mathrm{cm}^2\quad ,
\end{equation}
where $E_e=E_{\anu}-(M_n - M_p)c^2$  is the positron energy, when the (small)
neutron recoil is neglected, and $p_e$
is the corresponding momentum. The numerical factor in Eq.~(\ref{eq:crossSection})
is  tied directly, see~\cite{Bemporad:2001qy}, to the neutron lifetime,
known to 0.1\%~\cite{Yao:2006px}.
This expression of the total cross section is shown in Fig.~\ref{fig:Xsection}.

Corrections to the cross section of order  $E_e/M_p$, which are
negligible for geo-neutrinos whereas should be considered at reactor energies,
and the angular distribution of the positrons are described by Vogel and Beacom in Ref.~\cite{Vogel:1999zy};
see also Ref.~\cite{Bemporad:2001qy}.

A more general discussion of the neutrino/nucleon cross-section for
energies from threshold up to several hundred MeV can be found
in~\cite{Strumia:2003zx}; in the same paper Strumia and Vissani give
a simple approximation which agrees with their full result within
few per-mille for $E_{\nu}\lesssim 300 \mathrm{\ MeV}$,
\begin{equation}
    \sigma(\anue p) \approx 10^{-43} [\mathrm{cm}^{2}] p_e E_e
    E_{\nu}^{-0.07056+0.02018\ln E_\nu -0.001953 \ln^3 E_\nu}, \quad\quad E_e = E_{\nu}-\Delta,
\end{equation}
where $\Delta$ is the neutron-proton mass difference and all
energies are in MeV. They conservatively estimate an overall
uncertainty of the cross section at low energy of 0.4\%. At the
energy relevant for geo-neutrinos ($\leq 3.27$~MeV)
Eq.~(\ref{eq:crossSection}) overestimates the full result of Strumia
and Vissani by less than 1\% and it is already identical at about
2~MeV.

The geo-neutrino event rate
from the decay chain of element X = \U[238] or \Th[232] is:
\begin{equation}\label{eq:signalDefine}
    S(X) = N_p  \int dE_{\anu} \varepsilon(E_{\anu})
    \sigma(E_{\anu})\phi_{X}^{\mathrm{(arr)}}(E_{\anu})\quad ,
\end{equation}
where $N_p$ is the number of free protons in the target, $\varepsilon$
is the detection efficiency, $\sigma(E_{\anu})$ is the cross section
for reaction~(\ref{eq:inverseBetaSenzaE}), and:

\begin{equation}\label{eq:fluxArriving}
 \phi_{X}^{\mathrm{(arr)}}(E_{\anu})=\int_{V_{\oplus}}d\vec{r}\frac{\rho(\vec{r})}{4\pi|\vec{R}-\vec{r}|^2}
 \frac{a_X(r)C_X}{\tau_X m_X} f_X(E_{\anu}) p\left(E_{\anu},|\vec{R}-\vec{r}|\right)
\end{equation}
is the differential flux of antineutrinos from \U[238] or \Th[232]
arriving into the detector, $\rho$ is the density, $a_X$ is the elemental mass
abundance, $C_X$, $\tau_X$, and $m_X$ are the isotopic concentration,
lifetime and mass of nucleus $X$. The energy distribution of antineutrinos $f_X(E_{\anu})$ is
 normalized to the number of antineutrinos $n_X$ emitted per decay
 chain:
\begin{equation}
    n_X=\int dE_{\anu} f_X(E_{\anu}) \quad;
\end{equation}

 $p\left(E_{\anu},|\vec{R}-\vec{r}|\right)$ is the survival probability for \anu\  with
 energy $E_{\anu}$ produced at $\vec{r}$ to reach the
 detector at  $\vec{R}$.

 In view of the values of the oscillation length one can average the survival
 probability over a short distance, see Ref.~\cite{Mantovani:2003yd},
 and bring out of the integral the averaged survival probability:
 \begin{equation}\label{eq:averageProba}
    \langle P_{ee}\rangle = 1 - \frac{1}{2}\sin^2{2\theta} = \frac{1+\tan^4\theta}{(1+\tan^2\theta)^2} \quad .
 \end{equation}
In this way we are left with:
\begin{equation}
  \label{eq:ottieniSegnale1}
    S(X) = N_p \langle P_{ee}\rangle \int dE_{\anu} \varepsilon(E_{\anu})
    \sigma(E_{\anu}) f_X(E_{\anu})
    \int_{V_{\oplus}}d\vec{r}\frac{\rho(\vec{r})}{4\pi|\vec{R}-\vec{r}|^2}
 \frac{a_X(r)C_X}{\tau_X m_X}
\end{equation}
The second integral is proportional to the (angle integrated)
produced flux of anti-neutrinos
\begin{equation}
    \Phi(X) = \frac{n_X C_X}{4\pi \tau_X m_X}\int_{V_{\oplus}}d\vec{r}\frac{\rho(\vec{r}) a_X(r)}{|\vec{R}-\vec{r}|^2}
    \quad .
\end{equation}
Note that this quantity is different from the flux normal to earth surface.
Note also that ``produced'' essentially means the flux which one would observe in the absence of oscillations.

One can also assume the detection efficiency as approximately
constant over the small ($<2$~MeV) energy integration region. Then
Eq.~(\ref{eq:ottieniSegnale1}) becomes
\begin{equation}
    S(X) = N_p \langle P_{ee}\rangle  \varepsilon \Phi(X)
    \int dE_{\anu}
    \frac{\sigma(E_{\anu}) f_X(E_{\anu})}{n_X}  \quad .
\end{equation}
 It can be useful to introduce an average cross
section:
\begin{equation}
   \langle\sigma\rangle_{X} = \int dE_{\anu} \sigma(E_{\anu}) f_X(E_{\anu})
  / \int dE_{\anu} f_X(E_{\anu})
   \quad .
\end{equation}
This is computed by using equation (\ref{eq:crossSection}) for the cross section
$\sigma(E_{\anu})$ and the spectrum $f_X(E_{\anu}) $ obtained in
the previous section. Thus one finds
$\langle\sigma\rangle_{\U[238]}=0.404\times 10^{-44}$cm$^2$ and
$\langle\sigma\rangle_{\Th[232]}=0.127\times 10^{-44}$cm$^2$.

The event number can thus be written as the product of a few terms:
\begin{equation}
    S(X) = N_p \langle P_{ee}\rangle  \varepsilon \Phi(X) \langle\sigma\rangle_{X}
\end{equation}
The result is
\begin{eqnarray}
  S(\U[238]) &=& 4.04\times 10^{-7}\mathrm{\ s}^{-1} \times \langle P_{ee}\rangle \varepsilon
   \left(\frac{N_p}{10^{32}}\right) \left(\frac{\Phi(\U[238])}{10^6\mathrm{cm}^{-2}\mathrm{s}^{-1}}\right)\\
  S(\Th[232])&=& 1.27\times 10^{-7}\mathrm{\ s}^{-1} \times \langle P_{ee}\rangle \varepsilon
   \left(\frac{N_p}{10^{32}}\right) \left(\frac{\Phi(\Th[232])}{10^6\mathrm{cm}^{-2}\mathrm{s}^{-1}}\right) \quad .
\end{eqnarray}
This is the way in which Eqs.~(\ref{eq:signU}) and (\ref{eq:signTh})
were derived. Our goal in the rest of the paper will be to provide
calculations of the produced fluxes based on geological models.

\begin{figure}[p]
\includegraphics[width=0.5\textwidth,angle=-90]{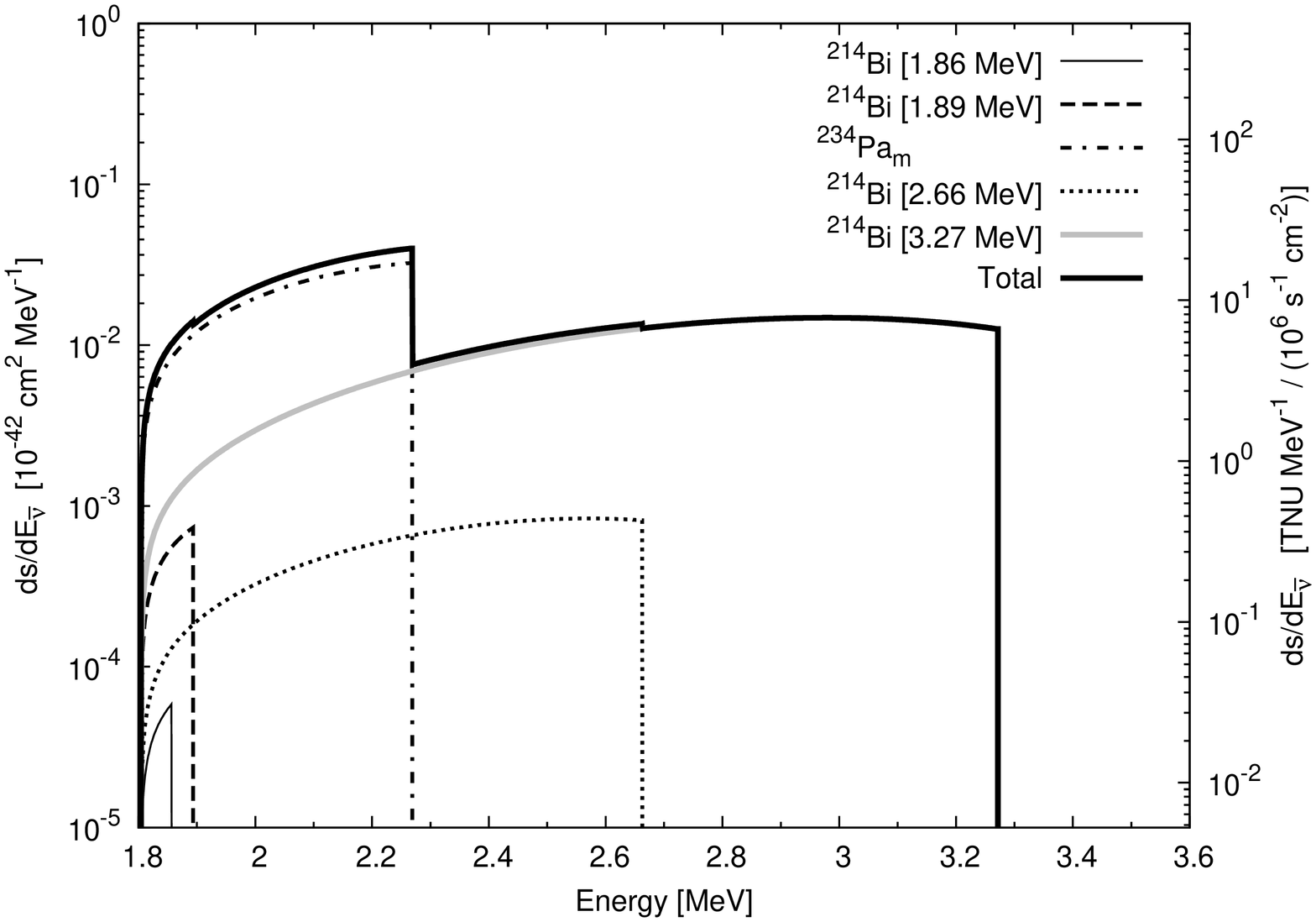}
 \caption[ggg]{Geo-neutrino differential signal per unit flux from
 the five main $\beta$ decays of the \U[238] chain, see Eq.~(\ref{eq:differentialSignal}).
  \label{fig:uSignal} }
\vspace{2cm}
\includegraphics[width=0.5\textwidth,angle=-90]{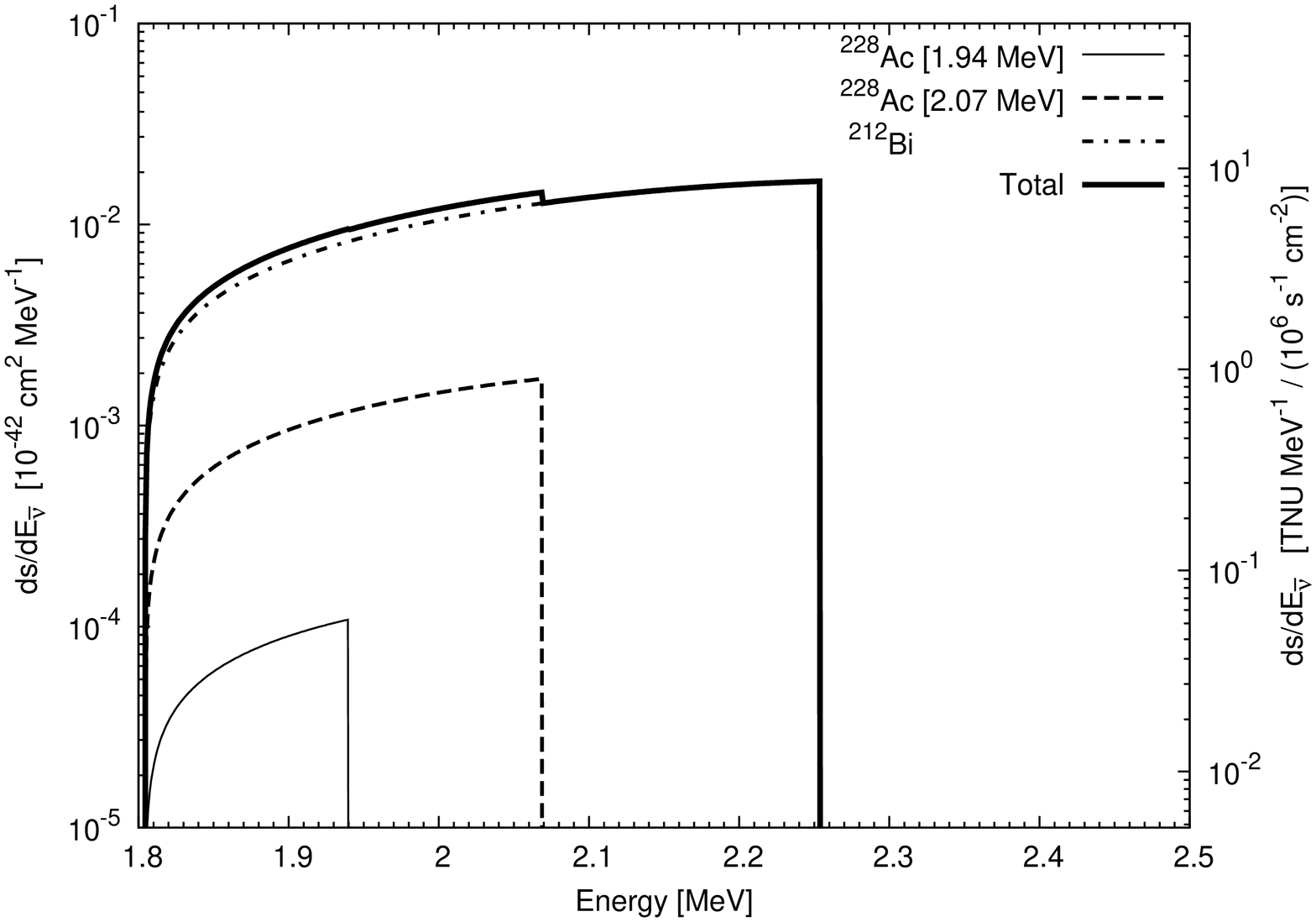}
 \caption[lll]{Geo-neutrino differential signal per unit flux from
 the three main $\beta$ decays of the \Th[232] chain, see Eq.~(\ref{eq:differentialSignal}).
 \label{fig:thSignal}}
\end{figure}

It is interesting to examine the differential geo-neutrino signal
per unit flux as a function of the energy:
\begin{equation}
\label{eq:differentialSignal}
 \frac{ds_X}{dE_{\anu}}
  =  \sigma(E_{\anu}) f_X(E_{\anu})
  / \int dE_{\anu} f_X(E_{\anu})
   \quad .
\end{equation}
This quantity is shown in Figs.~\ref{fig:uSignal} and
\ref{fig:thSignal} for uranium and thorium, respectively. Note that
most of the geo-neutrino flux originates from very few transitions.

\section{\label{sec:histo}A historical perspective}

\begin{figure}[p]
\includegraphics[width=0.4\textwidth]{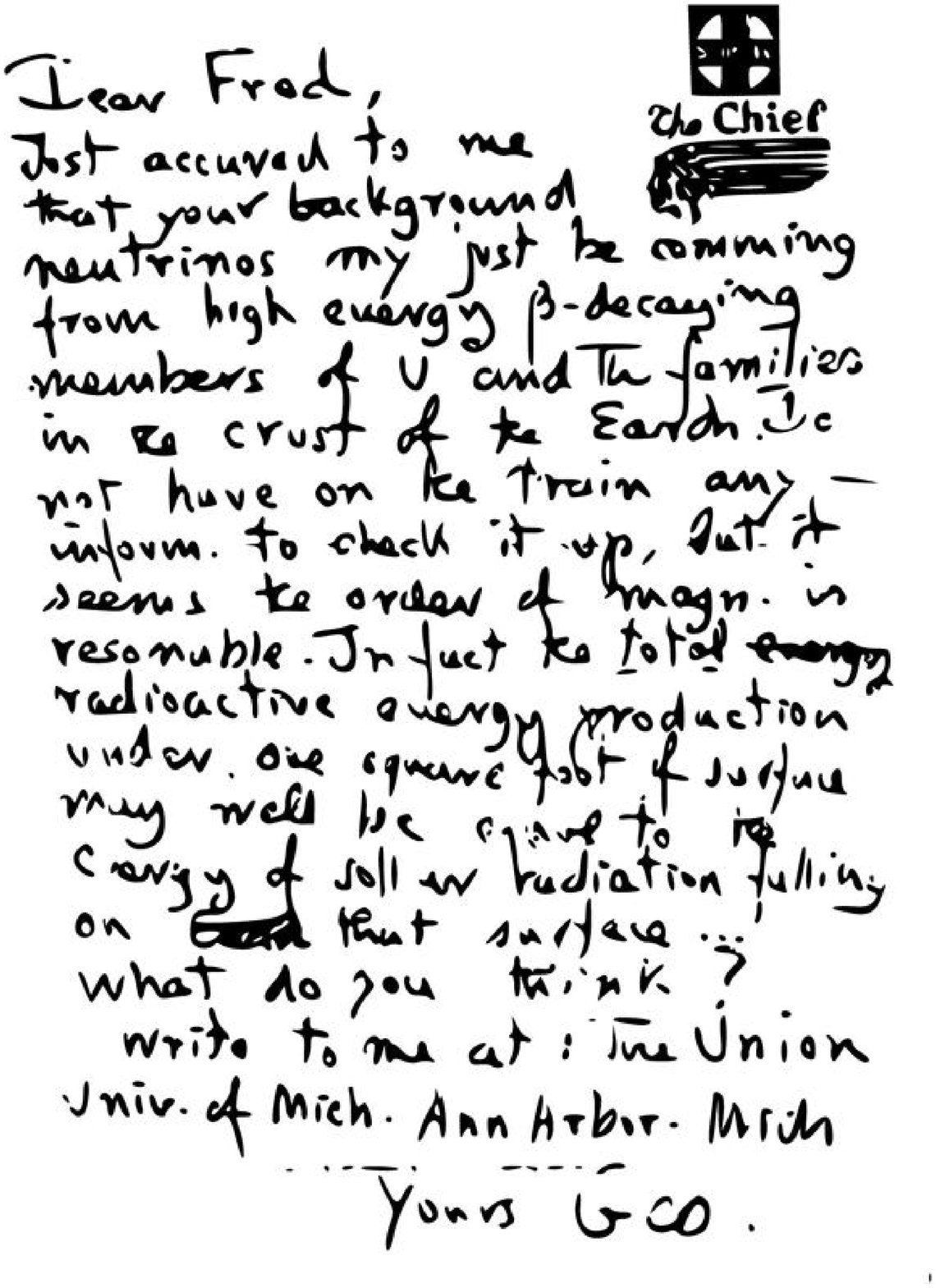}
 \caption[ppp]{The message from Georg Gamow to Fred Reines.}
\label{fig:gamowMessage}
\vspace{3cm}
\includegraphics[width=0.6\textwidth]{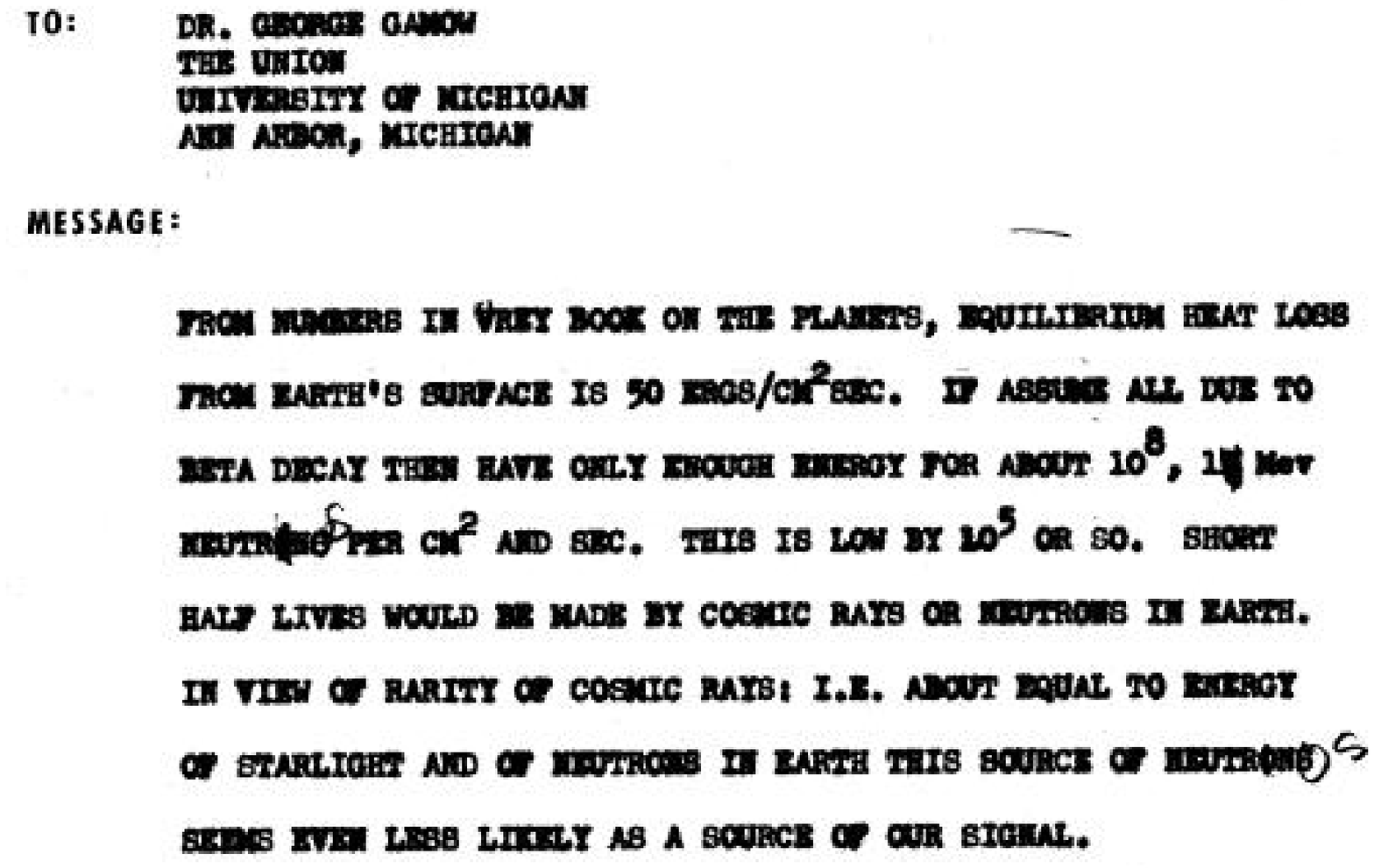}
 \caption[qqq]{The teletype message from Reines to Gamow.}
\label{fig:reinesAnswer}
\end{figure}

Geo-neutrinos have been conceived during the very first attempts of
neutrino detection, performed at the Hanford nuclear reactor by
Reines and Cowan in 1953. Experimental results showed an unexpected
and unexplained background\footnote{Actually the background was due
to cosmic radiation.}. While on board of the Santa Fe Chief Train,
Georg Gamow wrote to Fred Reines (see Fig.~\ref{fig:gamowMessage}):
\begin{quotation}
It just occurred to me that your background may just be coming from high energy
beta-decaying members of \U\ and \Th\ families in the crust of the Earth.
\end{quotation}
The first estimate of geo-neutrino flux was given in a
teletype message by Reines (Fig.~\ref{fig:reinesAnswer}) in response to the letter of Gamow:
\begin{quotation}
Heat loss from Earth's surface is 50~erg cm$^{-2}$ s$^{-1}$. If
assume all due to beta decay than have only enough energy for about
$10^8$ one-MeV neutrinos per cm$^2$ and s.
\end{quotation}

In the scientific literature, geo-neutrinos were introduced by
Eder~\cite{Eder:1966} in the sixties and Marx~\cite{Marx:1969be} soon realized their relevance.
In the eighties Krauss \et discussed their potential as probes of the
Earth's interior in an extensive publication~\cite{Krauss:1983zn}. In the nineties the
first paper on a geophysical journal was published by Kobayashi \et~\cite{Kobayashi:1991}.
Of particular interest, in 1998, Raghavan \et~\cite{Raghavan:1997gw} and
Rothschild \et~\cite{Rothschild:1997dd} pointed out the potential of KamLAND and
Borexino for geo-neutrino detection.

\begin{figure}[hptb]
\includegraphics[width=0.5\textwidth,angle=-90]{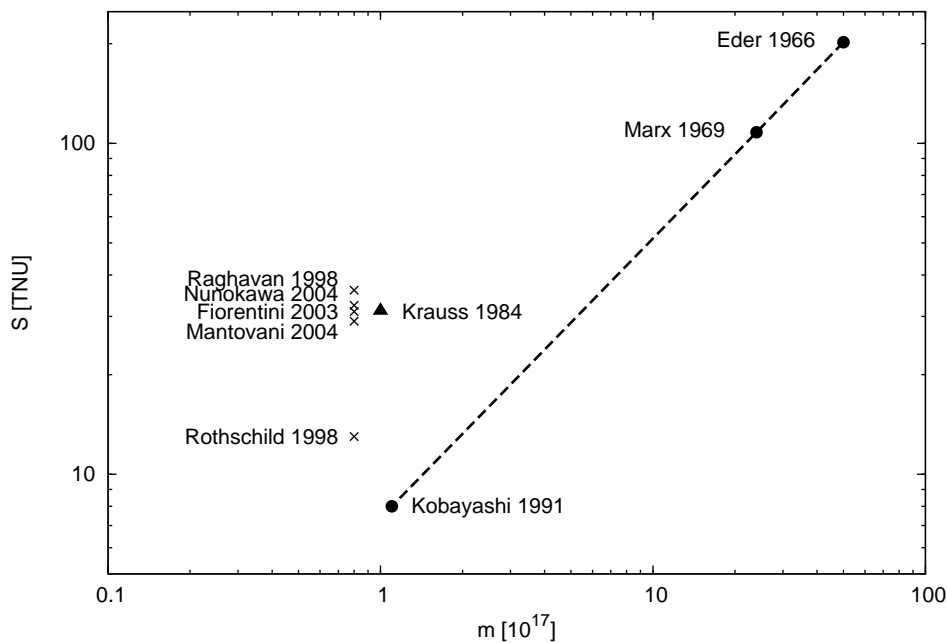}
 \caption[rrr]{Previous estimates of the geo-neutrino signal $S$,
 renormalized to the average survival probability $\langle P_{ee}\rangle = 0.59$,
 and the corresponding estimated uranium mass $m$. The signal is in
 Terrestrial Neutrino Units (1 TNU = 1 event/year/$10^{32}$ proton). From~\cite{Fiorentini:2005cu}.}
\label{fig:History}
\end{figure}

In the last few years more papers appeared than in the previous decades:
in a series of
papers~\cite{Fiorentini:2005mr,Fiorentini:2002bp,Fiorentini:2005zc,Mantovani:2003yd,Fiorentini:2003ww,Fiorentini:2005cu,Fiorentini:2004rj}
Fiorentini \et  discussed the role of geo-neutrinos for determining the
radiogenic contribution to the terrestrial heat flow and for discriminating among
different models of Earth's composition and origin. A reference model for geo-neutrino production,
based on a compositional map of the Earth's crust and on geochemical modeling of the mantle,
was presented in~\cite{Mantovani:2003yd}. Similar calculations were performed by Enomoto \et~\cite{Enomoto:2005mb}
and by Fogli \et~\cite{Fogli:2006}. The claim~\cite{Eguchi:2002dm} of an indication of
geo-neutrino events in the first data release of KamLAND stimulated several theoretical
investigations~\cite{Fogli:2005qa,Eguchi:2002dm,Nunokawa:2003dd,Miramonti:2003hw,Domogatsky:2004be,McKeown:2004yq,Fields:2004tf,Fiorentini:2005ma}.
A summary of the theoretical predictions is presented in Fig.~\ref{fig:History}.
Early models~\cite{Eder:1966,Marx:1969be,Kobayashi:1991} (full circles) assumed a
uniform uranium distribution in the
Earth and different values of the uranium mass. In fact these predictions are almost proportional
to the estimated mass of heat generating elements. The huge signals predicted by Eder and by Marx
were obtained by assuming that the uranium density in the whole Earth is about
the same as that observed in the continental crust; Marx (Eder) assumed thus an uranium
mass 30 (60) times larger than that estimated within the BSE model (see Section~\ref{subsec:radioHeat}).

Krauss \et~\cite{Krauss:1983zn} distributed about $10^{17}$~kg of uranium
uniformly over a 30~km crust. The other estimates (crosses) are all
obtained by using the BSE value for the uranium mass ($\approx 10^{17}$~kg)
as an input and different models for distributing the uranium content between
crust and mantle. In this class, Rothschild \et~\cite{Rothschild:1997dd}
obtained the minimal prediction by assuming for the crust a very small
uranium abundance, definitely lower than the values reported in more recent and detailed estimates.

In July 2005 the KamLAND collaboration presented the first evidence of a signal
truly originating from geo-neutrinos, showing that the technology for geo-neutrino
detection is now available. KamLAND reported~\cite{Araki:2005qa} data from an exposure
of $N_p = (0.346 \pm 0.017) \times 10^{32}$ free protons over a time of 749 days.
In the energy region where geo-neutrinos are expected, there are
152 counts. After subtracting several backgrounds, there remain about
25 true geo-neutrino events. This indicates the difficulties of this experiment:
a signal rate of one geo-neutrino event per month, to be distinguished over a
five times larger background, mostly originating from the surrounding nuclear
power plants. The implication of KamLAND result on radiogenic terrestrial heat
have been discussed in~\cite{Fiorentini:2005ma}.

Following the important KamLAND result, a meeting specifically devoted to
study the potential of geo-neutrinos in Earth's science was gathered at Hawaii
in December 2005~\cite{honolulu:2005}. It provided a first opportunity for a joint
discussion between the communities of particle physics and of geo-science.

In a few years KamLAND should provide definite evidence of the
geo-neutrino signal, after accumulating a much larger statistics and
reducing background. In the meanwhile other projects for
geo-neutrino detection are being developed. Borexino at Gran Sasso,
which is expected to take data soon, will benefit from the absence
of nearby reactors. Domogatski \et~\cite{Domogatsky:2004be} are
proposing a one-kton scintillator detector in Baksan, again very far
from nuclear reactors. A group at the Sudbury Neutrino Observatory
in Canada is studying the possibility of using liquid scintillator
after the physics program with heavy water is
completed~\cite{Chen:2006}. The LENA proposal envisages a 30-kton
liquid scintillator detector at the Center for Underground Physics
in the Pyhasalmi mine (Finland)~\cite{Undagoitia:2006qs}. Due to the
huge mass, it should collect several hundreds of events per year.
The proposal of a geo-neutrino directional detector at Curacao has
been advanced~\cite{deMeijer:2006}.The possibility of a detector
located at Hawaii islands has been presented by Dye~\cite{Dye:2006}.
In conclusion, one can expect that within ten years the geo-neutrino
signal from uranium and thorium will be measured at a few points on
the globe.
\section{\label{sec:radio}Radioactivity in the Earth}
\subsection{A first  look at Earth's interior}
A global look at Earth's interior is useful before entering a
detailed discussion on the element distributions. The amount of
information which we (assume to) have on Earth's interior is somehow
surprising, if one considers that the deepest hole which has ever
been dug is only about twelve kilometers deep.

\begin{figure}[p]
\includegraphics[width=0.6\textwidth]{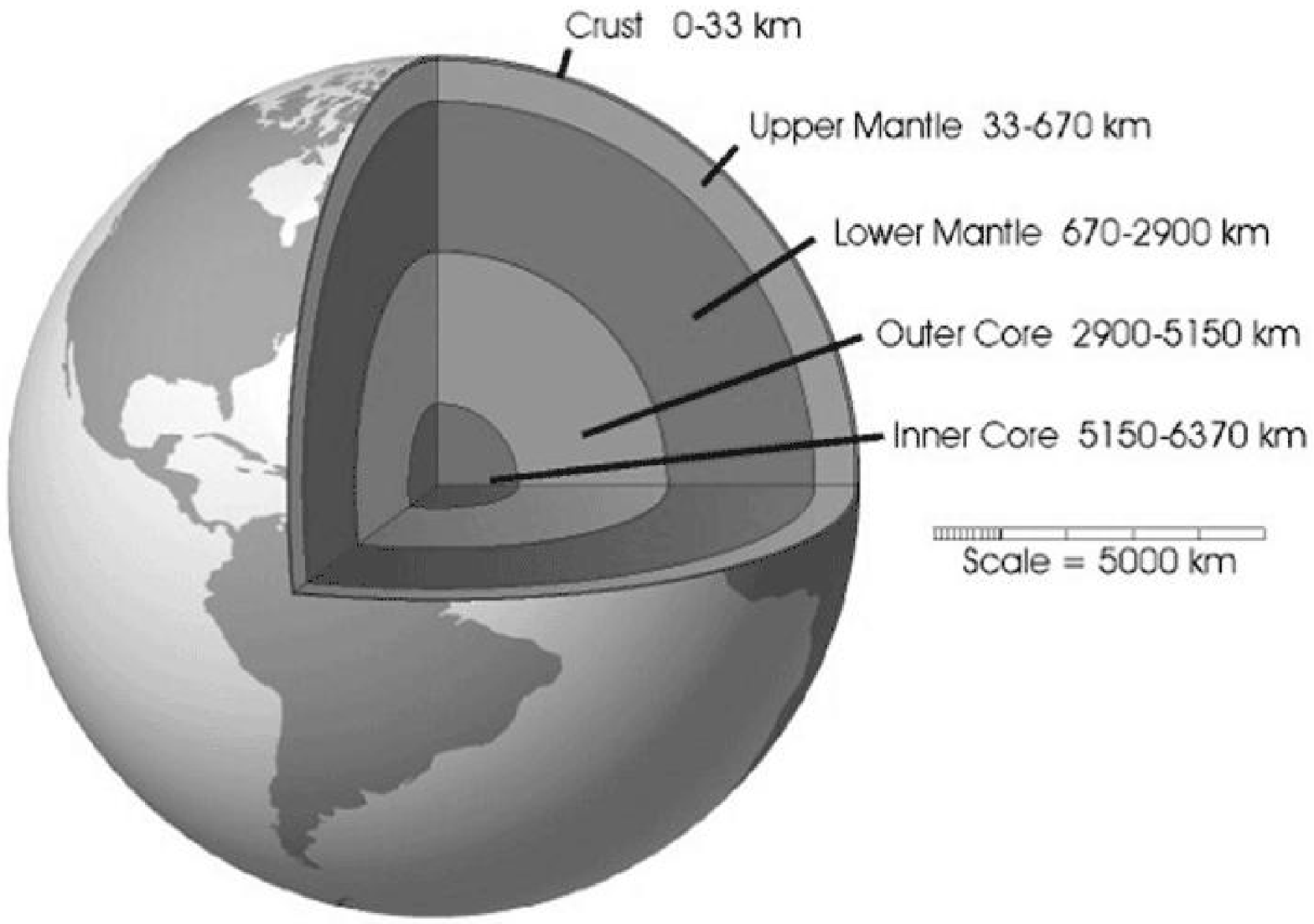}
 \caption[sss]{A sketch of the Earth's interior.}
\label{fig:modelEarth}
\vspace{3cm}
\includegraphics[width=0.5\textwidth,angle=-90]{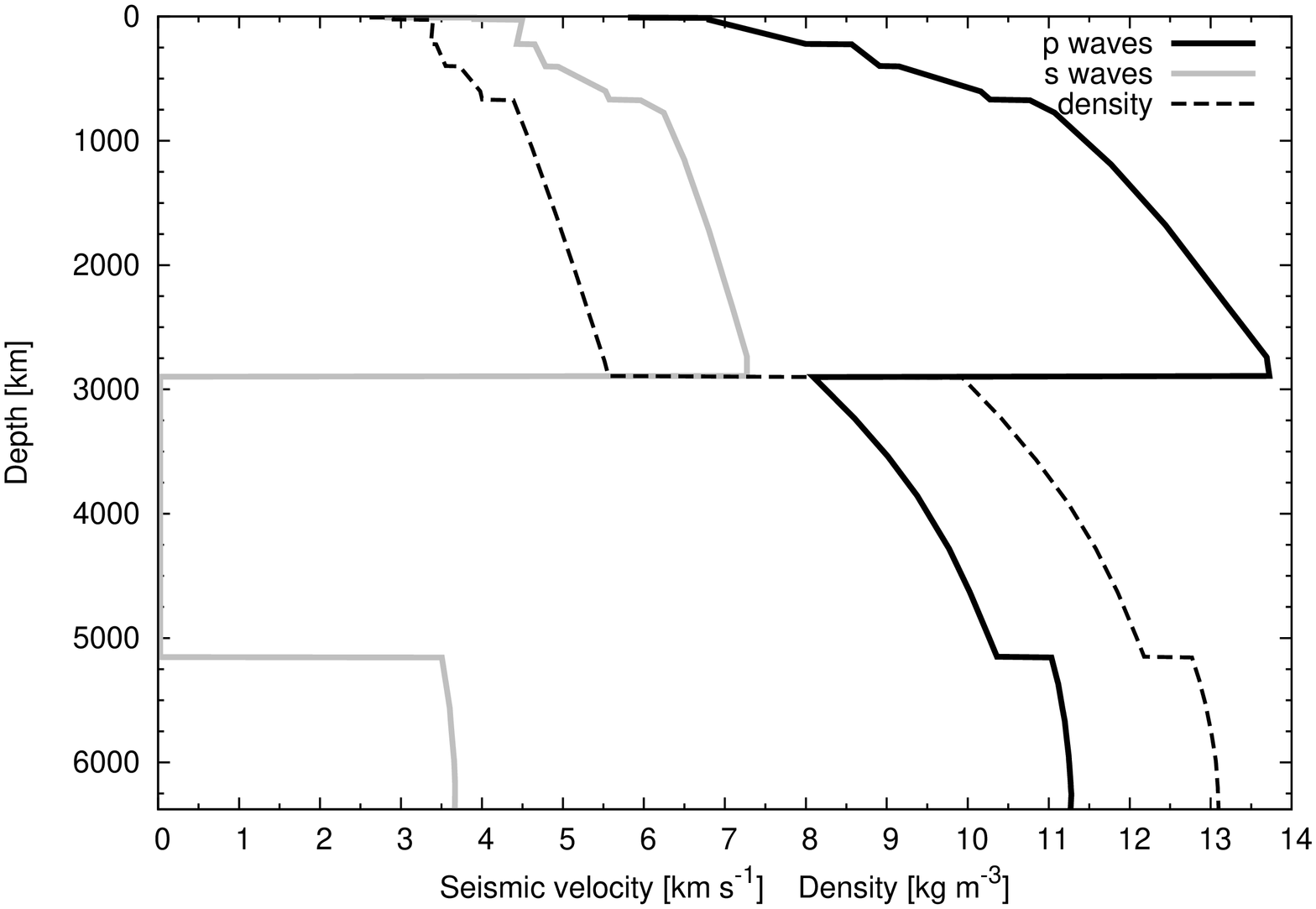}
 \caption[ttt]{PREM (Preliminary Reference Earth
 Model)~\cite{Dziewonski:1981}
  velocity structure trough
 the Earth: $\rho$ = density, $\alpha$ = seismic P-waves velocity, $\beta$ = S-waves
 velocities~\protect\footnote{Figure taken from
 \texttt{http://shadow.eas.gatech.edu/\~{}anewman/classes/geodynamics/random/prem\_earth.pdf}} .}
\label{fig:PREM}
\end{figure}

Seismology has shown that Earth is divided into several layers,
which can be distinguished from discontinuities in the sound speed,
see Figs.~\ref{fig:modelEarth} and \ref{fig:PREM}. The outer layer
is the relatively thin crust which accounts for 0.47\% of the Earth
mass; it is divided in two types, continental crust (CC) and oceanic
crust (OC). The former averages 38~km in thickness, varying around
the globe from 20 to 70 km, and it is made primarily of light
elements such as potassium, sodium, silicon, calcium, and aluminum
silicates. The oceanic crust is much thinner, from about 6 to 8 km.

Inside this crustal skin is Earth's mantle which is 2900~km deep over all.
Largely made up of iron and magnesium silicates, the mantle as a whole accounts for about 68\%
of Earth's mass. One distinguishes the upper mantle (UM) from the lower mantle (LM),
however, the seismic discontinuities between the two parts do not necessarily divide the
mantle into layers. The main questions about the mantle are: does it move as a single layer or
as multiple layers? Is it homogeneous in composition or heterogeneous?
How does it convect? These questions sound simple, but the answers are complex,
possibly leading to more questions, see~\cite{Davies:2002}.

Inside the mantle is Earth's core, which accounts for about 32\% of
Earth's mass. Based on comparison with the behavior of iron at high
pressures and temperatures in laboratory experiments, on the seismic
properties of the core, and on the fact that iron is the only
sufficiently abundant heavy element in the universe, the core is
generally believed to be made primarily of iron with small amounts
of nickel and other elements. Over thirty years ago, however, it was
suggested that a significant amount of potassium could be hidden in
Earth's core, thus providing a large fraction of the terrestrial
heat flow through \K[40] decay. This controversial possibility has
been revived recently in~\cite{RamaMurthy:2003}, see, however,
Ref.~\cite{Corgne:2007}.

Concerning the density profile of our planet, a classical reference
is the Preliminary Reference Earth Model (PREM) of Dziewonski and
Anderson~\cite{Dziewonski:1981}. This one-dimensional spherically
symmetric model is at the basis of all calculations for geo-neutrino
production from the mantle. In the last twenty years seismic
tomography has progressed so as to provide three dimensional views
of the mantle. Density differences with respect to the
one-dimensional model (typically of order of few percent) are most
important for understanding mass circulation inside the mantle;
however, they are too small in order to affect significantly the
calculated geo-neutrino production.

From seismic studies one can derive the density profile of our planet and the aggregation
state of the different layers; however, one cannot reconstruct its composition.

Earth global composition is generally estimated from that of CI chondritic meteorites
by using geochemical arguments which account for loss and fractionation
during planet formation. Along these lines the Bulk Silicate Earth (BSE)
model is built, which describes the element composition of the  ``primitive mantle'',
 \ie the outer portion of the Earth after core separation and before the differentiation
 between crust and mantle, see Table~\ref{tab:compSiliEarth}. The model is believed to
 describe the present crust plus mantle system.
 It provides the total amounts of \U, \Th, and \K\ in the Earth, as these lithophile
 elements should be absent in the core. Estimates from different
 authors~\cite{McDonough:2003} are concordant within 10-15\%, extensive reviews
 being provided in Refs.~\cite{Palme:2003,McDonough:1995}. From the mass, the present
 radiogenic heat production rate and neutrino luminosity can be immediately
 calculated by means of Eqs.~(\ref{eq:heatNatural}), (\ref{eq:heatIsoto}), (\ref{eq:lumNatural})
 and (\ref{eq:lumIsoto}), and are shown in the Table~\ref{tab:UthKbse}.

\begin{table}[htb] \caption[uuu]{The composition of the silicate Earth.
Abundances are given in  $\mu\mathrm{g}\cdot \mathrm{g}^{-1}$ (ppm),
unless stated as ``\%'' which are given in weight percentage. Data
from~\cite{McDonough:2003}. \label{tab:compSiliEarth} }
\renewcommand{\tabcolsep}{1pc} 
\begin{tabular}{llllll}
 \hline\hline
H   &   100 &   Zn  &   55  &   Pr  &   0.25    \\
Li  &   1.6 &   Ga  &   4   &   Nd  &   1.25    \\
Be  &   0.07    &   Ge  &   1.1 &   Sm  &   0.41    \\
B   &   0.3 &   As  &   0.05    &   Eu  &   0.15    \\
C   &   120 &   Se  &   0.075   &   Gd  &   0.54    \\
N   &   2   &   Br  &   0.05    &   Th  &   0.1 \\
O (\%)  &   44  &   Rb  &   0.6 &   Dy  &   0.67    \\
F   &   15  &   Sr  &   20  &   Ho  &   0.15    \\
Na (\% )    &   0.27    &   Y   &   4.3 &   Er  &   0.44    \\
Mg (\%) &   22.8    &   Zr  &   10.5    &   Tm  &   0.068   \\
Al (\%) &   2.35    &   Nb  &   0.66    &   Yb  &   0.44    \\
Si (\%) &   21  &   Mo  &   0.05    &   Lu  &   0.068   \\
P   &   90  &   Ru  &   0.005   &   Hf  &   0.28    \\
S   &   250 &   Rh  &   0.001   &   Ta  &   0.037   \\
Cl  &   17  &   Pd  &   0.004   &   W   &   0.029   \\
K   &   240 &   Ag  &   0.008   &   Re  &   0.0003  \\
Ca (\% )    &   2.53    &   Cd  &   0.04    &   Os  &   0.003   \\
Sc  &   16  &   In  &   0.01    &   Ir  &   0.003   \\
Ti  &   1200    &   Sn  &   0.13    &   Pt  &   0.007   \\
V   &   82  &   Sb  &   0.006   &   Au  &   0.001   \\
Cr  &   2625    &   Te  &   0.012   &   Hg  &   0.01    \\
Mn  &   1045    &   I   &   0.01    &   Tl  &   0.004   \\
Fe (\%) &   6.26    &   Cs  &   0.021   &   Pb  &   0.15    \\
Co  &   105 &   Ba  &   6.6 &   Bi  &   0.003   \\
Ni  &   1960    &   La  &   0.65    &   Th  &   0.08    \\
Cu  &   30  &   Ce  &   1.68    &   U   &   0.02    \\
   \hline \hline
\end{tabular}\\[2pt]
\end{table}

\begin{table}[htb] \caption[vvv]{Mass, heat production, and geo-neutrino luminosity of \U, \Th, \K[40]
according to BSE.\label{tab:UthKbse} }
\newcommand{\dg}{\hphantom{$0$}}
\renewcommand{\tabcolsep}{1pc} 
\begin{tabular}{cccc}
 \hline\hline
           &  $m$             &   $H_R$           &   $L_{\anu}$             \\
           &  [$10^{17}$~kg]  &    [$10^{12}$~2]  & [$10^{24}$~s$^{-1}$]   \\
 \hline
 \U        &    0.8           &         8.0       &       \dg6.2            \\
 \Th       &    3.2           &         8.7       &       \dg5.3            \\
 \K[40]    &    1.1           &         3.6       &         26.5            \\
   \hline \hline
\end{tabular}\\[2pt]
\end{table}

The BSE is a fundamental geochemical paradigm. It is consistent with most
observations, which however regard the crust and the uppermost portion of the mantle only.
Its prediction for the present radiogenic production is 19~TW.

Concerning the distribution of heat generating elements, estimates for
uranium in the (continental) crust based on observational data are in the range:
\begin{equation}
    m_C(U) = (0.3\div 0.4)\times 10^{17}\mathrm{\ kg} \quad .
\end{equation}
The crust -- really a tiny envelope -- should thus contain about one half of
uranium in the Earth. For the mantle, observational data are scarce and restricted
to the uppermost part, so the best estimate for its uranium
content $m_M(U)$ is obtained by subtracting the crust contribution from
the BSE estimate:
\begin{equation}
    m_M(U) = m_{\mathrm{BSE}}(U) -  m_{C}(U)\quad .
\end{equation}
We remark that this estimate is essentially based on a cosmo-chemical argument and
there is no direct observation capable of telling how much uranium is in the mantle,
and thus on the whole Earth.

Similar considerations hold for thorium and potassium, the relative mass abundance with
respect to uranium being globally estimated as
$a(\Th)$ : $a(\U)$ : $a(\K) \approx $ 4 : 1 : 12000. Geochemical arguments are against the
presence of radioactive elements in the (completely unexplored) core, as
discussed by McDonough in an excellent review of compositional models of the
Earth~\cite{McDonough:2003}.

A comprehensive review about the knowledge of Earth's interior is given in
volumes 2 and 3 of~\cite{Holland:2003}.

The following subsections are devoted to present, in some more detail,
the available information on the amounts of heat generating elements in the whole
Earth and within its separate layers.

\subsection{The BSE model and heat generating elements in the interior of the Earth}
In the BSE frame, the amount of heat/neutrino generating material inside
Earth is determined through the following steps:
\begin{enumerate}
  \item[(a)]
  From the compositional study of selected samples emerging from the mantle,
after correcting for the effects of partial melting, one establishes the absolute primitive
abundances in major elements with refractory and lithophile character, \ie elements with high
condensation temperature (so that they do not escape in the processes leading to Earth formation)
and which do not enter the metallic core. In this way primitive absolute abundances of elements such
as Al, Ca and Ti are determined, a factor about 2.8 times CI chondritic abundances.
  \item[(b)]
  It is believed, and supported by studies of mantle samples, that refractory lithophile elements
  inside Earth are in the same proportion as in chondritic meteorites. In this way, primitive abundances
  of \Th\ and \U\ can be derived by rescaling the chondritic values.
  \item [(c)]
  Potassium, being a moderately volatile elements, could have escaped in the planetesimal
  formation phase. Its absolute abundance is best derived from the practically constant mass
  ratio with respect to uranium observed in crustal and mantle derived rocks.
\end{enumerate}
There are several calculations of element abundances in the BSE model, all consistent with each
other to the level of 10\%. By taking the average of results present in the literature,
in~\cite{Mantovani:2003yd} the following values were
adopted\footnote{We shall always refer to element abundances in mass and we remind the reader
that the natural isotopic composition is $\U[238]/\U = 0.993$, $\Th[232]/\Th = 1$, and $\K[40]/\K = 1.2 \times 10^{-4}$}:
for the uranium abundance $a_{\mathrm{BSE}}(\U) = 2 \times 10^{-8}$, for the ratio of elemental
abundances $\Th/\U \equiv a_{\mathrm{BSE}}(\Th)/a_{\mathrm{BSE}}(\U) = 3.9$,
 and $\K/\U \equiv a_{\mathrm{BSE}}(\K)/a_{\mathrm{BSE}}(\U) = 1.14\times 10^{4}$.
 For a comparison, a recent review~\cite{Palme:2003}  -- subsequent to~\cite{Mantovani:2003yd} --
 gives
$a_{\mathrm{BSE}}(\U) = 2.18 \times 10^{-8}$, $\Th/\U = 3.83 $, and
$K/U = 1.2 \times 10^4$.
\subsection{The crust}
Earth is the only planet, in our solar system, that has both liquid water and a topographically
bimodal crust, consisting of low-lying higher-density basaltic oceanic crust (OC) and
high-standing lower-density andesitic continental crust (CC)~\cite{Rudnick:1995}.

Although the continental crust is insignificant in terms of mass (about half of a percent
of the total Earth), it forms an important reservoir for many of the trace elements on our planet,
including the heat producing elements. It also provides us with a rich geologic history:
the oldest dated crustal rocks formed within 500~Ma (million years) of Earth accretion,
whereas the oceanic crust records only the last 200~Ma of Earth history.

The crust extends vertically from the Earth's surface to the
Mohorovicic (Moho) discontinuity, a jump in compressional wave
speeds from $\approx 7$ to $\approx 8$~km/s which occurs, on the
average, at a depth of $\approx 40$~km for the continental crust and
at a depth of about $ 8$~km for the oceanic crust.

The Conrad discontinuity  separates the continental crust into two
parts. Actually, based on additional seismic information several
authors, \eg \cite{Rudnick:1995}, identify three components in the
crust, the upper-, middle-, and lower-crustal layers (which we shall
refer to as UC, MC and LC, respectively). The upper crust is readily
accessible to sampling and robust estimates of its composition are
available for most elements, whereas the deeper reaches of the crust
are more difficult to study, so that the estimated elemental
abundances are more uncertain. The observations show that the crust
becomes more mafic\footnote{Mafic is used for silicate minerals,
magmas, and rocks which are relatively high in the heavier elements.
The term is derived from ``magnesium,, and ``ferrum'' (Latin for
iron), but mafic magmas are also relatively enriched in calcium and
sodium. } with depth and the concentration of heat producing
elements drops rapidly from the surface downwards. Not only the
crust is vertically stratified in terms of its chemical composition,
but it is also heterogeneous from place to place. This makes it
difficult to determine the average composition of such a
heterogeneous mass.

\subsubsection{Abundances of heat generating elements}
For each component of the crust, one has to adopt a value for the
abundances\footnote{Throughout the paper the term abundance refers
to abundance in mass.} $a(\U)$, $a(\Th)$, and $a(\K)$ and to
associate it with an uncertainty. In the literature of the last
twenty years one can find many estimates of abundances for the
various components of the crust (upper, middle, lower crust = UC,
MC, LC and oceanic crust = OC), generally without an error value,
two classical reviews being~\cite{Taylor:1995,Wedepohl:1995}.
Average elemental abundances in the continental crust, and their
vertical distribution in the three main identifiable layers have
been presented in a recent comprehensive review~\cite{Rudnick:2003},
together with a wealth of data and with a critical survey of earlier
literature on the subject. A most useful and continuously updated
source is provided by the GERM Reservoir database.
Table~\ref{tab:UabundCompare} presents the uranium abundances used
for geo-neutrino calculation in a few studies.

\begin{table}[htb] \caption[www]{Uranium mass abundances in the Earth's
reservoirs and in the silicate Earth (BSE) used in recent
geo-neutrino studies. Units are $\mu g/g$ (ppm) for the crust, $n
g/g$ (ppb) otherwise.\label{tab:UabundCompare} }
\renewcommand{\tabcolsep}{1pc} 
\begin{tabular}{cccccccc}
& & \multicolumn{3}{c}{Mantovani \et} &\multicolumn{2}{c}{Fogli \et}& Enomoto \\
& & \multicolumn{3}{c}{\cite{Mantovani:2003yd}} &\multicolumn{2}{c}{\cite{Fogli:2005qa}} & \cite{Enomoto:2005}\\
Reservoir & Units & Adopted value & $\frac{(a_{\mathrm{max}}-a_{\mathrm{min}})}{2}$ & $\sigma$ & Adopted value &  $ \sigma $&  Adopted value  \\
 \hline\hline
   UC & ppm & 2.5  &  0.3 & 0.13  & 2.7  & 0.6  &  2.8  \\
   MC & ppm & 1.6  &   -  &   -   & 1.3  & 0.4  & 1.6  \\
   LC & ppm & 0.63 & 0.45 & 0.23  & 0.2  & 0.08 & 0.2  \\
   OC & ppm & 0.1  &  -   &   -   & 0.1  & 0.03 & 0.1  \\
   UM & ppb & 6.5  & 1.5  & 1.5   & 3.95 & 1.2  & 12  \\
  BSE & ppb & 20   & 2.5  & 1.0   & 17.3 & 4.7  & 20  \\
   \hline \hline
\end{tabular}\\[2pt]
\end{table}

Earlier paper on geo-neutrinos adopted abundances from some review papers,
without tackling the problem of the associated uncertainties. A similar approach
is taken in the recent paper by Enomoto \et~\cite{Enomoto:2005mb} where the values
from~\cite{Rudnick:1995} are used directly, without any estimate of the associated uncertainties.

Our group adopted as reference values for the abundances the average
$\langle a \rangle$ of values which were available in the
GERM database\footnote{Geochemical Earth Reference Model (GERM) available
online at {\tt http://earthref.org/GERM/}.} in 2003, considering the
spread of the reported abundances
$(a_{\mathrm{max}}-a_{\mathrm{min}})/2$ as indication of the
corresponding uncertainty. In Table~\ref{tab:UabundCompare} we also
presents the standard deviation of the average:
\begin{equation}
    \sigma = \sqrt{\frac{1}{N(N-1)}\sum_{i=1}^N \left( a_i - \langle a\rangle\right)^2} \quad .
\end{equation}
Fogli \et~\cite{Fogli:2005qa} basically adopt the results of a
recent and comprehensive review~\cite{Rudnick:2003} for the UC, MC,
LC abundances of (\U, \Th, \K) and the uncertainties quoted in that
paper. Since no error estimates are given in~\cite{Rudnick:2003} for
the LC, Fogli \et  assume a fractional $1\sigma$ errors of 40\%. A
more extensive comment on uncertainties  can be found in
Appendix~\ref{sec:appenErrors}.

From the table one sees that the values adopted by different groups
are generally in agreement within the quoted uncertainties, however
with the following remarks:
\begin{itemize}
  \item
   A major difference lies in the abundance for the (poorly constrained) lower portion
  of the crust. In the literature, values as low as 0.2~ppm and as high as 1.1~ppm have been reported,
  as a consequence of different assumptions for the fraction of (uranium poor)
  metaigneous rocks and (uranium rich) metapelitic rocks.
  \item
  Concerning the upper crust, the values quoted by different authors using different methods (surface
  exposure data, sedimentary data and loess correlations with La) are consistent within
  about 10\%. From a study of loess\footnote{Loess is a deposit of silt
  (sediment with particles 2-64 microns in diameter) that have been laid down by wind action.}
   correlations with La, in~\cite{Rudnick:2003}
 a concordant average value has being obtained, however with a $1\sigma$ uncertainty of 21\%
 mainly due to the variability of the \U/La correlation.
\end{itemize}
From the table, it emerges that the contributions to geo-neutrino production from different
portions of the Earth's crust are markedly different, the continental crust being an
order of magnitude richer in heat generating material than the oceanic part.
Relative uncertainties, as natural, increase with depth and their
assessment is at the moment somehow tentative.

\subsubsection{The distribution of heat generating elements}
The earlier geo-neutrino studies considered the distribution of heat generating elements
as spherically symmetrical over the Earth's crust. Actually one has to distinguish
between continental and oceanic crust since they have quite different contents of heat
generating elements. In addition the thickness of the crust significantly differs from place
to place. More recent studies, since~\cite{Rothschild:1997dd}, take into account
the actual inhomogeneity of Earth crust.

A global crustal model on a $2^{\circ}\times 2^{\circ}$  degree grid, available
at~\cite{Laske:2001}, has been widely used in recent years. Data
gathered from seismic experiments were averaged globally for similar
geological and tectonic settings (such as Archean, early
Proterozoic, rifts, etc.). The sedimentary thickness is based on the
recent compilation by~\cite{Laske:2001,Bassin:2000}. Bathymetry and
topography is that of ETOPO5.

Within each $2^{\circ}\times 2^{\circ}$
 degree tile, one distinguishes oceans and seawater,
the continental crust, subdivided into three sub-layers (upper, middle, and lower),
sediments and oceanic crust.
For all these layers values of density and depth are given over the
globe\footnote{Note that additional useful databases are available at \texttt{http://mahi.ucsd.edu/Gabi/rem.html}.}.

\subsection{The mantle: data, models and debate}
\label{subsec:mantleDataModels}
Sandwiched between Earth's crust and metallic core, the mantle is
a 2900~km layer of pressurized rock at high temperature.
As reviewed by Hofmann in~\cite{Hofmann:1997,Hofmann:2003}, mantle models can be
divided into two broad classes, essentially corresponding to the presently contradictory
geochemical and geophysical evidence of Earth's interior.

\subsubsection{Geochemical and geophysical evidences}
\begin{figure}[htbp]
\includegraphics[width=0.6\textwidth]{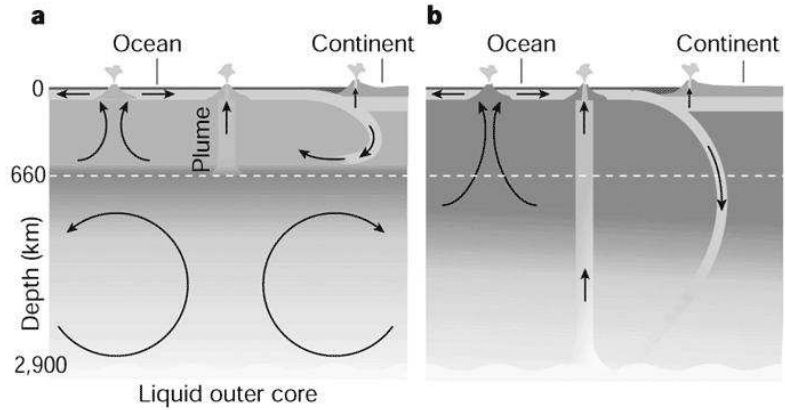}
\nobreak\caption[yyy]{Models
of mantle circulation, adapted from~\cite{Hofmann:2003}:
(a) is the traditional two-layer model with demarcation at 670 km and
nearly complete isolation between upper and lower layers; (b) is a fully mixed model.}
\label{fig:mantleCircul}
\end{figure}

Geochemists have long insisted on a two-layer model, in which the mantle
consists of a relatively primitive layer below a depth of about 670~km and an upper layer
that is highly depleted of heat producing elements (panel (a) in Fig.~\ref{fig:mantleCircul}).
The two layers are viewed as separate sources of the Mid-Ocean-Ridge Basalts (MORB),
which come from mantle regions that have been already depleted in incompatible elements by
extraction of the continental crust, and of Ocean Island Basalts (OIB),
which form by melting of deeper, less depleted or even enriched mantle sectors.
Also, a more primitive deep layer is needed from global constraints, otherwise the amount of radiogenic
elements inside Earth is much too small with respect to that estimated within the Bulk Silicate Earth (BSE) paradigm.

On the other hand, over the past several years seismic tomography has provided
increasingly detailed images of apparently cold slab descending into the deep mantle,
below the 670~km boundary. If cold slabs descend into the deep mantle,
there must be a corresponding upward flow of deep-mantle material to
shallow levels (panel (b) in Fig.~\ref{fig:mantleCircul}). If this circulation reaches
the bottom of the mantle (whole mantle convection), it would destroy any
compositional layering below the crust in a few hundred million years
(at a typical speed of 3~cm~yr$^{-1}$ it takes about $10^8$~yr to move down to 2900 km).

In brief, the composition and circulation inside Earth's mantle is the subject of a
strong and so far unresolved debate between geochemists and geophysicists.
Geochemical evidence has been used to support the existence of two compositionally distinct
reservoirs in the mantle, the borders between them being usually placed at a depth
near $h_0 = 670$~km, whereas geophysics presents evidence of mantle convection
extending well beyond this depth. If this convection involves the whole mantle,
it would have destroyed any pre-existing layering, in conflict with geochemical evidence.

When building their respective reference models for geo-neutrino production,
our group in~\cite{Mantovani:2003yd}, as well as Fogli \et~\cite{Fogli:2005qa},
used a two-reservoir mantle model. Observational values are adopted for the upper
mantle whereas the lower mantle abundances are inferred from the BSE mass balance constraint.
On the other hand, Enomoto \et~\cite{Enomoto:2005mb} prefer a wholly mixed mantle, with uniform abundance
within it derived from the BSE constraint, see Table~\ref{tab:UabundCompare}.

\subsubsection{A class of two-reservoir models}
More generally, new views on mantle convection models overcome the widely diffused model
of two-layer mantle convection, namely an outgassed and depleted upper layer overlying
a deeper, relatively primordial and undegassed mantle layer.
The ensemble of geochemical and geophysical evidence along with terrestrial heat flow-heat
production balance argues against both whole mantle convection and layering at 670~km depth models,
suggesting the existence of a transition between the two reservoirs (outgassed and depleted -- degassed and primordial)
at 1600-2000 km depth~\cite{Hilst:1999,Kellogg:1999,Albarede:1999}.
In the numerical simulation of their mantle convection model, Kellogg \et~\cite{Kellogg:1999}
located this boundary at a depth of about 1600~km.

\begin{figure}[p]
\includegraphics[width=0.3\textwidth]{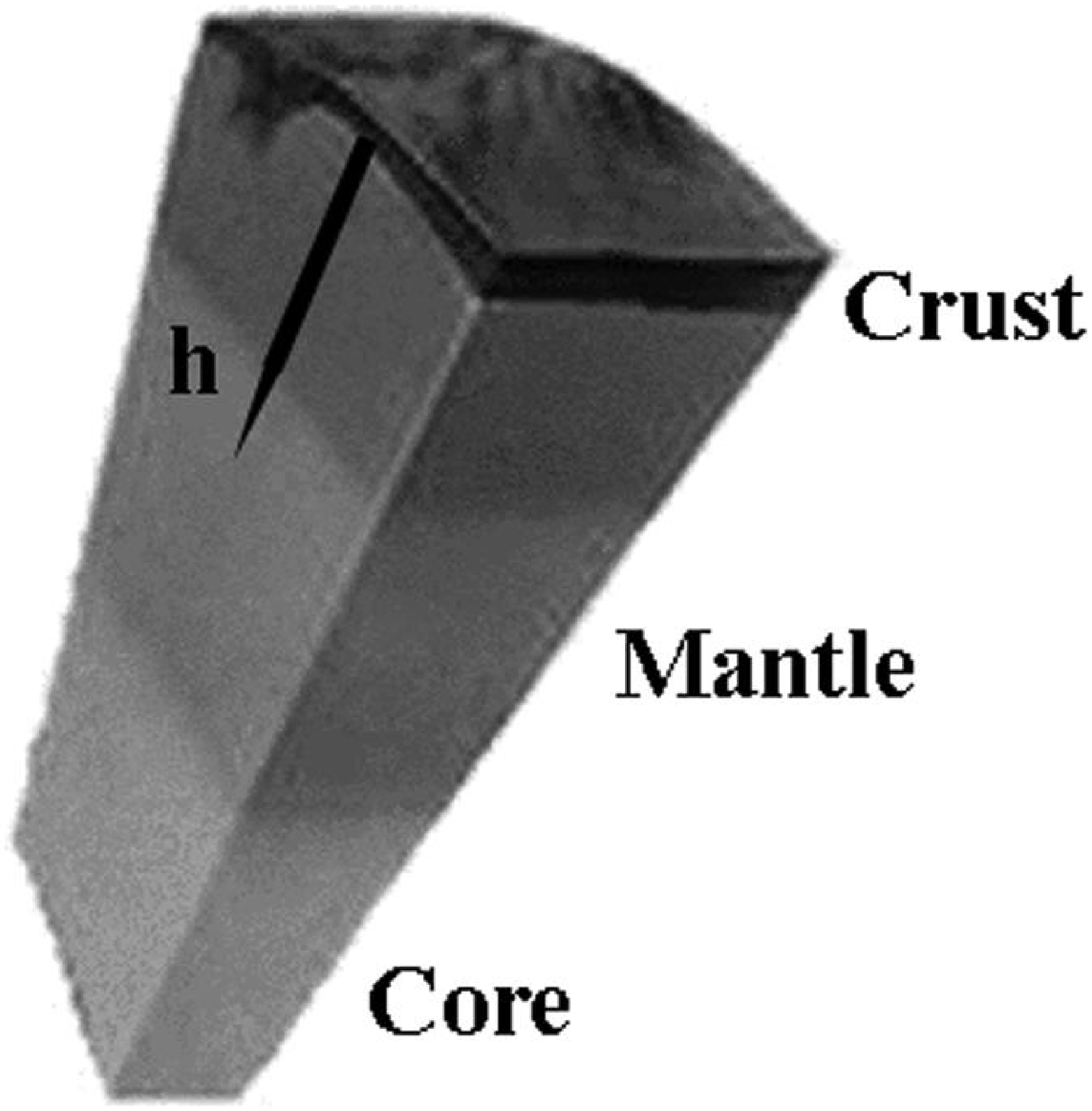}
 \caption[zzz]{Generic two-reservoir mantle model: the critical depth $h$ is a free parameter.}
\label{fig:twoReservModel}
\vspace{4cm}
\includegraphics[width=0.5\textwidth,angle=-90]{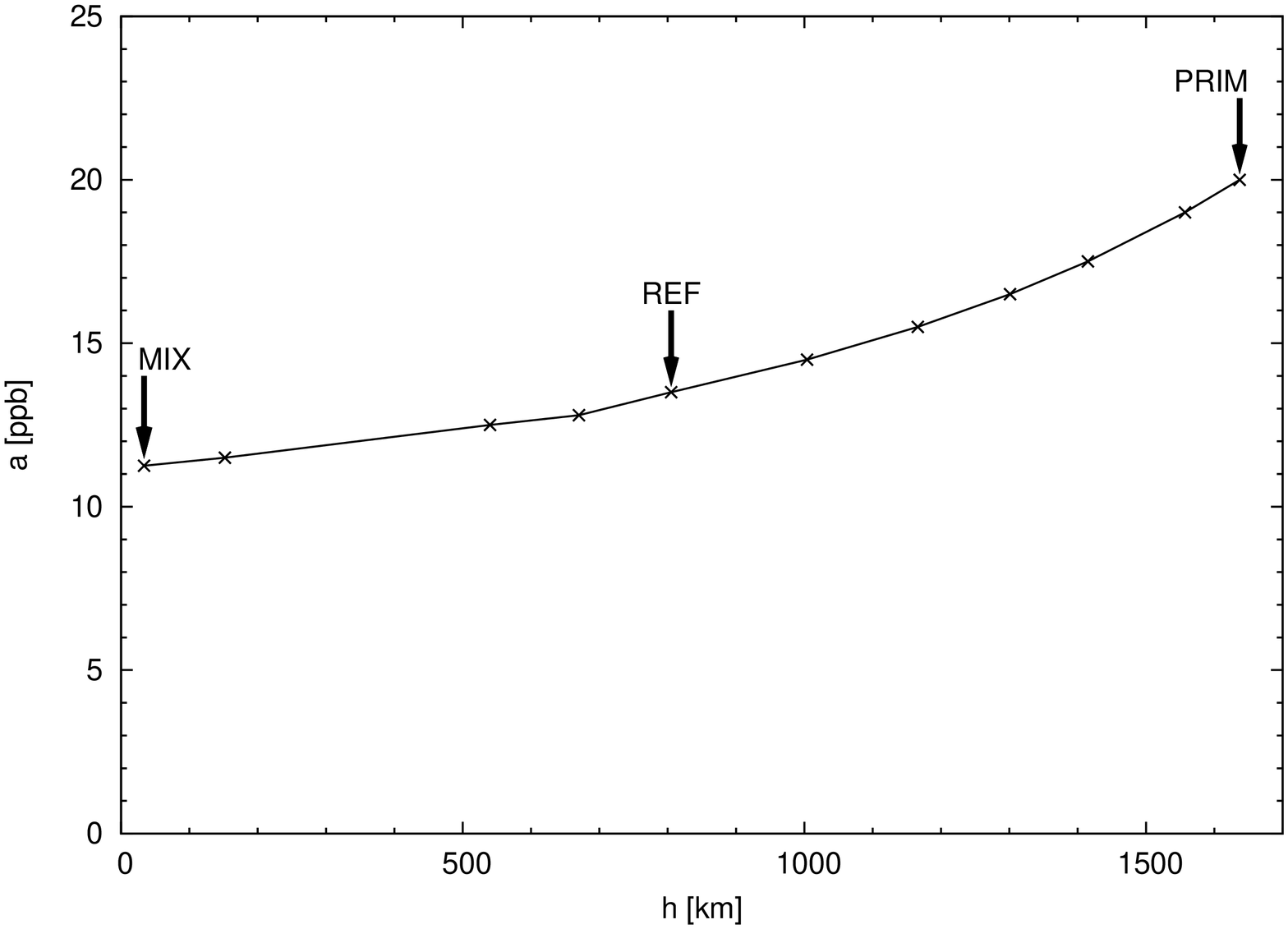}
 \caption[abb]{Uranium abundance in the lower part of the mantle as a
 function of the critical depth $h$ from Earth's surface, from~\cite{Fiorentini:2005mr}.
 } \label{fig:lowMantleUabu}
\end{figure}

In order to consider the implications of the present  debate on
mantle circulation and composition on the predicted geo-neutrino
fluxes, our group has also considered~\cite{Fiorentini:2005mr} the
uranium distribution in a wider class of models, including the
extreme geochemical and geophysical models, in terms of just one
free parameter, the depth $h$ marking the borders between the two
hypothetical reservoirs (see Fig.~\ref{fig:twoReservModel}):
\begin{itemize}
  \item[(i)]
   above $h$ one assumes uniform uranium abundance in the
     range from 2 to 7.1 ppb, as deduced from measures of the depleted upper mantle.
  \item[(ii)]
   Below $h$ one assumes an uniform abundance, determined by requiring mass balance
  for the whole Earth. This means that uranium mass below the critical depth, $m_{>h}$,
  is obtained by subtracting from the total BSE estimated mass ($m_{\mathrm{BSE}}$)
  the quantity observationally determined in the crust ($m_C$) and that contained in
  the mantle above $h$ ($m_{<h}$):
\end{itemize}
\begin{equation}
    m_{>h} = m_{\mathrm{BSE}} - m_C - m_{<h} \quad .
\end{equation}

For each model, the calculated uranium abundance in the lower
 portion of the mantle is shown in Fig.~\ref{fig:lowMantleUabu}.

This class of models includes a fully mixed mantle (MIX), which is obtained
for $h = 25$~km (\ie just below a mean crust thickness obtained averaging the vales for continental
and oceanic crust) so that the strongly impoverished mantle has a negligible thickness.
The traditional geochemical model (REF) corresponds to $h = h_0$. As $h$ increases,
the depleted region extends deeper inside the Earth and -- due to mass balance --
the innermost part of the mantle becomes richer and closer in composition to the primitive mantle.

Among all possible models, the case $h = 1630$~km is particularly interesting.
Below this depth the resulting uranium abundance is 20~ppb, corresponding to the BSE estimate.
The innermost part of the mantle is thus primitive in its trace element composition and the
crust enrichment is obtained at expenses of the mantle content above $h$.
Again following~\cite{Fiorentini:2005mr}, we shall refer to this model as PRIM.

Concerning geo-neutrino fluxes from the mantle, all the models proposed above have the
same amount of heat/anti-neutrino sources and only the geometrical distribution is varied.
The largest flux corresponds to the model with sources closest to the surface, \ie to the MIX model.
On the other hand, the minimal prediction is obtained when the sources are concentrated at
larger depth, which corresponds to the PRIM case\footnote{This is a part of the
\emph{proximity argument}, which we shall exploit extensively later, see Section~\ref{sec:beyond}.}.

\section{\label{sec:heat}Terrestrial heat}
Earth emits a tiny heat flux with an average value $q \approx
(60\div90)$ mW/m$^2$, definitely smaller than the radiation coming
from the Sun, $ 1.4$~kW/m$^2$, larger, however, than the energy
deposited by cosmic rays, about $ 10^{-8}$~W/m$^2$. When integrated
over the Earth's surface, the tiny flux translates into a huge heat
flow, $H \approx (30\div45)$~TW, the equivalent of ten thousand
nuclear power plants. In this section we briefly review the
estimates of terrestrial heat flow and discuss the sources which can
sustain it.
\subsection{Heat flow from the Earth: data and models}
A frequently quoted value for the total heat release from Earth's
surface is the estimate by Pollack \et~\cite{Pollack:1993}, $H =
44.2 \pm 1.0$~TW. In spite of the small quoted error, uncertainties
seem to be much bigger: a recent revisitation of the problem by
Hofmeister and Criss~\cite{Hofmeister:2005} yields $31 \pm 1$~TW
with a central value close to that quoted in the seventies.

The heat flux $q$ is determined by using the conduction law (if one
assumes that conduction is the main mechanism for heat transport):
one measures the temperature gradient $\nabla T$ in near-surface
rocks and their thermal conductivity $k$, and derives:
\begin{equation}
\label{eq:gradT}
    q = - k\nabla T \quad .
\end{equation}
In this way one obtains a mean heat flux of 65~mW/m$^2$ from the
continents. This commonly agreed value yields a continental
contribution to terrestrial heat flow of 13~TW. Direct measurements
from  the oceanic crust give comparable values for the flux, so
that, when weighted with the surface, the oceanic contribution would
be about 20~TW. By summing these contributions one finds a total
heat flow of about 30~TW, a value commonly found in the \emph{old}
literature.

However, the estimate of the actual heat flow for the ocean is more difficult
and controversial. The point is that the heat flow determined from Eq.~(\ref{eq:gradT})
is a lower bound for its actual value, since in porous and permeable rocks heat
may be also carried out by convective flow of interstitial fluid (water).
This convection lowers the temperature gradient below the value it would have
if the rocks were dry or impervious to water. On these grounds, other attempts
have been developed for estimating the oceanic heat flow.

Instead of using the distribution of conductive heat measurements,
Pollack \et have based the estimate of the heat flux from the oceanic crust
on theoretical \emph{thermal models}, such as the Half Space Cooling (HSC)
model and its variants. The models aim at a description of both ocean
depths and heat flow versus age data. In the HSC model, as an example,
depth and heat flow vary as the square root of age and the reciprocal
of the square root of age, respectively. The models reproduce data at
a semi-quantitative level; however, the predicted heat fluxes come
out to be larger than those provided by the conductive heat measurements,
particularly for very young ages. As a result, the mean heat flow from the
oceans amounts to 101~mW/m$^2$, with the oceans giving a contribution
to the terrestrial heat flow of about 31~TW, so that the total flow is near 45~TW.

Note that with respect to heat flow inferred from conductive heat measurements in the ocean,
there is a difference of 11~TW which is attributed to hydrothermal flow. As already mentioned,
however, this procedure has been criticized by Hofmeister and Criss, with their paper
opening a debate, see~\cite{vonHerzen:2005,Hofmeister:2005b}.

In conclusion, it seems to us that for the global heat flow 30~TW is a sound lower limit
based on direct observations, whereas 45~TW is a reasonable upper limit, as it corresponds
to the highest estimate available in the literature.

\subsection{Energy sources}

Coming to the sources of heat flow, the situation is even more complex.
A comparison between the Sun and Earth energy inventories may be useful for illustrating the
differences between the two cases and for appreciating the difficulties when discussing Earth's energetics.
Clearly, a heat flow $H$ can be sustained for a time $t$ provided that an energy source of
at least $U=H\times t$   is available. For the Sun $U_{\odot}=H_{\odot} t_{\odot}\approx 5\times 10^{43}$~J:
clearly neither gravitation, $U_G\sim GM^2_{\odot}/R_{\odot}\approx 4\times 10^{41}$~J,
nor chemical reactions, $U_{\mathrm{chem}} \sim 0.1\mathrm{\ eV}\times N_{\odot}\approx 2\times 10^{37}$~J
(where $N_{\odot}$ is the number of nucleons in the Sun) are sufficient;
only nuclear energy, $U_{\mathrm{nucl}}\sim  1\mathrm{\ MeV}\times N_{\odot}\approx 2\times 10^{44}$~J,
can sustain the solar luminosity over the solar age, as beautifully demonstrated
by Gallium experiments in the previous decade~\cite{Hampel:1998xg,Abdurashitov:2002nt,Altmann:2000ft}.
On the other hand, for the Earth one finds $U_G\approx 4\times 10^{32}$~J,
$U_{\mathrm{chem}} \approx 6\times 10^{31}$~J, and $U_{\mathrm{nucl}}  \approx 6\times 10^{30}$~J,
(assuming that some $10^{-8}$ of Earth mass consists of radioactive nuclei), so that any of the
previous mechanisms can sustain the present heat flow for the Earth's age:
$U_{\oplus}=H_{\oplus} t_{\oplus}\approx 5\times 10^{30}$~J.

In order to understand the energetics of the Earth one has to clarify the roles of the different
energy sources, their locations and when they have been at work. In 1980, at the end of a
review on the Earth energy sources Verhoogen~\cite{Verhoogen:1980} summarized the situation with the
following words:
\begin{quotation}
  What emerges from this morass of fragmentary and uncertain data is that
radioactivity by itself could plausibly account for at least 60 percent, if not 100 percent,
of the Earth's heat output. If one adds the greater rate of radiogenic heat production
in the past, \ldots possible release of gravitational energy (original heat,
separation of core, separation of inner core, tidal friction \ldots meteoritic impact \ldots),
the total supply of energy may seem embarrassingly large. \ldots Most, if not all of the
figures mentioned above are uncertain by a factor of at least two, so that disentangling
contributions from the several sources is not an easy problem.
\end{quotation}

Anderson~\cite{Anderson2005} opens a recent review, entitled
``Energetics of the Earth and the Missing Heat Source Mistery'',
with the following words:
\begin{quotation}
Global heat flow estimates range from 30 to 44~TW. Estimates of the
radiogenic contribution (from the decay of \U, \Th, and \K\ in the mantle), based on cosmochemical
considerations, vary from 19 to 31~TW. Thus, there is either a good balance between current input and
output, \ldots or there is a serious missing heat source problem, up to a deficit of 25~TW.
\end{quotation}

Anderson summarizes in Table~\ref{tab:ThermalEnergySources} the
potential contributions to the terrestrial energy budget. Similar to
Verhoogen, he notes that the potential supply from radiogenic and
non radiogenic sources, up to 66~TW, can even exceed the observed
heat flow, so that \emph{paradoxes such as the missing heat source
problem can be traced to non-realistic assumptions and initial and
boundary conditions} and \emph{The bottom line is that there appears
to be no mismatch between observed heat flow and plausible sources
of heating}. He also notes, and we agree, that uncertainties on the
different ingredients of the energy balance (total outflow, amounts
of radiogenic material in the Earth, \ldots ) are much larger than
it was estimated in the past.

\begin{table}[htb] \caption[acc]{Sources of thermal energy in the Earth's
interior, adapted
from~\cite{Anderson2005}.\label{tab:ThermalEnergySources} }
\begin{tabular}{lcc}
 \hline\hline
Energy supply (potential contributions) & \multicolumn{2}{c}{TW} \\
\hline\hline
\multicolumn{3}{l}{Non radiogenic:}\\
Conducted from core &  8.6 & \\
Mantle differentiation & 0.6 & \\
Thermal contraction & 2.1 & \\
Earthquake induced gravitational energy  & 2  & \\
Radiated seismic energy & 0.3 & \\
Tidal friction & $1 \div 2$  & \\
\hline
Total (non radiogenic) & & $15 \div 16$ \\
\hline
\multicolumn{3}{l}{Radiogenic:}\\
Present radiogenic  & $19 \div 31$ & \\
Delayed radiogenic  & & \\
(1 to 2 Ga delay between production and arrival at surface) &  5 & \\
\hline
Total (radiogenic) &  & $24 \div 36$  \\
\hline
Secular cooling ($0 \div 100$ K/Ga) & & $0 \div 14$ \\
\hline
Total input & & $39 \div 66$ \\
   \hline \hline
\end{tabular}\\[2pt]
\end{table}
\subsection{Radiogenic heat: the BSE, unorthodox and even heretical Earth models}
\label{subsec:radioHeat}
We recall that the canonical Bulk Silicate Earth model predicts a present radiogenic
heat production of 20~TW. The BSE is a fundamental geochemical paradigm:
it is consistent with most observations, which however regard the crust and the
uppermost portion of the mantle only. On the grounds of available geochemical and/or
geophysical data, however, one cannot exclude that radioactivity in the present Earth
provides a larger contribution to the terrestrial heat flow, sufficient to account for
even the highest estimate of terrestrial heat flow.

For a comparison, let us summarize some -- less orthodox or even heretical --
alternatives to the BSE.

\begin{itemize}
  \item[(a)]
  It is conceivable that the original material from which the Earth formed is not exactly
  the same as inferred from CI-chondrites. A model with initial composition as that of
 enstatite chondrites could account for a present production of some 30~TW~\cite{Javoy:1995,Hofmeister:2005}.
  \item[(b)]
  A model where the BSE abundances of \U, \Th, and \K\ are proportionally rescaled by a factor
  of 2.2 cannot be excluded by the observational data, if one assumes that the missing radiogenic material
  is hidden below the upper mantle. This model gives a present radiogenic heat production
  of 44~TW, which matches the highest estimate of the present heat flow.
  \item[(c)]
  Starting with~\cite{Lewis:1971,Hall:1971}, several authors have been considering the possibility
  that a large amount of potassium is sequestered into the Earth's core, where it could provide the light
  element that accounts for the right core density, the energy source for driving the terrestrial dynamo,
  and -- more generally  -- an additional contribution to Earth energy budget.
  This possibility has been recently revived in~\cite{RamaMurthy:2003}, where from high-pressure and
  high-temperature data it was shown that potassium can enter iron sulphide melts in a strongly
  temperature-dependent fashion so that \K[40] could be as a substantial heat source in the core of the Earth.
  \item[(d)]
  Herndon~\cite{Herndon:2003} has proposed that a large drop of uranium has been collected at
  the center of the Earth, forming a natural 3-6~TW breeder reactor, see also~\cite{Raghavan:2002eh}.
  In this case nuclear fission should provide the energy source for terrestrial magnetic field, a contribution
  to missing heat, and the source of the anomalous \nucleus[3]{He}/\nucleus[4]{He} flow from Earth.
\end{itemize}
In summary, an unambiguous and observationally based determination of the radiogenic heat
production would provide an important contribution for understanding Earth's energetics.
It requires to determine how much uranium, thorium and potassium are on the Earth,
quantities which are strictly related to the anti-neutrino luminosities from these elements.

\section{\label{sec:refmod}The reference model}
A reference model for geo-neutrino production is a necessary starting point for studying the
potential and expectations of detectors at different locations.

By definition, it should incorporate the best available geochemical and geophysical information
on our planet. In practice, it has to be \emph{based on selected geophysical and geochemical data and
models (when available), on plausible hypotheses (when possible), and admittedly on arbitrary
assumptions (when unavoidable)}. These duly cautious words from~\cite{Fogli:2005qa}
explain the difficulties and to some extent the arbitrariness when building such models.
They also mean that estimates of uncertainties on the predicted geo-neutrino fluxes are at least
as important as the predicted values.

\subsection{Comparison among different calculations}
Recently a few such models have been presented in the
literature~\cite{Mantovani:2003yd,Enomoto:2005mb,Fogli:2005qa}.
Predictions by different authors for a few locations are compared in
Table~\ref{tab:predicRateCompare}.

\begin{table}[htb] \caption[add]{Predicted geo-neutrino rate from $\U+\Th$ at various locations.
Rates are in TNU. All calculations are normalized to a survival
probability $ \langle P_{ee}\rangle=0.57$. For Mantovani \et the
uncertainties are estimated as
$(N_{\mathrm{high}}-N_{\mathrm{low}})/6$, see Table XII
of~\cite{Mantovani:2003yd}. \label{tab:predicRateCompare} }
\renewcommand{\tabcolsep}{2pc} 
\begin{tabular}{lccc}
 \hline\hline
    Location  &  Mantovani \et &   Fogli \et &  Enomoto~\protect\footnote{Private communication.} \\
              & \cite{Mantovani:2003yd}  &   \cite{Fogli:2005qa}  &   \\
 \hline
Hawaii      & $ 12.5 \pm 3.6 $ & $ 13.4 \pm 2.2 $ &  13.4  \\
Kamioka     & $ 34.8 \pm 5.9 $ & $ 31.6 \pm 2.5 $ &  36.5  \\
Gran Sasso  & $ 40.5 \pm 6.5 $ & $ 40.5 \pm 2.9 $ &  43.1  \\
Sudbury     & $ 49.6 \pm 7.3 $ & $ 47.9 \pm 3.2 $ &  50.4  \\
Phyasalami  & $ 52.4 \pm 7.6 $ & $ 49.9 \pm 3.4 $ &  52.4  \\
Baksan      & $ 51.9 \pm 7.6 $ & $ 50.7 \pm 3.4 $ &  55.0  \\
   \hline \hline
\end{tabular}\\[2pt]
\end{table}

All these models rely on the geophysical $2^{\circ}\times 2^{\circ}$
crustal map of~\cite{Laske:2001,Bassin:2000} and on the density
profile of the mantle as given by PREM~\cite{Dziewonski:1981}.

Concerning the adopted abundances in the crust layers, Mantovani \et~\cite{Mantovani:2003yd}
use average values from results available in the literature in 2002,
Fogli \et~\cite{Fogli:2005qa} refer to the values of the recent review by
Rudnick and Gao~\cite{Rudnick:2003}, whereas Enomoto \et~\cite{Enomoto:2005mb}
adopt the values reported in the 1995 review by Rudnick and Fontaine~\cite{Rudnick:1995}.
Mantovani \et and Fogli \et assume a
chemically layered mantle, with abundances in the upper mantle
from \cite{Jochum:1983,Zartman:1988} and \cite{Salter:2004,Workman:2005}, respectively;
whereas Enomoto \et consider a chemically homogeneous mantle. The adopted uranium abundances in
the various reservoirs are compared in Table~\ref{tab:UabundCompare}. All papers use the BSE mass
constraint in order to determine the adopted abundances in the lower portion of the mantle.

Concerning uncertainties on the abundances in the crust and in the
upper mantle, Mantovani \et~\cite{Mantovani:2003yd} estimate them
from the spread of published values, whereas Fogli
\et~\cite{Fogli:2005qa} use the errors quoted by Rudnick and
Gao~\cite{Rudnick:2003}, where available,  see
Table~\ref{tab:UabundCompare}. The uncertainties of the abundances
in the lower mantle are obtained by Fogli \et by propagating in the
mass constrain also the uncertainties estimated for the BSE.
Mantovani \et release the BSE constraint and take a more
conservative attitude, including as extreme values the possibility
that the lower mantle has the same small abundances observed in the
upper mantle and, on the other side, the possibilities that the
amount of heat generating elements can sustain a fully radiogenic
44~TW heat flow, with most of the material being hidden in the
unexplored lower mantle. The BSE mass constraint used by Fogli \et
fixes the total amount of heat generating elements to the level of
$\pm 14\%$ ($1\sigma$), whereas the range considered in Mantovanti
\et is much wider. This is at the origin of the differences in the
uncertainties quoted in Table~\ref{tab:predicRateCompare}. Indeed,
if the calculation of Mantovani \et is restricted to the BSE range,
the uncertainties become comparable to those of Fogli \et.

 One has to remark\footnote{A more extensive comment on the
 treatment of  uncertainties is presented in
Appendix~\ref{sec:appenErrors}.}
 that Fogli \et also present a systematic approach to the ubiquitous
issue of covariances in geo-neutrino analyses. In fact, for the only
(at the moment) relevant case, the correlation between \U\ and \Th,
they use the value $\rho =0.94$, estimated by the errors on the
$\U/\Th$ ratio, very close to  $\rho = 1$, which was used in
\cite{Mantovani:2003yd}.

\subsection{The contribution of the various reservoirs}
\begin{figure}[p]
\includegraphics[width=0.7\textwidth,angle=0]{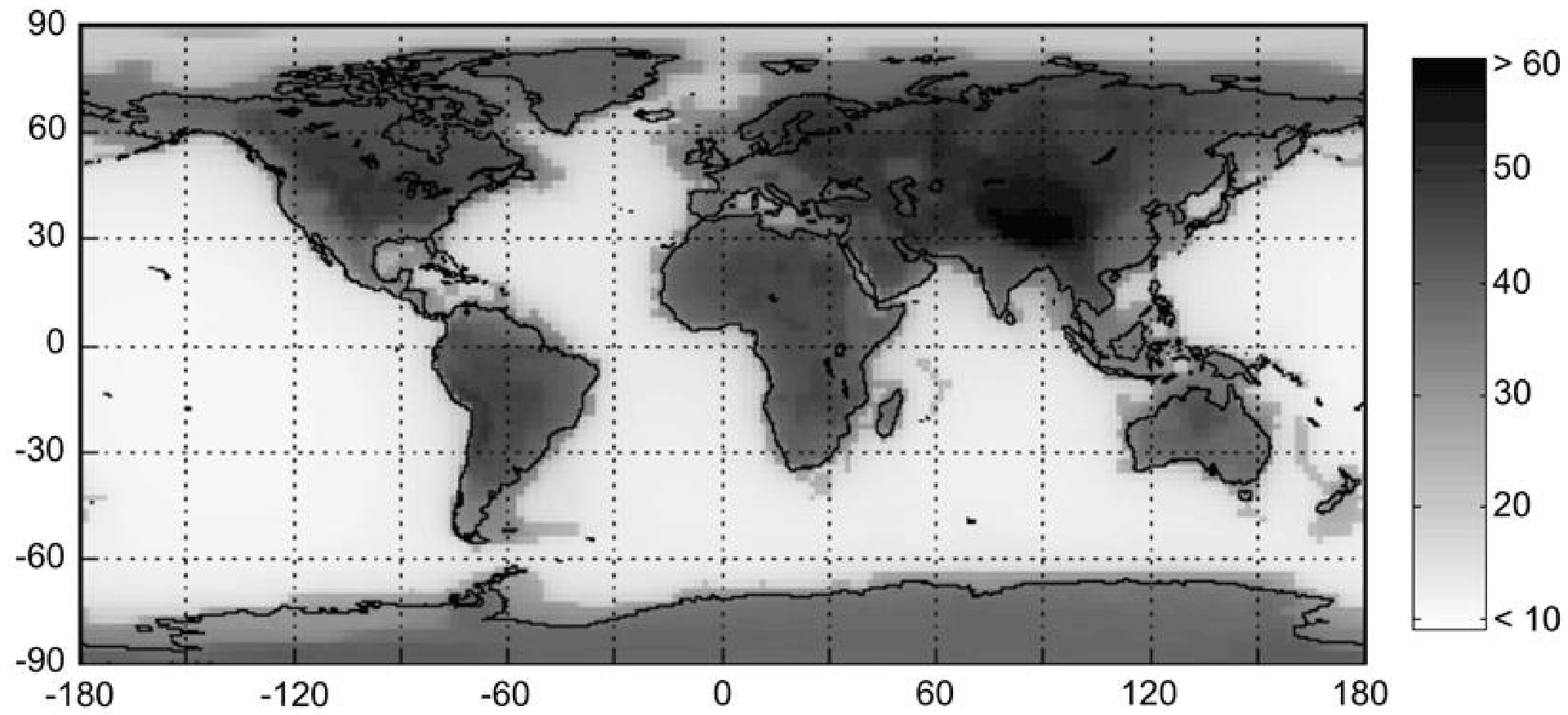}
 \caption[aee]{Predicted geo-neutrino events from uranium and thorium decay
chains, normalized to $10^{32}$~proton~yr and 100\% efficiency, from~\cite{Fiorentini:2005mr}.}
\label{fig:geonuEventsMap}
\vspace{3cm}
\includegraphics[width=0.5\textwidth,angle=-90]{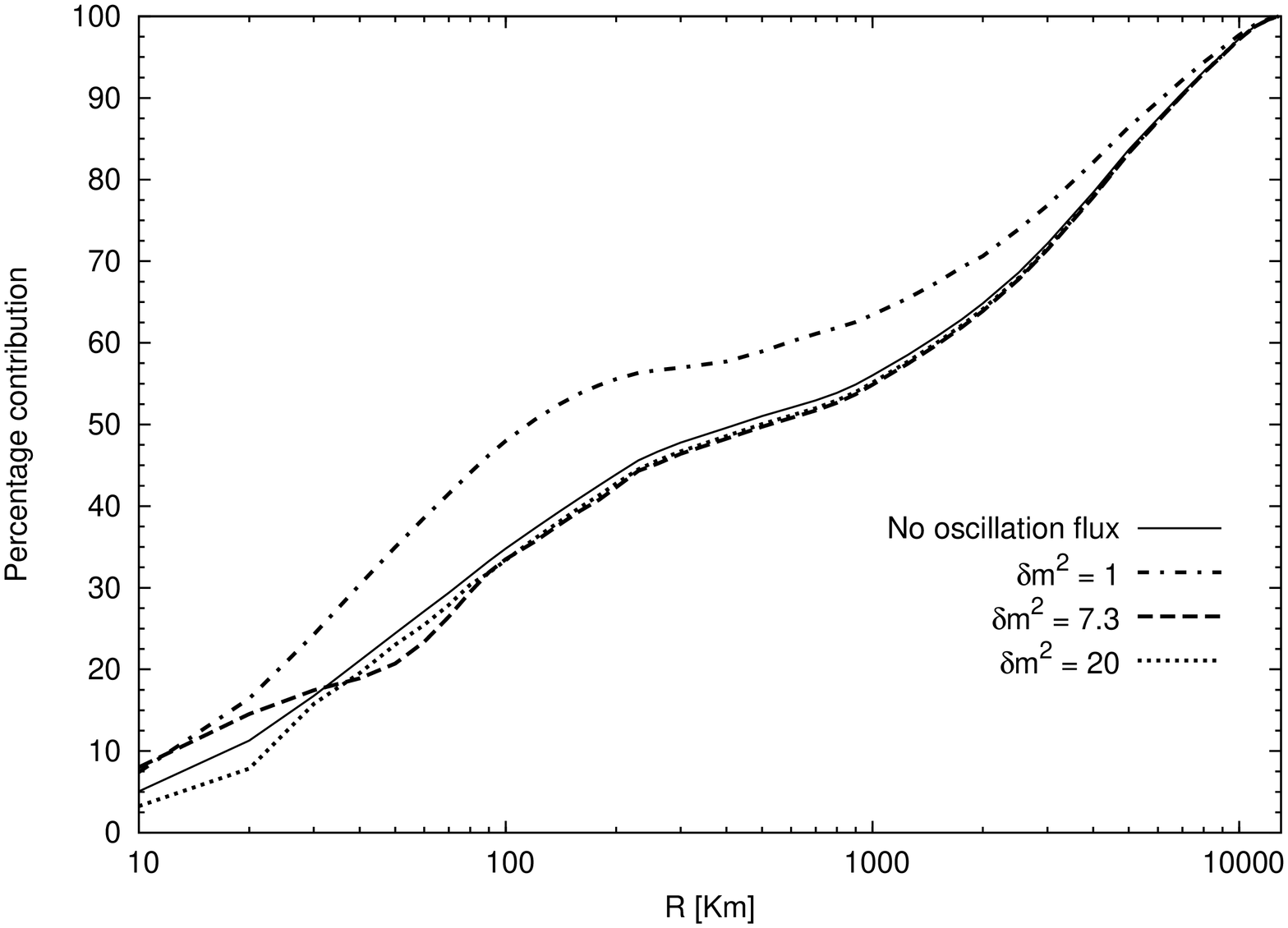}
 \caption[agg]{Contributed signal as a function of distance. The percentage contribution to the event
yield at Kamioka originating from sources within a distance $R$ is
shown for the indicated values of $\Delta m^2$ in units of
$10^{-5}$~eV$^2$ at fixed $\sin^2 2\theta = 0.863$. The percentage
contributed neutrino flux without oscillation is also shown for
comparison. From~\cite{Mantovani:2003yd}.}
\label{fig:signalVsDistance}
\end{figure}

The predicted signal all over the world is shown in Fig.~\ref{fig:geonuEventsMap},
taken from~\cite{Fiorentini:2005mr}. It presents the Earth as it shines in geo-neutrinos.
The more intense signals arise from regions with a thick continental crust,
whereas over the oceans the signal essentially originates from the mantle.

The separate contributions of the different reservoirs to the produced flux of geo-neutrinos
from the uranium decay chain are analyzed in Table~\ref{tab:fluxU}, from~\cite{Mantovani:2003yd}.
\begin{table}[htb]
\caption[aff]{\label{tab:fluxU} Uranium: masses, radiogenic heat,
and predicted fluxes. Units are $10^{17}$~kg, TW and
$10^6$~cm$^{-2}$~s$^{-1}$, respectively. The reference values, lower
and upper limits are labeled as ref, low, and high, respectively.
Crust summarizes CC and OC; UM (LM) denotes upper (lower) mantle.
Data from~\cite{Mantovani:2003yd}. }
\begin{ruledtabular}
\begin{tabular}{lcccccc}
      &           &         & Himalaya  & Gran Sasso & Kamioka   & Hawaii   \\
      &           &         & $33^{\circ}$~N  $85^{\circ}$~E
                                      & $42^{\circ}$~N  $14^{\circ}$~E
                                      & $36^{\circ}$~N  $137^{\circ}$~E
                                      & $20^{\circ}$~N  $156^{\circ}$~W                  \\
\hline
      &   $M(\text{U})$ &  $H(\text{U})$ & \multicolumn{4}{c}{$\Phi_{\text{U}}$} \\
\hline Crust low & 0.206\protect\footnote{This value corresponds to
an uranium abundance in the continental crust equal to 0.91~ppm as
estimated in Ref.~\cite{Taylor:1985}. Starting from
Ref.~\cite{Fiorentini:2004rj} we dismiss this estimate, since it is
inconsistent with data from all other authors, and we use 0.3 as the
lower limit for the uranium mass in the crust, see
Ref.~\cite{Fiorentini:2004rj} for a discussion. The values of the
fluxes from the crust corresponding to  0.3 are 4.92, 2.84, 2.35,
and 0.33 for the four locations, respectively.}
& 1.960 & 3.337 & 1.913 & 1.594 & 0.218 \\
\textbf{Crust ref} & \textbf{0.353} & \textbf{3.354} & \textbf{5.710 }& \textbf{3.273} & \textbf{2.727} & \textbf{0.373} \\
Crust high & 0.413 & 3.920 & 6.674 & 3.826 & 3.187 & 0.436 \\
\hline
UM low & 0.048 & 0.455 & 0.146 & 0.146 & 0.146 & 0.146 \\
\textbf{UM ref} & \textbf{0.062} & \textbf{0.591} & \textbf{0.189} & \textbf{0.189} & \textbf{0.189} & \textbf{0.189} \\
UM high & 0.077 & 0.727 & 0.233 & 0.233 & 0.233 & 0.233 \\
\hline
LM low & 0.147 & 1.399 & 0.288 & 0.288 & 0.288 & 0.288 \\
\textbf{LM ref} & \textbf{0.389 }& \textbf{3.695} & \textbf{0.760} & \textbf{0.760 }& \textbf{0.760} & \textbf{0.760} \\
LM high & 1.177 & 11.182 & 2.299 & 2.299 & 2.299 & 2.299 \\
\hline
Total low & 0.401 & 3.814 & 3.770 & 2.346 & 2.027 & 0.651 \\
\textbf{Total ref} & \textbf{0.804 }& \textbf{7.639} & \textbf{6.659} & \textbf{4.222} & \textbf{3.676} & \textbf{1.322} \\
Total high & 1.666 & 15.828 & 9.206 & 6.358 & 5.720 & 2.968 \\
\end{tabular}
\end{ruledtabular}
\end{table}
At Himalaya, a site which maximizes the crust contribution, the prediction
is $\Phi(U) = 6.7\times 10^6$~cm$^{-2}$s$^{-1}$ whereas at Hawaii,
a site which minimizes the crust contribution, the prediction is
$\Phi(U) = 1.3\times 10^6$~cm$^{-2}$s$^{-1}$, originated mainly from the mantle.
For the Kamioka mine, where the KamLAND detector is in operation,
the predicted uranium flux is $\Phi(U) = 3.7\times 10^6$~cm$^{-2}$s$^{-1}$.
Within the reference model, about 3/4 of the flux is generated from material
in the crust and the rest mainly from the lower mantle. At Gran Sasso laboratory,
where Borexino~\cite{Alimonti:1998aa} is in preparation, the prediction is
$\Phi(U) = 4.2\times 10^6$~cm$^{-2}$s$^{-1}$, this larger flux arising from a
bigger contribution of the surrounding continental crust. A similar calculation
for Sudbury, the place which hosts the SNO detector,
gives $\Phi(U) = 4.4\times 10^6$~cm$^{-2}$s$^{-1}$. The crust contribution exceeds 80\%.

The contribution to the signal as a function of the distance from the detector is
shown in Fig.~\ref{fig:signalVsDistance} from Mantovani \et~\cite{Mantovani:2003yd}
for the specific case of Kamioka. One notes that some 15\% of the total signal originates from
a region within 30~km from the detector, whereas half of the signal is generated within
some 600 km.  We remind the typical linear dimension of each tile in the
$2^{\circ}\times 2^{\circ}$
crustal map is of order 200 km, so that any information on a smaller scale is lost.
A better geological and geochemical description of the region surrounding the detector
is needed for a more precise estimate of the geo-neutrino signal.

\subsection{The effect of uncertainties of the oscillation parameters}
\label{sec:effectUncertOscill}
In this review we have fixed $\Delta m^2 = 8.0\times  10^{-5}$~eV$^2$ and $ \tan^2 \theta = 0.45$,
which gives an asymptotical survival probability $\langle P_{ee}\rangle = 0.57$, following the best fit
of Strumia and Vissani~\cite{Strumia:2005tc}. The same paper gives a 99\% CL range
$ 7.2 \times 10^{-5}\mathrm{eV}^2 < \Delta m^2 < 8.9 \times 10^{-5}\mathrm{eV}^2 $ and $0.33< \tan^2 \theta <0.61$,
with the corresponding range for the average survival probability $0.53<\langle P_{ee}\rangle<0.63$.

The effect of these uncertainties on the predicted signal is presented in Table~\ref{tab:oscillationParameters}.

\begin{table}[htb] \caption{
Effect of the oscillation parameters on the predicted $\U +\Th$ signal at
Kamioka. The
relative/absolute variation is computed with respect to the
prediction for the best fit values  ($\Delta m^2 = 8.0\times
10^{-5}$~eV$^2$ and $\tan^2\theta = 0.45$). }
\label{tab:oscillationParameters}
\newcommand{\dg}{\hphantom{$0$}}
\newcommand{\cc}[1]{\multicolumn{1}{c}{#1}}
\renewcommand{\tabcolsep}{2pc} 
\begin{tabular}{lc}
 \hline\hline
Parameter & Signal variation \\
\hline
$\Delta m^2 = 7.2\times 10^{-5}$~eV$^2$ &  $+0.11$~TNU   \\
$\Delta m^2 = 8.9\times 10^{-5}$~eV$^2$ &  $-0.09$~TNU   \\
 \hline
$\tan^2\theta = 0.61$                  & $-7.5\%$ \\
$\tan^2\theta = 0.33$                  & $+9.3\%$ \\
\hline \hline
\end{tabular}\\[2pt]
\end{table}
The predicted signal is practically unaffected by the uncertainty on
 $\Delta m^2$: when this is varied within its 99\% CL interval the signal changes by less than
one tenth of TNU. This holds for any value of the total uranium and thorium mass, since the precise
value of $\Delta m^2$ only matters in the region near the detector.
In addition, we observe that the predictions computed for the best
value ($\Delta m^2 = 8.0\times  10^{-5}$~eV$^2$) and for the limit  $\Delta m^2=\infty $ differ by $+0.3$~TNU.

The uncertainty on the mixing angle is most important: at the  99\% CL the relative error on the
signal $ \Delta S / S \approx 9\%$, somehow smaller (but not negligible)
in comparison with the geological uncertainties.
\section{\label{sec:refine}Refinements of the reference model: the regional contribution}
The geo-neutrino signal depends on the total amount of heat generating elements in the Earth and
on the geochemical and geophysical properties of the region around the detector.
For KamLAND, we estimated that about one half of the signal generated in the crust comes from
a region within 200~km from the detector (half of the total signal is originated from within~600 km).
This region, although containing a globally negligible amount of heat generating elements,
produces a large contribution to the signal as a consequence of its proximity to the detector.
This contribution has to be determined on the grounds of a detailed geochemical and geophysical study of the region.

The study of the region around Kamioka based on a detailed analysis
of the six tiles depicted in Fig.~\ref{fig:UraniumAbundanceJapan},
including the possible effects of the subducting plates below the
Japan Arc and a discussion of the contribution from of the Japan
Sea, has been presented in~\cite{Fiorentini:2005cu} and
in~\cite{Enomoto:2005mb,Enomoto:2005}. The result
of~\cite{Fiorentini:2005cu} for this regional contribution to the
signal from uranium geo-neutrinos is:
\begin{equation}
\label{eq:UregContri}
    S_{\mathrm{reg}}(\U) = (15.41\pm 3.07) \mathrm{\ TNU}\quad\quad (3\sigma)\quad .
\end{equation}
This result is obtained by including several effects, discussed in the next subsections.
These refinements increase the signal by about 1~TNU. The global error is obtained by
adding in quadrature the individual independent uncertainties.

The results of that paper, which only considered geo-neutrinos from uranium decay chains,
have been extended to include geo-neutrinos from thorium.

\subsection{The six tiles near Kamland}
The depth distribution of the Conrad and Moho discontinuities beneath the whole of the Japan
Islands are derived in~\cite{Zhao:1992}, with an estimated standard error of $\pm 1$~km over most of Japan territory.
This allows distinguishing two layers in the crust: an upper crust extending down to the
Conrad and a lower part down to the Moho discontinuity.

\begin{figure}[hptb]
\includegraphics[width=0.6\textwidth,angle=0]{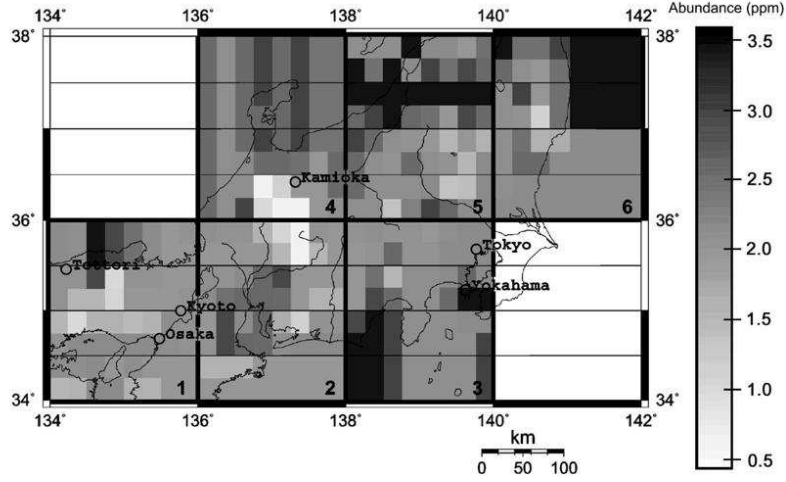}
\caption{Uranium abundance in the upper crust of Japan from~\cite{Fiorentini:2005cu}.}
\label{fig:UraniumAbundanceJapan}
\end{figure}

In Fiorentini \et~\cite{Fiorentini:2005cu} a map of uranium abundance in the upper crust has been built,
under the important assumption that the composition of the whole upper crust is the same as that
inferred in Togashi \et~\cite{Togashi:2004} from the study of the exposed portion, see Fig.~\ref{fig:UraniumAbundanceJapan}.
There is no specific study of the lower part of the Japan crust, however, it is well known
that there are similarities between the composition of the Japanese crust and that of the Sino-Korean block.
In an extensive compositional study of East China crust~\cite{Gao:1998}, the uranium abundance
in the lower part is estimated between 0.63 and 1.08~ppm. On these grounds
Fiorentini \et~\cite{Fiorentini:2005cu}  adopt for the abundance in the lower crust of
Japan $a_{\mathrm{LC}} (U)= (0.85 \pm 0.23)$~ppm.

We remind that for building the reference model, the entire Earth's crust was subdivided
into $2^{\circ} \times 2^{\circ}$ tiles. In Fiorentini \et~\cite{Fiorentini:2005cu} a more detailed
grid was used where each tile is subdivided into 64 $(1/4)^{\circ} \times (1/4)^{\circ}$ cells,
thus with a linear scale of about 20~km. This size is chosen since the sampling density for
the study of the upper crust in the region near Kamioka is about one specimen per
400~km$^2$ and also, concerning the vertical distribution of abundances in the crust,
it is presently impossible to have information on the chemical composition on a scale smaller
than the Conrad depth, generally lying at about 20~km.

The contributions from the six tiles to the uranium signal is
$S_6(\U) = 12.74$~TNU. The calculation for the same region within the reference model gave
$S_6(\U) = 14.10$~TNU. A similar reduction was also found in~\cite{Enomoto:2005}.
The difference is understood in terms of the depletion of \U\ concentration in the Japanese
islands with respect to the average continental crust, already noted in Togashi \et~\cite{Togashi:2004}.

With respect to the prediction of the reference model, the six tiles contribute 45\% of the total signal.
In more detail, the tile hosting Kamioka generates 30\% of the total produced signal.
Note that the uranium mass contained in the six tiles is about $m_6 = 3.3 \times 10^{13}$~kg,
really negligible (less then 0.05\%) with respect to that estimated for the whole Earth.

Fiorentini \et~\cite{Fiorentini:2005cu} consider several sources of the uncertainties
affecting this estimate: measurement errors of the chemical analysis,
discretization of the upper crust, chemical composition of the lower crust and crustal depth.
Their effects are summarized in Table~\ref{tab:regionalErrors}.

\begin{table}[htb] \caption{Errors from the regional geophysical
and geochemical uncertainties.} \label{tab:regionalErrors}
\newcommand{\dg}{\hphantom{$0$}}
\newcommand{\cc}[1]{\multicolumn{1}{c}{#1}}
\renewcommand{\tabcolsep}{2pc} 
\begin{tabular}{lcl}
\hline
Source & $\Delta S$~(TNU) & Remarks \\
 \hline
 Composition of
upper-crust samples                    &  0.96   & $3\sigma$ error\\
Upper-crust discretization              &  1.68   &            \\
Lower-crust composition                &  0.82   & Full range \\
Crustal depths                         &  0.72   & $3\sigma$ error           \\
Subducting slab                        &  2.10   & Full range \\
Japan Sea                              &  0.31   & Full Range \\
\hline Total                           &  \bf{3.07}   &            \\
 \hline
\end{tabular}\\[2pt]
\end{table}
\subsection{Effect of the subducting slab beneath Japan}
\label{subsec:subductingSlab}
The Japan arc, at the crossing among the Eurasian, Philippine and Pacific plates, is the theater of
important subduction processes. The Philippine plate is moving towards the Eurasia plate at about 40 mm/yr and is
subducting beneath the southern part of Japan. The Pacific Plate is moving in roughly the same direction
at about 80 mm/yr and is subducting beneath the northern half of Japan.

In order to estimate the effect of the subducting slab on geo-neutrino production,
Fiorentini \et~\cite{Fiorentini:2005cu} considered two extreme cases:
(a) one assumes that the slab keeps its trace elements while subducting;
(b) at the other extreme, it is possible that, as the slab advances,
all uranium from the subducting crust is dissolved in fluids during dehydration
reactions and accumulates in the lower part of the continental crust of Japan,
thus strongly enriching it.

As there is no argument for deciding which of the extreme cases (a) or (b)
is closer to reality and in order to encompass both of them,
the contribution from the subducting slab was estimated in~\cite{Fiorentini:2005cu}
as\footnote{Enomoto \et~\cite{Enomoto:2005mb} only consider
case (a) and get a smaller correction.}:
\begin{equation}
    S_{\mathrm{slab}}(\U) = (2.3\pm 2.1) \mathrm{\ TNU}\quad\quad (3\sigma)\quad .
\end{equation}
\subsection{The crust below the Japan Sea}
The morphology of the Japan Sea is characterized by three major basins (Japan, Yamato, and Ulleung Basins).
The crust of the Japan basin is generally considered as oceanic, whereas the nature of other
basins is controversial and debated. Again in~\cite{Fiorentini:2005cu}] two extreme models are considered:
\begin{itemize}
  \item[(a)]
 following Ref.~\cite{Laske:2001,Bassin:2000} all the basins are formed with
oceanic crust, extending down to 7 km below 1 km of sediments.
  \item[(b)]
 Deeper crustal depths (up to 19 km for the Oki bank) and thicker
sediments layers (up to 4 km for the Ulleung basin) are reported in
the literature, see Table~\ref{tab:vertExtensCrustLayers}. By taking
these values and assigning the abundances typical of continental
crust, one maximizes geo-neutrino production.
\end{itemize}

\begin{table}[htb]
\caption{The vertical extensions  (km) of crustal layers in the
Yamano basin (YB), Oki bank (OK), and Ulleung basin (UB) used for
model (b).} \label{tab:vertExtensCrustLayers}
\newcommand{\dg}{\hphantom{$0$}}
\newcommand{\cc}[1]{\multicolumn{1}{c}{#1}}
\renewcommand{\tabcolsep}{2pc} 
\begin{tabular}{lccc}
\hline
       & YB            & OK & UB \\
\hline
Sediments   &  1.2 & \dg0.3 & 4 \\
Upper       &  2.8 & \dg8.7 & 2 \\
Lower       &  8.5 &   10.5 & 8 \\
\hline
\end{tabular}\\[2pt]
\end{table}

In order to encompass these two extreme cases, Fiorentini \et~\cite{Fiorentini:2005cu}
 fix the contribution to the signal from the Japan Sea as:
\begin{equation}
    S_{\mathrm{JS}}(\U) = (0.37\pm 0.31) \mathrm{\ TNU}\quad\quad (3\sigma)\quad .
\end{equation}

\subsection{Thorium contribution and the total geo-neutrino regional signal}
By adding the above contributions, and summing in quadrature independent uncertainties
one obtains equation (\ref{eq:UregContri}).
The same analysis repeated for \Th\ gives a regional contribution:
\begin{equation}
\label{eq:ThRegContri}
    S_{\mathrm{reg}}(\Th) = (3.66\pm 0.68) \mathrm{\ TNU}\quad\quad (3\sigma)\quad .
\end{equation}
Assuming a complete correlation for the errors, one has:
\begin{equation}
    S_{\mathrm{reg}}(\U+\Th) = (19.1\pm 3.8) \mathrm{\ TNU}\quad\quad (3\sigma)\quad .
\end{equation}

\section{\label{sec:beyond}Beyond the reference model}

\subsection{Overview}
As discussed in the preceding sections, masses of heat generating elements in the Earth are
estimated on the grounds of cosmochemical arguments, based on the compositional similarity
between Earth and carbonaceous chondrites. Measurements of samples from the Earth's crust
imply that the crust contains about one half of this global estimate, whereas the mantle
-- which should contain the rest -- is practically unexplored in this respect. A direct
determination of the mass of heat generating elements in the globe is clearly an important test
of the origins of the Earth and will fix the radiogenic contribution to the terrestrial heat flow,
 which is a presently a debated issue, see Sect.~\ref{sec:heat}.

The geo-neutrino signal depends on the total mass of heat generating elements in the
Earth and on the geochemical and geophysical properties of the region around the detector.
The region close to the detector, although containing a globally negligible amount of uranium,
produces a large contribution to the signal as a consequence of its proximity to the detector.
This contribution has to be determined on the grounds of a detailed geochemical and geophysical
study of the region, if one wants to extract from the total signal the remaining part which carries
the relevant information on the mass of heat generating elements. Such a study of the region around
Kamioka has been presented in the previous section.

The contribution to the geo-neutrino signal from the rest of the
world depends on the total amount of heat generating elements
\emph{as well as on their distribution inside the Earth}, since the
closer is the source to the detector the larger is its contribution
to the signal. For each value of the total mass, we shall construct
distributions of abundances which provide the maximal and minimal
signals, under the condition that \emph{they are consistent with
geochemical and geophysical information on the globe}.

This will bring us beyond the reference model. Essentially, we shall build models of the Earth which
respect the observational data available, concerning the abundances in the crust and the density profile
of the Earth, and we shall release the BSE constraint on the global amounts of heat generating elements.

In practice, we shall keep the masses of heat generating elements in
the unexplored lower mantle as free parameters. One can still vary
the abundances along the mantle depth (which are generally believed
to increase with increasing depth) and obtain different geo-neutrino
signal for the same total amount of heat generating elements. The
important point~\cite{Fiorentini:2005cu} is the following: the
assumption that the abundances are spherically symmetrical and
non-decreasing with depth will be enough to provide rather tight
constraints on the mantle contribution to the geo-neutrino signal.

This has to be further combined with the contribution from the crust, which also can be
maximized/minimized by varying the abundances in the range allowed by observational data.

The combination of this information with the regional study allows to find the connection between
the geo-neutrino signal and the masses of heat generating elements in the Earth.

This relationship will be developed in the following subsections, where we elucidate
the proximity argument and then we combine the regional contribution for Kamioka found
in the previous section with that of the rest of the world.

In principle, this argument can be developed separately for uranium and thorium, \ie
connecting the geo-neutrino signals $S(\U)$ and $S(\Th)$ with the
respective masses $m(\U)$ and $m(\Th)$ all over Earth.

\begin{figure}[htbp]
\includegraphics[width=0.5\textwidth,angle=-90]{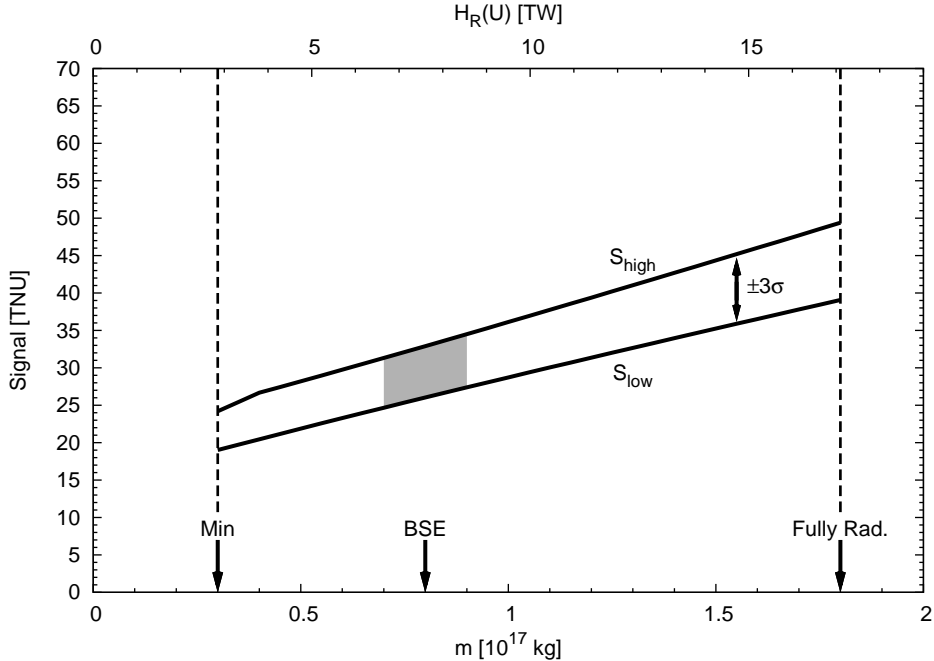}
\caption{The predicted signal from uranium geo-neutrinos at KamLAND,
adapted from~\cite{Fiorentini:2005cu}.}
\label{fig:KamlandsignalmassU}
\end{figure}

The main result is shown in Fig.~\ref{fig:KamlandsignalmassU},
which presents for Kamioka the connection among the uranium mass,
heat generation, and geo-neutrino signal.

By assuming the BSE mass constraint, our prediction is
\begin{equation}
    S(\U) = (29.5\pm1.6) \mathrm{\ TNU}\quad (1\sigma,\mathrm{\
    BSE})\quad.
\end{equation}
If the chondritic \Th/\U\ ratio is assumed, one finds
\begin{equation}
    S(\U+\Th) = (32\pm 2) \mathrm{\ TNU}\quad (1\sigma,\mathrm{\
    BSE})\quad.
\end{equation}
However, if the BSE constraint is released we get
\begin{equation}
    S(\U+\Th) = (32^{+8}_{-4}) \mathrm{\ TNU}\quad (1\sigma,\mathrm{\
    no BSE})\quad.
\end{equation}
\subsection{The proximity argument}
The main question is to build models which, for a given total uranium mass in the Earth,
$m(\U)$, provide the minimal and maximal signals, with the additional constraint that these models
be consistent with available geochemical and geophysical observational data.
This result can be accomplished by means of what we call the ``proximity argument'':
\emph{the minimal (maximal) contributed flux is obtained by placing heat radiogenic elements as far
(close) as possible to the detector}~\cite{Fiorentini:2004rj}.

This argument can be used in several steps.

\begin{itemize}
   \item[(i)]
For a place on or near the continental crust, since the continental
crust lies on the average closer than the mantle to the detector the
maximal (minimal) signal is obtained by putting as much (little)
radiogenic material in the crust, as consistent with the observed
values\footnote{The opposite holds for places very far from the
continents as the Hawaii Islands, see
Section~\ref{sec:predOtherLocat}.}. This determines the mass of
uranium mass in the crust, $m_{C}(\U)$, which is constrained by
observational data to lie in the interval $(0.3 \div 0.4) \times
10^{17}$~kg.
   \item[(ii)]
This leaves us with the problem of distributing the remaining mass,
$m(\U) - m_{C}(\U)$, inside the mantle.
Under the assumptions that the abundances in this reservoir are non-decreasing functions
of the depth, the extreme predictions for the signal are obtained by:
(a) placing heat generating elements in a thin layer at the bottom, or
(b) distributing it with uniform abundance over the mantle.
   \item[(iii)]
One can then combine the extreme cases so as to obtain lower and upper limits to the global contribution to the signal.
\end{itemize}

\subsection{The case of KamLAND\label{sec:caseKamLAND}}
In the case of Kamioka, after excluding the region where a separate
geo-chemical and geophysical investigation has been performed (see
Sec.~\ref{sec:refine}), the proximity argument provides for the
signal from the rest of the world:
\begin{equation}\label{eq:restOfWorld}
S_{\mathrm{RW}}(\U) = (2.25+14.76 \times m(\U)  )\pm( -0.55 +2.61 \times m(\U)) \quad ,
\end{equation}
where the signal is in TNU, the mass is in units of $10^{17}$~kg and the interval within
the $\pm$ sign corresponds to the full range of models which have been considered.

By combining this results with the regional contribution, calculated in the
previous section,
$  S_{\mathrm{reg}}(\U) = (15.41\pm 3.07) \mathrm{\ TNU} $,
we get the uranium geo-neutrino signal as a function of uranium mass in the Earth:
\begin{equation}\label{eq:signalTotal}
    S(\U)=S_0(\U) \pm \Delta(\U) \quad ,
\end{equation}
 where:
\begin{eqnarray}
  S_0(\U)    &=&  17.66 + 14.76 \times m(\U)
\label{eq:totalGeoSignal}\\
  \Delta^2(\U) &=& (3.07)^2 +\left(2.61\times m(\U) - 0.55\right)^2
 \quad . \label{eq:DeltaTotalGeoSignal}
\end{eqnarray}
This error is obtained by combining in quadrature all geochemical and geophysical
uncertainties discussed in the preceding paragraphs. All of them have been estimated
so as to cover $\pm 3\sigma$ intervals of experimental measurements and total ranges of theoretical predictions.

However, this error does not account for present uncertainties on
neutrino oscillation parameters. We remind (see
Section~\ref{sec:effectUncertOscill}) that the uncertainty on the
mixing angle implies a 99\% CL relative error on the signal  $\Delta
S / S \approx 9\%$, which is somehow smaller (but not negligible) in
comparison with the geological uncertainties.

For the sake of discussing the potential of geo-neutrinos, we shall
ignore in the following  the  error originating from uncertainties
on the mixing parameter, which however should be measured more
accurately.

The expected signal from uranium geo-neutrinos at KamLAND is presented as a function of the total
uranium mass $m(\U)$ in Fig.~\ref{fig:KamlandsignalmassU}. The predicted signal as a function
of $m(\U)$ is between the two lines denoted as $S_{\mathrm{high}}$ and $S_{\mathrm{low}}$,
which correspond, respectively, to $S_0(\U) \pm \Delta(\U)$.

We remark that the extremes of the band correspond to the whole
range of uncertainty, which is estimated according to the following
criteria:
\begin{itemize}
  \item[(i)]
for statistical errors we consider a $\pm 3\sigma$ interval;
  \item[(ii)]
for systematic uncertainties of geochemical and geophysical
parameters we determine an interval such as to cover all modern
estimates which we found in the literature;
  \item[(iii)]
independent errors are combined in quadrature.
\end{itemize}
We remark that the ``proximity argument'', combining global mass
balance with geometry, is very powerful in constraining the range of
fluxes: in the allowed band are enclosed all the models consistent
with geological data.

Since the minimal amount of uranium in the Earth is $0.3 \times 10^{17}$~kg (corresponding to the
minimal estimate for the crust and the assumption of negligible amount in the mantle),
we expect $S(\U)$ to be at least 19 TNU. On the other hand, the maximal amount of uranium
tolerated by Earth's
energetics\footnote{For an uranium mass $m(\U) = 1.8 \times 10^{17}$~kg and relative abundances
as in Eq.~(\ref{eq:heatNatural}), the present radiogenic heat production rate from \U, \Th, and
\K\ decays equals the maximal estimate for the present heat flow from Earth,
$H_{\mathrm{max}} = 44$~TW~\cite{Pollack:1993}.},
$1.8 \times 10^{17}$~kg, implies S(U) not exceeding 49 TNU.

For the central value of the BSE model, $m(\U) = 0.8 \times 10^{17}$~kg,
it was found in~\cite{Fiorentini:2005cu} $ S(\U) = 29.5 \pm 3.4$~TNU, \ie with an accuracy of 12\% at $3\sigma$.

We remark that estimates by different authors for the uranium mass within the BSE are
all between $(0.7 \div 0.9) \times 10^{17}$~kg. This implies that the uranium signal has to be
in the interval $(24.7 \div 34.5)$~TNU. The measurement of geo-neutrinos can thus provide a
direct test of an important geochemical paradigm.

\begin{figure}[htbp]
\includegraphics[width=0.5\textwidth,angle=-90]{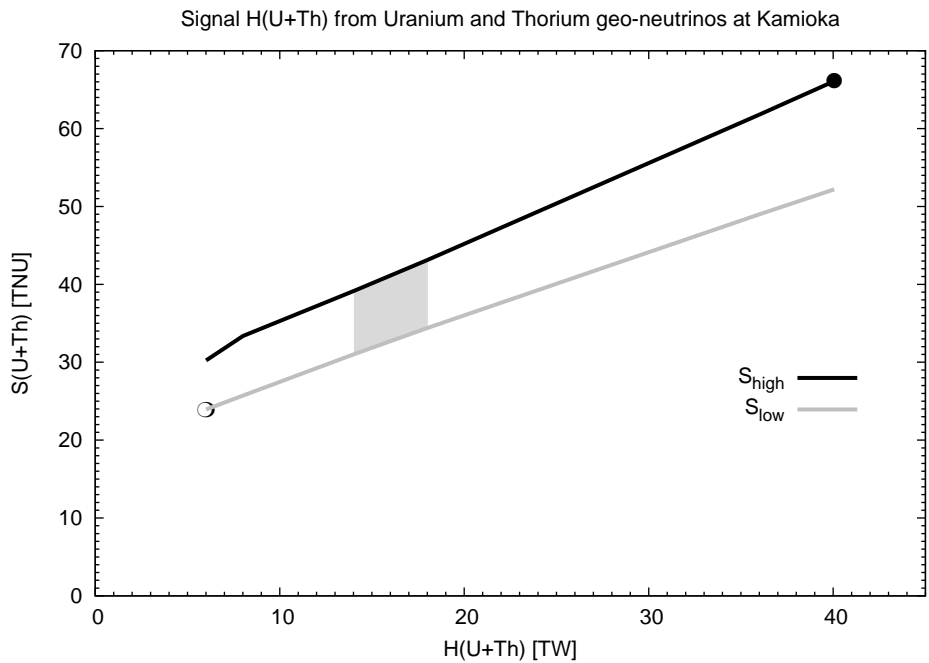}
\caption[appp]{Predictions on the combined signal $S(\U+\Th)$ from uranium and thorium
geo-neutrinos at Kamioka as a function of the radiogenic heat production rate $H(\U+\Th)$.
The area between the two lines ($S_{\mathrm{high}}$ and $S_{\mathrm{low}}$)
denotes the region allowed by geochemical and geophysical constraints.
The shaded area denotes the region allowed by the BSE constraint.
Earth energetics implies the signal does not exceed 62 TNU, and \U\ and \Th\
measured in the crust imply a signal of at least 24 TNU.} \label{fig:KamlandsignalmassUplusTh}
\end{figure}

We do not expect that the next generation of experiments can collect enough statistics
so as to clearly separate the two components (\U\ and \Th) in the signal
(see Section~\ref{sec:geonu}) and will have to rely on the chondritic estimate
for the ratio for the global abundances of these two elements. On these grounds,
we shall also assume this value of the ratio and rescale the results calculated for
uranium to get the total signal as a function of the radiogenic heat from $\U + \Th$.
In this way one gets Fig.~\ref{fig:KamlandsignalmassUplusTh}.

\subsection{Predictions at other locations \label{sec:predOtherLocat}}
In order to understand the potential of detectors at other locations one should
 perform a study similar to the one  just outlined for Kamioka.
In the absence of a detailed geochemical and geophysical study of the region surrounding
the detector, we can tentatively address the problem by using the prediction for
the crust as derived by the $2^{\circ} \times 2^{\circ}$ crustal
map\footnote{We remind that for Kamioka the more detailed calculation changes
the signal by just about 1 TNU; however the uncertainty of the regional
contribution is about $\pm 3$~TNU (at $3\sigma$).}.

We resort again to the proximity argument, by making it more general,
so as to cover also locations which lie far from the continents.

\paragraph{Contribution from the crust.}
The contribution, as a function of the uranium mass contained in the crust $m_{\mathrm{C}}(\U)$,
can be obtained by rescaling the predictions of the reference model.
This gives at the $i$-th location:
\begin{equation}
    S_{\mathrm{C}}^i = \alpha_i \times  m_{\mathrm{C}}(\U) \quad ,
\end{equation}
where the response coefficients $\alpha_i$ are presented in Table~\ref{tab:responCoeff}
for a few locations.
\begin{table}[htb]
\caption{The crust response coefficients $\alpha_i$ for a few locations.} \label{tab:responCoeff}
\newcommand{\dg}{\hphantom{$0$}}
\newcommand{\cc}[1]{\multicolumn{1}{c}{#1}}
\renewcommand{\tabcolsep}{2pc} 
\begin{tabular}{lr}
\hline
     Site         & $\alpha$ \\
\hline
Hawaii       &      8   \\
Kamioka      &      57  \\
Gran Sasso   &      69  \\
Sudbury      &      93  \\
\hline
\end{tabular}\\[2pt]
\end{table}

\paragraph{Contribution from the mantle.}
We remind that, under the assumptions that the abundances in this reservoir
are radial and non decreasing function of the depth, the extreme predictions for the signal are obtained by:
\begin{itemize}
  \item[(i)]
  placing uranium in a thin layer at the bottom;
  \item[(ii)]
  distributing it with uniform abundance over the mantle.
\end{itemize}
For a mass $m_{\mathrm{M}}(\U)$ in the mantle, the two cases give, respectively,
 and independently of the location:
\begin{eqnarray}
  S_{\mathrm{M, low}}(\U) &=& \beta_{\mathrm{low}} \times m_{\mathrm{M}}(\U)\mathrm{\ TNU} \quad , \\
   S_{\mathrm{M, high}}(\U) &=& \beta_{\mathrm{high}} \times m_{\mathrm{M}}(\U)\mathrm{\ TNU} \quad ,
\end{eqnarray}
with  $\beta_{\mathrm{low}}= 12.15$~TNU and  $\beta_{\mathrm{high}}= 17.37$~TNU when the mass,
as here and in the following, is measured in units of $10^{17}$~kg.
\paragraph{Combining mantle and crust.}
At each location, the total uranium signal will be:
\begin{equation}
    S(\U) =  S_{\mathrm{C}}(\U) + S_{\mathrm{M}}(\U) =
    \alpha_i \times  m_{\mathrm{C}}(\U)  + \beta \times  m_{\mathrm{M}}(\U) =
    (\alpha_i - \beta ) \times  m_{\mathrm{C}}(\U) + \beta \times m(\U) \quad ,
\end{equation}
where the total uranium mass is $m(\U) =m_{\mathrm{C}}(\U) + m_{\mathrm{M}}(\U)$.

If one wants the largest signal, obviously one has to
put $\beta = \beta_{\mathrm{high}}$ and, if  $\alpha_i >  \beta_{\mathrm{high}}$,
$m_{\mathrm{C}}(\U)$ has to be as large as possible, consistently with the observations of the
crust, \ie $m_{\mathrm{C}}(\U) = \min (m(\U), 0.41)$.
In the opposite case,  $\alpha_i <  \beta_{\mathrm{high}}$, one takes instead $m_{\mathrm{C}}(\U)$
as small as possible, \ie  $m_{\mathrm{C}}(\U) = 0.3$.
Note that the first case is what occurs on or close to the continental crust,
whereas the second case corresponds to locations far from the continents.

Similarly, for minimizing the signal one puts $\beta = \beta_{\mathrm{low}}$ and then one
takes $m_{\mathrm{C}}(\U) = 0.3$, if  $\alpha_i >  \beta_{\mathrm{low}}$,
or $m_{\mathrm{C}}(\U) = \min (m(\U), 0.41)$ in the opposite case.

\begin{figure}[p]
\includegraphics[width=0.33\textwidth,angle=-90]{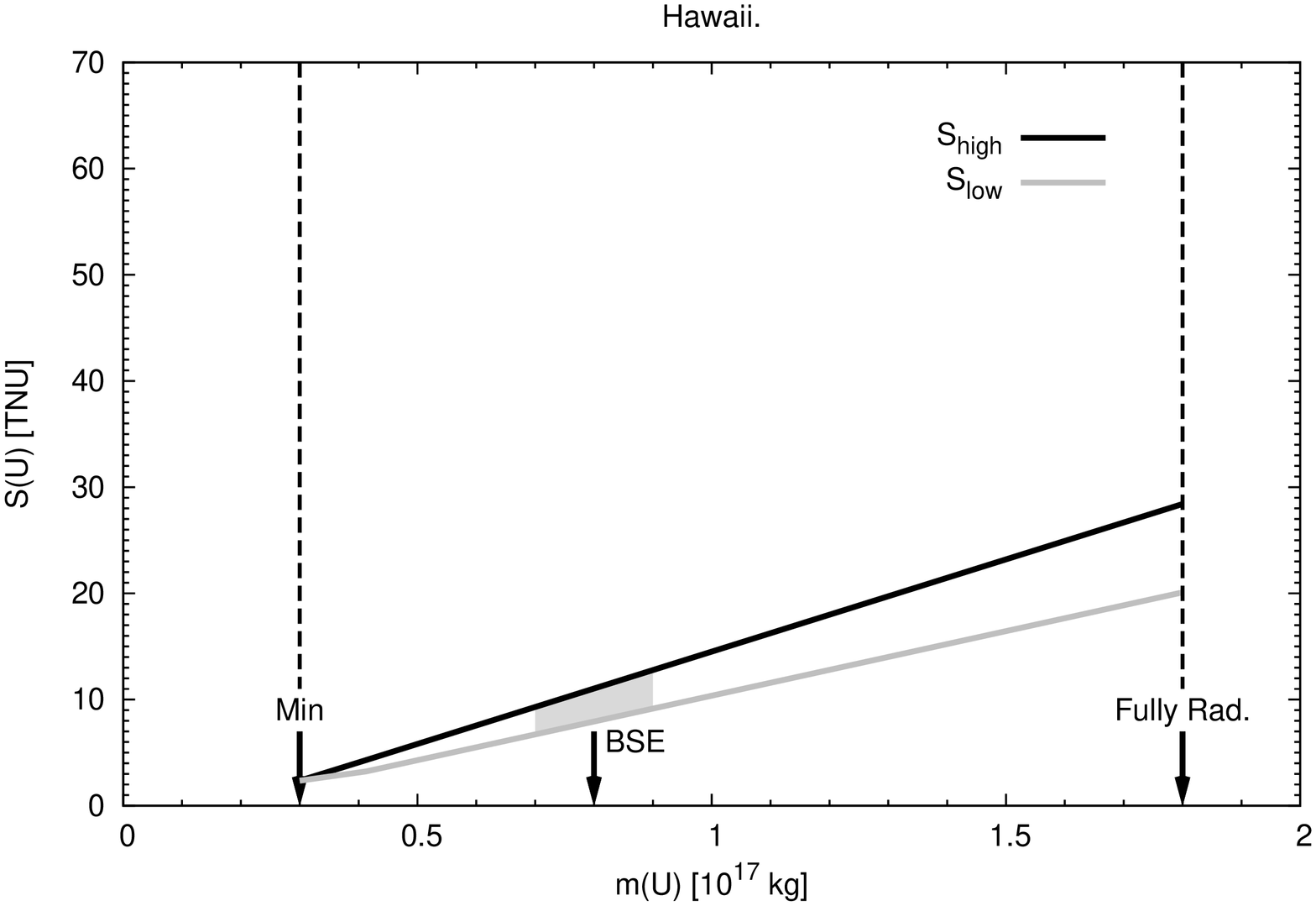} \hfill
\includegraphics[width=0.33\textwidth,angle=-90]{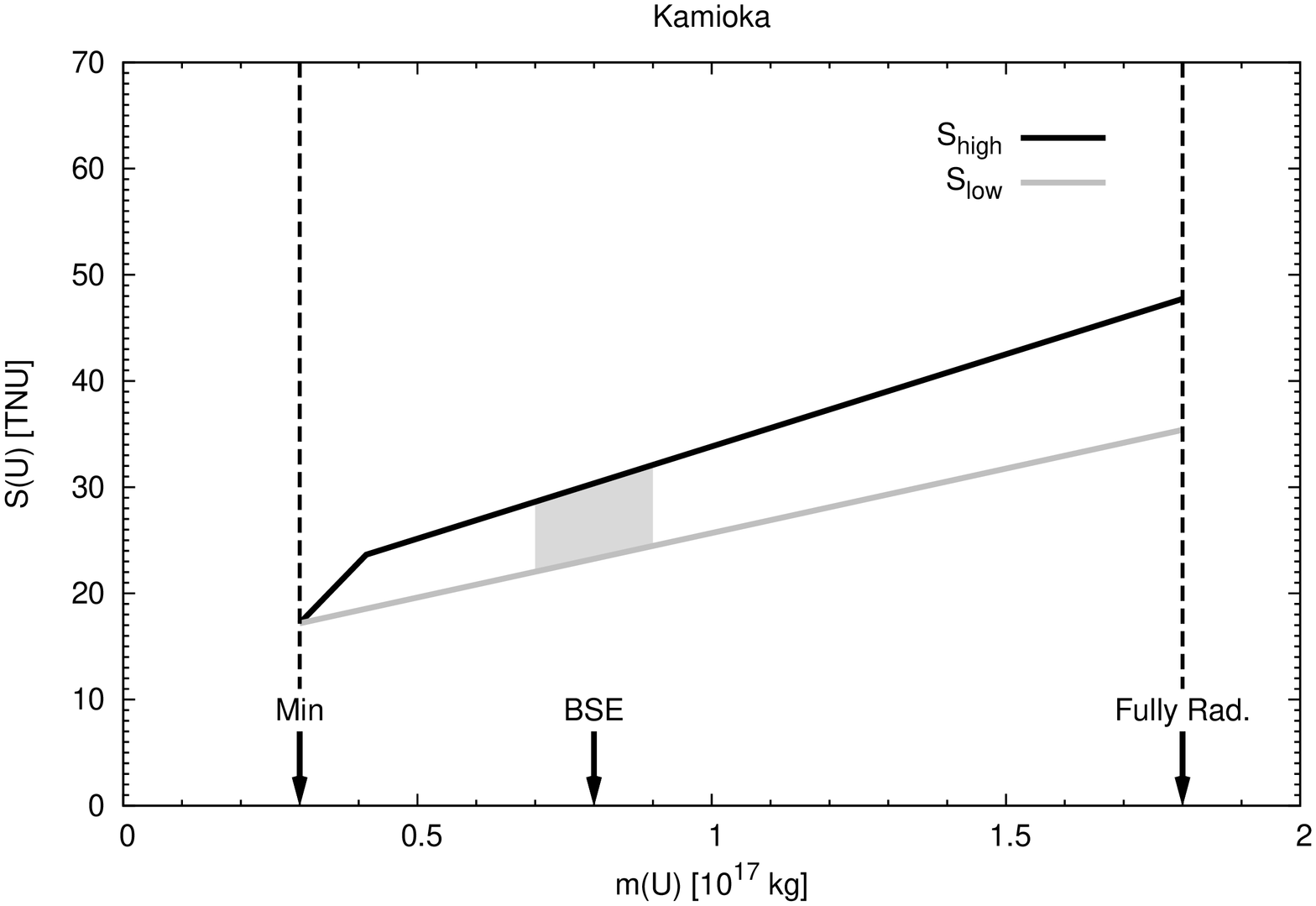} \\
\includegraphics[width=0.33\textwidth,angle=-90]{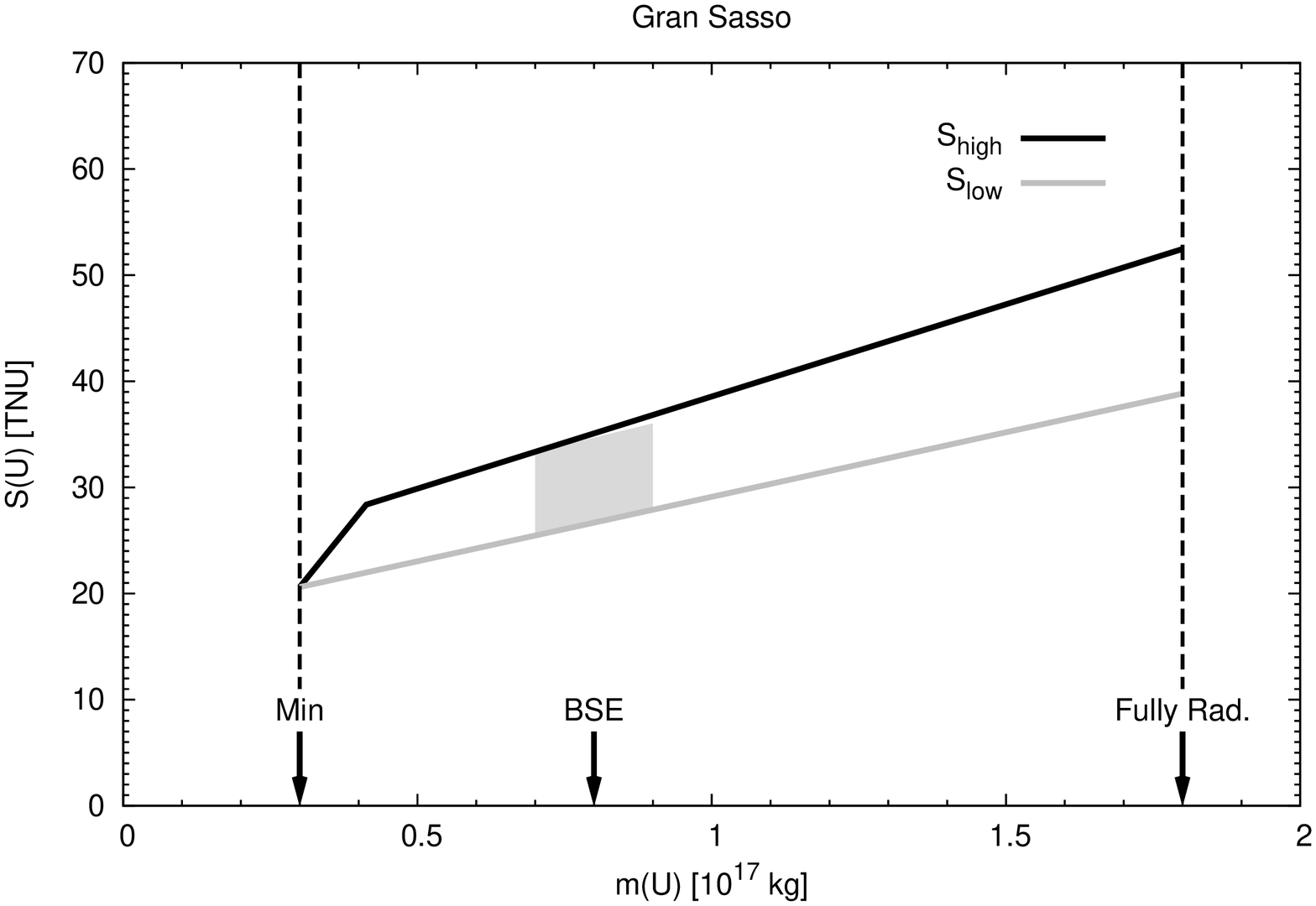} \hfill
\includegraphics[width=0.33\textwidth,angle=-90]{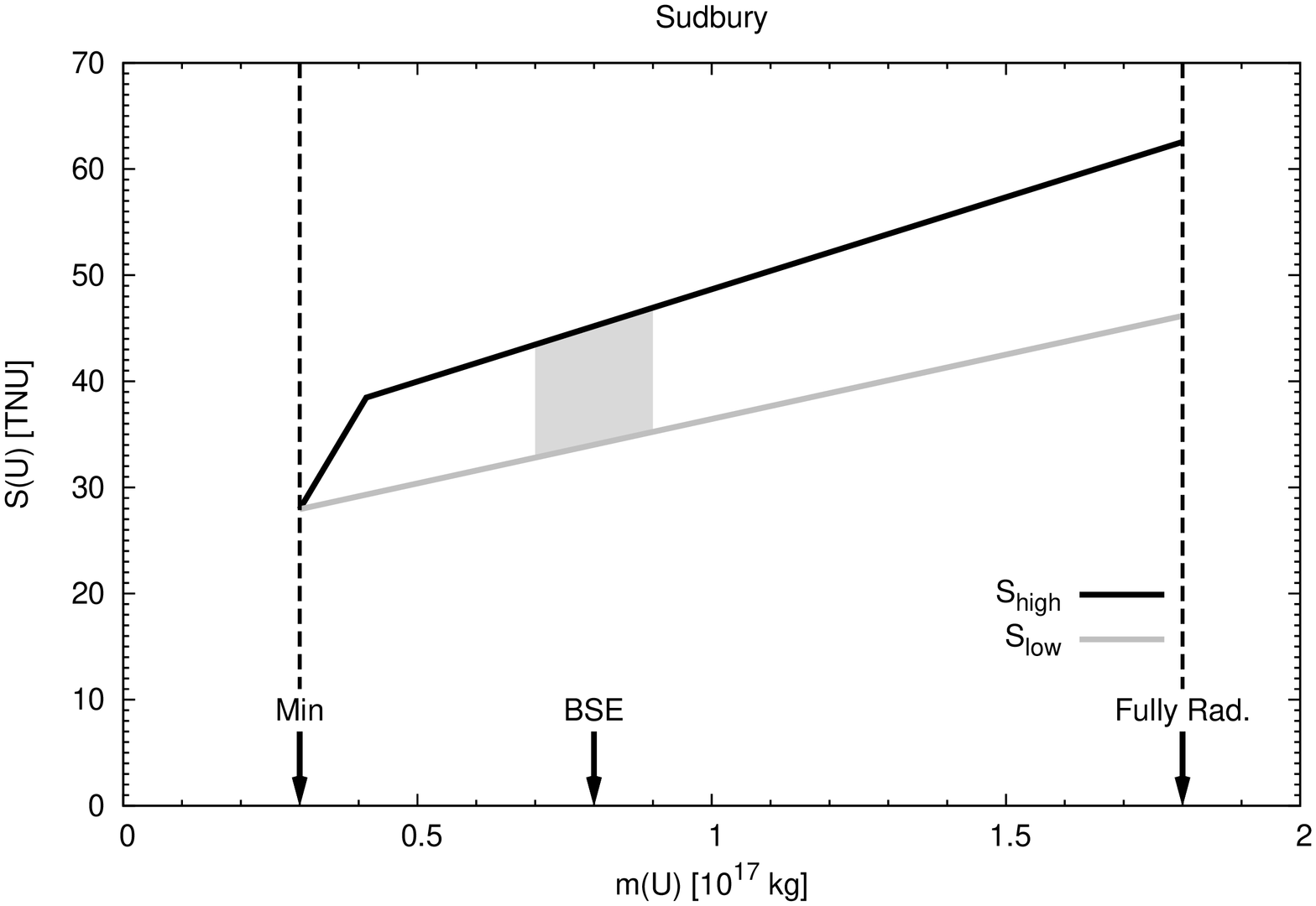}
\caption[aqqq]{The predicted signal from uranium geo-neutrinos at Hawaii (top-left),
Kamioka (top-right), Gran Sasso (bottom-left), and Sudbury (bottom-right).
The area between the black line ($S_{\mathrm{high}}$) and
the grey line ($S_{\mathrm{low}}$) denotes the region allowed by geochemical and geophysical
constraints.} \label{fig:SignalMassConstraint}
\end{figure}

The results for the extreme cases, shown in
Fig.~\ref{fig:SignalMassConstraint}, deserve the following comments.
The band is most narrow for the Hawaii, as natural since this place,
far from the continents, is most sensitive to the amount of
radioactivity hidden in the mantle. On the other hand, it is more
wide at Sudbury, where the signal is dominated from the contribution
of the crust. Concerning Kamioka, the contour defined by this
analysis are close (within four TNU or less) to those which were
derived by adding the geological and geochemical information on the
region, see Fig.~\ref{fig:KamlandsignalmassU}. This gives us some
confidence about the predictions for the other sites.

\section{\label{sec:kam}KamLAND results and their interpretation}
\subsection{Overview}
KamLAND (Kamioka Liquid scintillator Anti-Neutrino Detector) is so far the largest
low-energy antineutrino detector ever built and studies a wide range of science,
spanning particle physics, geophysics and
astrophysics\footnote{An overview of KamLAND science plan is found in~\cite{Gratta:1999ib}.}.

The KamLAND collaboration is made up of researchers from Japan, China, France and United States.
The detector is situated in the same cavern used by the original Kamiokande experiment,
where 2002 Nobel laureate Masatoshi Koshiba performed much of his ground-breaking research in neutrino physics.

Located on the island of Honshu in Japan, since 2002 KamLAND detects hundreds of anti-neutrinos per
 year from nuclear reactors hundreds of kilometers away, an enormous improvement over previous attempts
with any other detector. KamLAND has observed an anti-neutrino
deficit as well as energy spectral distortion confirming neutrino
oscillations and hence non vanishing neutrino
masses~\cite{Eguchi:2002dm,Araki:2004mb}. As a natural continuation
of the scientific program, KamLAND aims now at the direct
observation of \nucleus[7]{Be} solar neutrinos by detecting recoil
energy in neutrino-electron scattering processes.

KamLAND has also performed a geo-neutrino investigation. In 2005,
the KamLAND collaboration has published~\cite{Araki:2005qa}
experimental results, claiming some 28 geo-neutrino events from
uranium and thorium decay chains in a two-year exposure. This
important step shows that the technique for exploiting geo-neutrinos
in the investigation of the Earth's interior is now available. From
the KamLAND data, including new measurements of the
$\nucleus[13]{C}(\alpha,n)\nucleus[16]{O}$ cross section, one finds
$S(\U+\Th) = (63^{+28}_{-25})$~TNU, see~\cite{Fiorentini:2005ma}.
The central value is close to the prediction of a maximal and fully
radiogenic model (see Sect.~\ref{sec:caseKamLAND}), however the BSE
prediction is within $1\sigma$ from it.

In the future, with more statistics KamLAND should be capable of providing a three-sigma
evidence of geo-neutrinos, but discrimination between BSE and fully radiogenic models definitely
requires new detectors, with class and size similar to that of KamLAND, far away from nuclear power plants.

In the next subsections we shall present KamLAND results on
geo-neutrinos and  discuss their implications on the terrestrial
heat.

\subsection{The KamLAND detector}
Neutrinos are detected at
KamLAND\footnote{An extensive description of the detector can be found
in the PhD thesis of S.~Enomoto~\cite{Enomoto:2005}.}
by the inverse beta-decay reaction,
\begin{equation}
    \anue + p \to e^+ + n - 1.806 \mathrm{\ MeV}\quad ,
\end{equation}
with a large amount of organic liquid scintillator.
The liquid scintillator essentially consists of hydrocarbons ($C_nH_{2n}$)
which provide the hydrogen nuclei acting as the target for antineutrinos.
We remind that the energy threshold of the reaction, 1.806 MeV,
is low enough to detect a part of the \U-series and \Th-series geo-neutrinos.

The reaction produces two correlated signals. A prompt signal is
given by the slowing down positron and by the two 0.51 MeV gamma
rays from positron annihilation. The delayed signal consists of a
2.2 MeV gamma particle, which is emitted in the thermal neutron
capture on proton. The thermalization and capture processes take
place in about 200 microseconds, and neutron capture occurs
typically 30-50 cm apart from the neutrino reaction vertex. The time
and space correlations of the two signals are distinguishing
characteristics of electron-type antineutrino events. The delayed
coincidence of the two signals provides an effective way to select
antineutrino events with excellent separation of background events.

The KamLAND detector basically consists of 1000 tons of ultra-pure liquid scintillator (LS)
contained in a 6.5~m radius spherical balloon and of 1879 surrounding  photomultiplier tubes (PMT)
that cover 34\% of the sphere. The detector is located 1000~m underground in the Kamioka mine,
just beneath the Mt. Ikenoyama summit, Gifu, Japan ($36.42^{\circ}$~N, $137.31^{\circ}$~E).
The 2700~m water equivalent thickness of rock covering the detector sufficiently reduces cosmic muon flux,
resulting in 0.34~Hz of muon event rate.

\begin{figure}[htbp]
\includegraphics[width=0.7\textwidth]{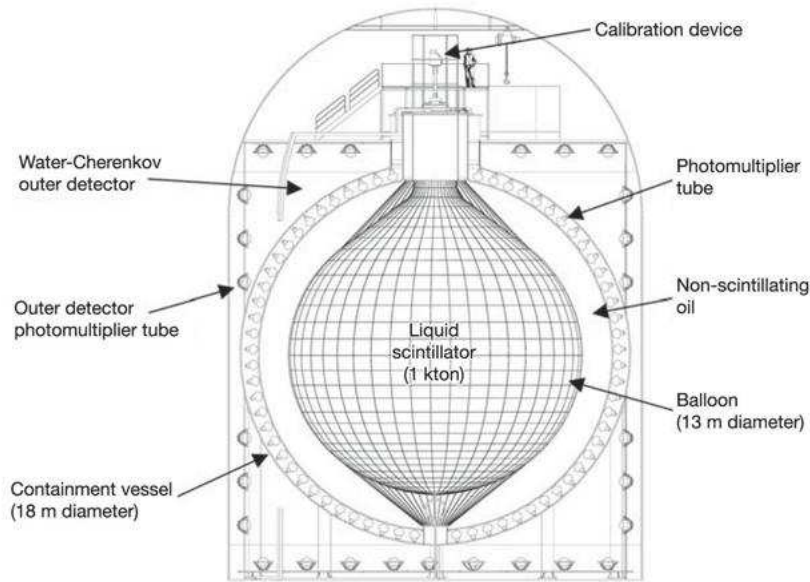}
\caption[asss]{An overview of the KamLAND detector
from~\cite{Enomoto:2005}.} \label{fig:kamLANDdetector}
\end{figure}

Figure~\ref{fig:kamLANDdetector} illustrates the KamLAND detector.
The LS, balloon and PMT's are contained in a 9~m radius spherical
stainless steel vessel. 1325 17-inch PMT and 554 20-inch PMT's are
mounted inside the stainless steel vessel viewing the center of the
LS sphere. The 6.5 m radius LS-containing balloon is positioned at
the center of the stainless steel vessel, being supported and
constrained by a network of Kevlar ropes. Non-scintillating mineral
oil (MO) is filled between the stainless steel vessel and the LS
containing balloon, providing gravity/buoyancy balance to the LS
sphere, and also acting as a buffer layer against radiations into
the LS from the stainless steel vessel, PMT, and everything
surrounding the vessel. The MO layer is further divided into two
spherical shells by 8.25~m radius transparent acrylic wall, to
isolate the balloon contacting MO from the PMT/vessel exposed MO and
reduce radioactive contamination around the LS.

The inner part of the 9~m radius stainless vessel is called the
inner detector (ID), whereas the portion outside of the 9 m radius
vessel is called the outer detector (OD). The space between the
vessel and the cave wall is filled with 3200 tons of pure water, and
viewed by 225  20 inch PMT's. Cosmic muons passing through the OD
are tagged by the OD PMT's, by detecting Cherenkov light.

\subsection{KamLAND results on geo-neutrinos}
The KamLAND collaboration has reported~\cite{Araki:2005qa} data from
an exposure of $N_p = (0.346 \pm 0.017) \times 10^{32}$ free protons
over a time $T = 749$ days with a detection efficiency  $\varepsilon
= 68.7\%$; the effective exposure is thus $\alpha= N_p\times T
\times \varepsilon = (0.487 \pm 0.025) \times
10^{32}$~proton$\cdot$yr.

\begin{figure}[p]
\includegraphics[width=0.6\textwidth]{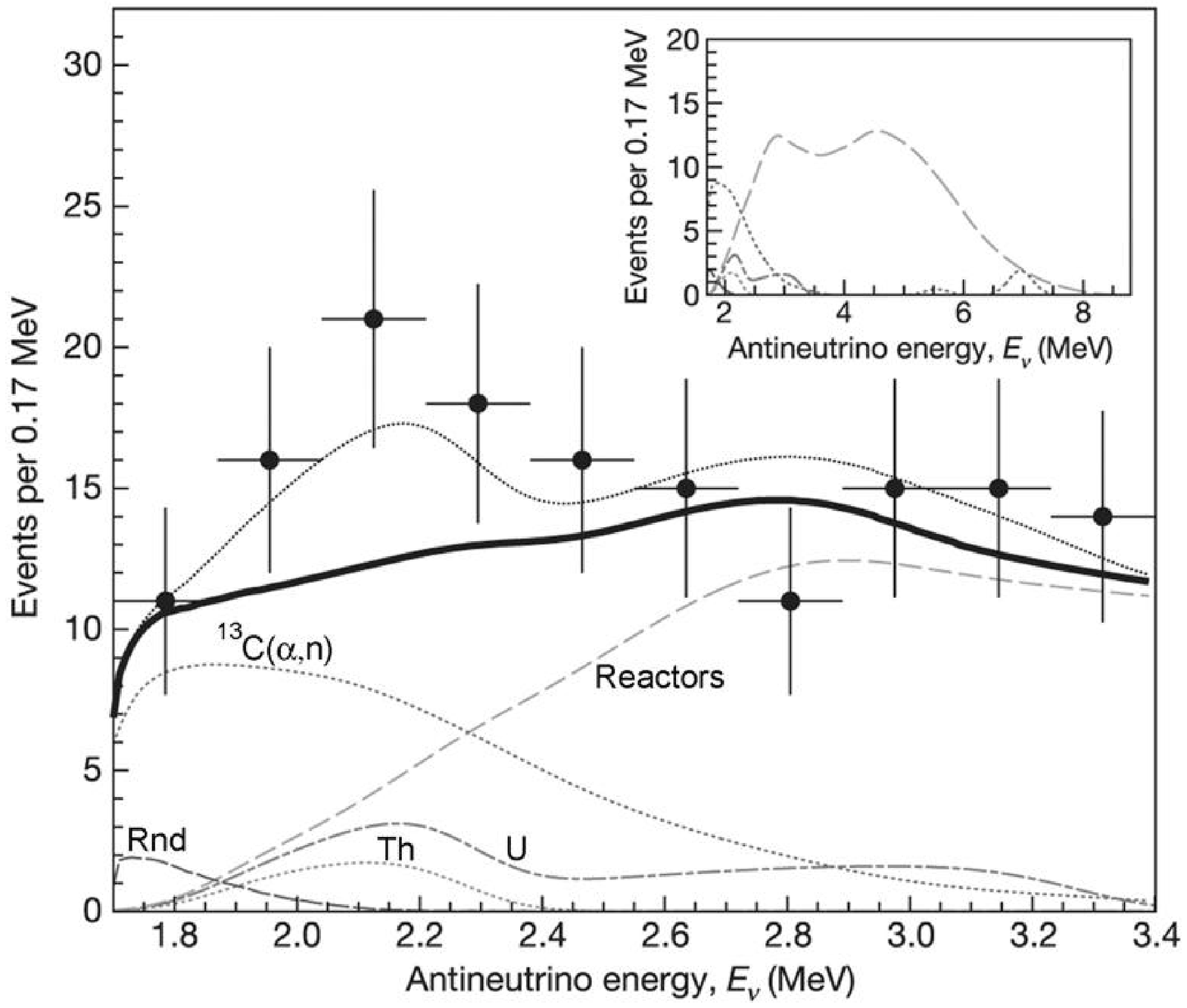}
\caption[asss]{The energy spectra in KamLAND, adapted
from~\cite{Araki:2005qa}. Main panel, experimental points together
with the total expectation (thin dotted black line). Also shown are
the total expected spectrum excluding the geo-neutrino signal(thick
solid black line), the expected signals from \U[238] (dot-dashed
line labeled \U) and \Th[232] (dotted line labeled Th)
geo-neutrinos, and the backgrounds due to reactor antineutrinos
(dashed line labeled Reactor),
$\nucleus[13]{C}(\alpha,n)\nucleus[16]{O}$ reactions (dotted line
labeled $\nucleus[13]{C}(\alpha,n)$), and random coincidences
(dashed line labeled Rnd). \label{fig:energySpectraKam}}
\vspace{2cm}
\includegraphics[width=0.5\textwidth]{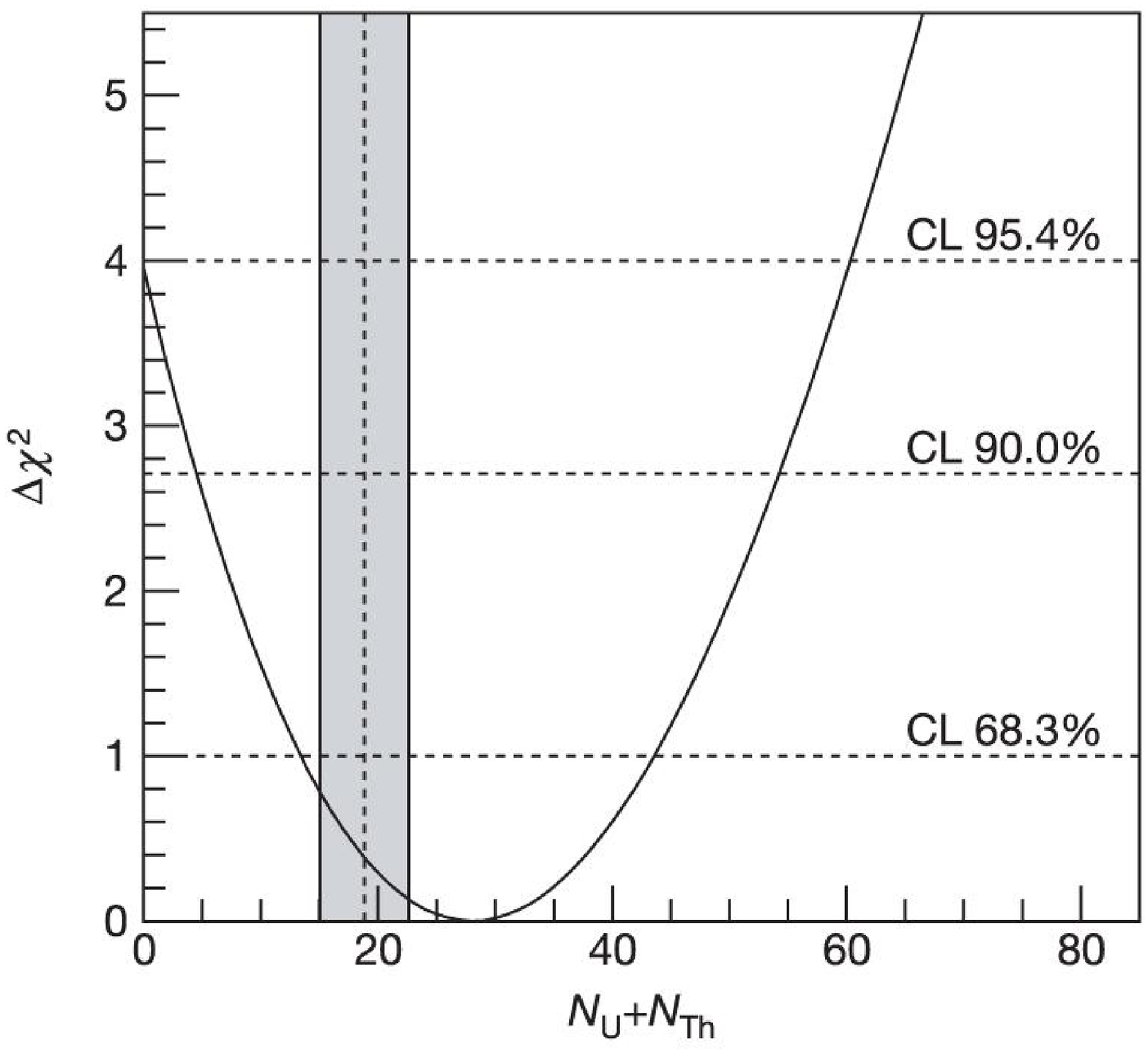}
\caption[asss]{Confidence intervals for the number of geo-neutrinos detected, from~\cite{Araki:2005qa},
assuming the mass ratio $\Th/\U = 3.9 $.
The shaded area represents the prediction of the reference model.} \label{fig:confIntervalsKam}
\end{figure}

In the energy region where geo-neutrinos are expected, see
Fig.~\ref{fig:energySpectraKam}, there are $C = 152$ counts,
implying a statistical fluctuation\footnote{In this section and in
the following ones the quoted errors correspond to $1\sigma$
interval.}
 of $\pm 12.5$. Of these counts, a
number $R = 80.4 \pm 7.2$ are attributed to reactor events, based on
an independent analysis of higher energy data. Fake geo-neutrino
events\footnote{See Section~\ref{sec:fakeAnuEvents} for a more
detailed discussion of this point.}, originating from
$\nucleus[13]{C}(\alpha,n)\nucleus[16]{O}$ reactions following the
alpha decay of contaminant \nucleus[210]{Po}, are estimated to be $F
= 42 \pm 11$, where the error is due to a 20\% uncertainty on the
$\nucleus[13]{C}(\alpha,n)\nucleus[16]{O}$ cross section and a 14\%
uncertainty on the number of  \nucleus[210]{Po} decays in the
detector. Other minor backgrounds account for $B = 4.6 \pm 0.2$
events.

A straight estimate by subtraction, $N(\U+\Th) = C-R-F-B$, with an
uncertainty obtained by combining quadratically errors gives:
$N(\U+\Th) = 25 \pm 18$.

KamLAND ``rate only'' analysis~\cite{Araki:2005qa}, which includes detection systematic errors, partially
correlated with background errors, gives $N(\U+\Th) = 25^{+19}_{-18}$;
the corresponding geo-neutrino signal is thus $S(\U+\Th) = N(\U+\Th) /\alpha  = 51^{+39}_{-36}$~TNU.

This ``rate only'' study has been improved in Ref.~\cite{Araki:2005qa} by exploiting the shape of the spectrum,
with the ratio of events  $N(\U)/N(\Th)$ being  fixed at the chondritic (BSE) prediction.
A likelihood analysis of the unbinned spectrum, see Fig.~\ref{fig:confIntervalsKam}, yields
$ N(\U+\Th) = 28^{+16}_{-15}$ which  implies $S(\U+\Th) = 57^{+33}_{-31}$~TNU.

\begin{figure}[htbp]
\includegraphics[width=0.5\textwidth]{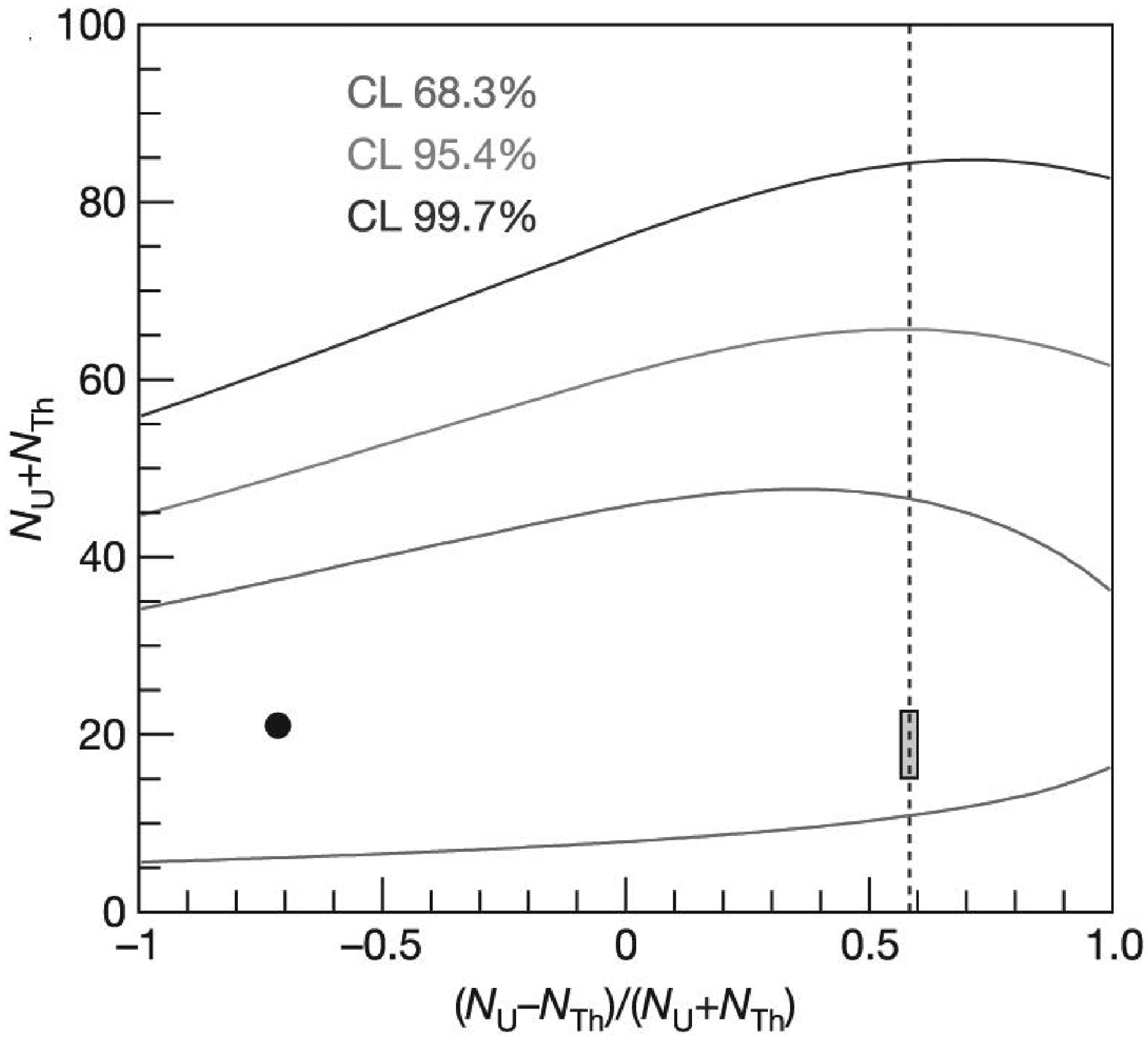}
\caption[asss]{Confidence intervals for the number of geo-neutrinos detected, from~\cite{Araki:2005qa},
when the mass ratio $\Th/\U$ is left as free parameter. The shaded area represents the prediction of the
reference model.} \label{fig:ThSuUliberoKam}
\end{figure}

As a curiosity, an analysis where both $N(\U)$ and $N(\Th)$ are left
unconstrained yields as a best fit $N(\Th)/N(\U) \approx 5.7$, which
looks far from the \emph{chondritic} value $\approx 0.25$; however,
both values are comfortably  consistent with the data already at the
$1 \sigma$ level. In fact, the statistics is so poor that one cannot
say anything from the KamLAND data concerning $N(\Th)/N(\U)$, see
Fig.~\ref{fig:ThSuUliberoKam}.

\subsection{Fake antineutrinos and a refinement of the analysis\label{sec:fakeAnuEvents}}
As a result of \nucleus[222]{Rn} contamination, \nucleus[210]{Pb},
which has half-life of 22~yr, is distributed throughout the detector.
It produces (see Fig.~\ref{fig:uChain})
\nucleus[210]{Po} which decays emitting  $\alpha$ particles with a kinetic energy
of 5.3~MeV. These act as a neutron source through $\nucleus[13]{C}(\alpha,n)\nucleus[16]{O}$
reactions occurring on the \nucleus[13]{C} nuclei which are present in the organic scintillator.
The neutrons in the $\nucleus[13]{C}(\alpha,n)\nucleus[16]{O}$ reaction are produced with kinetic energy
up to 7.3~MeV. Owing to scintillation-light quenching for high-ionization density, only about
one-third of this energy is converted into \emph{visible} energy as the neutrons thermalize.
The thermal neutrons are captured by protons with a mean capture time of 200~microseconds, producing a
delayed signal identical to that from neutron inverse  $\beta$-decay.
In summary, one has a fake antineutrino signal.

In order to extract the true geo-neutrino signal one has to subtract these events.
As already remarked, a major uncertainty originates from the $\nucleus[13]{C}(\alpha,n)\nucleus[16]{O}$
cross section\footnote{In fact, the claim of 9 geo-neutrino events in~\cite{Eguchi:2002dm}
should be dismissed: more than half of these events are to be considered as fake signals,
produced from the $\nucleus[13]{C}(\alpha,n)\nucleus[16]{O}$ reaction.}.

The number of \nucleus[13]{C} nuclei in the fiducial volume is determined from the measured
\nucleus[13]{C}/\nucleus[12]{C} ratio in the KamLAND scintillator.
On the basis of the $\nucleus[13]{C}(\alpha,n)\nucleus[16]{O}$
reaction cross-section, the  $\alpha$-particle energy loss in the scintillator, and the number of
\nucleus[210]{Po} decays, the total number of neutrons produced is expected to be $93 \pm 22$.
This error is dominated by the uncertainty in the total $\nucleus[13]{C}(\alpha,n)\nucleus[16]{O}$
reaction cross section. The values for the cross section used in~\cite{Araki:2005qa}
are taken from the JENDL~\cite{JENDL} compilation, which provides an R-matrix fit of relatively old data.
A 20\% overall uncertainty has been adopted in~\cite{Araki:2005qa}, corresponding to the accuracy claimed
in the original experimental papers (see, \eg~\cite{JENDL}).

\begin{figure}[htbp]
\includegraphics[width=0.5\textwidth,angle=-90]{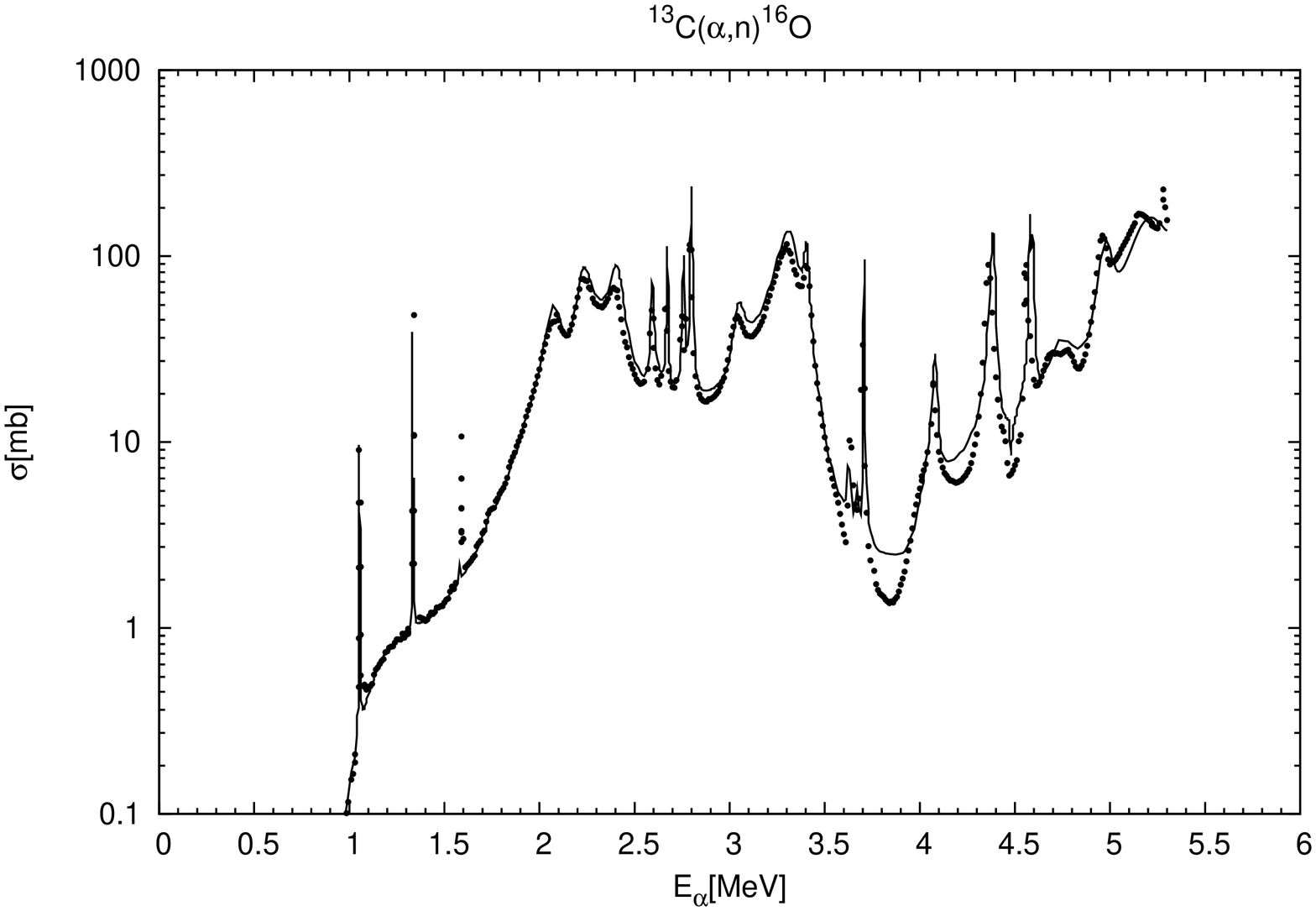}
\caption[asss]{Cross section of
$\nucleus[13]{C}(\alpha,n)\nucleus[16]{O}$. The solid line
corresponds to the JENDL compilation, dots are the experimental
points from~\cite{Harissopulos:2005cp}.} \label{fig:XsectionCO}
\end{figure}

Recently a series of high-precision measurements for this cross
section has been performed~\cite{Harissopulos:2005cp}. In the
relevant energy range $(1 \div 5.3)$~MeV, the absolute normalization
has been determined with a 4\% accuracy. The measured values are
generally in very good agreement with those recommended in JENDL,
see Fig.~\ref{fig:XsectionCO}; however, it was found
in~\cite{Fiorentini:2005ma} that the neutron yield per $\alpha$
particle is 5\% smaller. It follows that the number of fake
geo-neutrinos is lower, $F = 40 \pm 5.8$, and geo-neutrino events
obviously increase\footnote{Indeed Ref.~\cite{Araki:2005qa}
mentions that an alternative analysis including the time structure
of the scintillation light from different particles produced a
slightly larger geo-neutrino signal, which is consistent with the
result presented here.}.

The ``rate only'' analysis gives now $27^{+16}_{-15}$ geo-neutrino
events~\cite{Fiorentini:2005ma}, corresponding to $S(\U+\Th) = 55^{+33}_{-31}$~TNU.
An analysis of the binned spectrum has also been performed in~\cite{Fiorentini:2005ma} with the
result $N(\U+\Th) = 31^{+14}_{-13}$ counts, corresponding to $S(\U+\Th) = 63^{+28}_{-25}$~TNU.

These signals should be compared to $S(\U+\Th) = 51^{+39}_{-36}$~TNU
and $S(\U+\Th) = 57^{+33}_{-31}$~TNU, respectively, which were
obtained using the JENDL $\nucleus[13]{C}(\alpha,n)\nucleus[16]{O}$
cross section.

In summary, by using the new high-precision data on $\nucleus[13]{C}(\alpha,n)\nucleus[16]{O}$
 one extracts from KamLAND data a larger geo-neutrino signal with a smaller error.
This corroborates the evidence for geo-neutrinos in KamLAND data, which becomes close to $2.5 \sigma$.

\subsection{Implications of KamLAND results}

The geo-neutrino signal reported by KamLAND, $S(\U+\Th) =
57^{+33}_{-31}$~TNU, is well consistent with the BSE prediction,
$\approx 37$~TNU, as well as with the $\approx 56$~TNU prediction of
models for maximal and fully radiogenic heat flow, see
Fig.~\ref{fig:KamlandsignalmassUplusTh}.

\begin{figure}[htbp]
\includegraphics[width=0.5\textwidth,angle=-90]{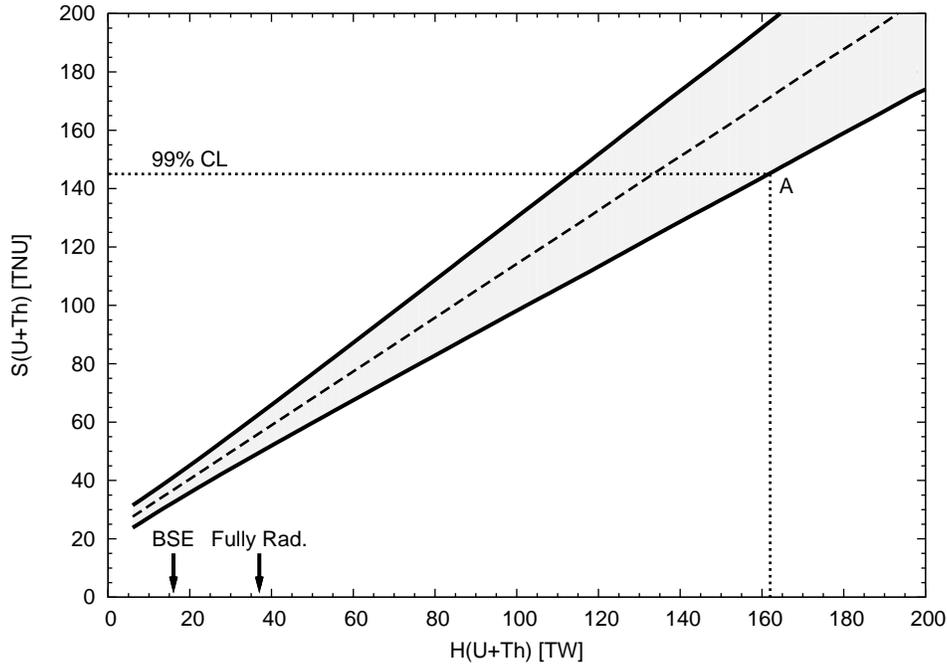}
\caption[attt]{The combined signal from uranium and thorium
geo-neutrinos and the radiogenic heat production rate. The shaded
area denotes the region allowed by geochemical and geophysical
constraints. The dashed median line represents our best estimate for
the relationship between signal and power.}
\label{fig:signHeatGeoConstraint}
\end{figure}

In order to extract some more quantitative information from the
data, we have to extend Fig.~\ref{fig:KamlandsignalmassUplusTh},
including models which produce even larger heat and signal. These
models have been built so that an arbitrary amount of uranium and
thorium in the chondritic proportion is hidden in the
mantle\footnote{We note that models with $H(\U+\Th) > 37$~TW are
essentially unrealistic; this portion of the graph is included just
for discussing KamLAND results.}. The allowed band in
Fig.~\ref{fig:signHeatGeoConstraint} is estimated by considering
\emph{rather extreme} models for the distributions of radioactive
elements, chosen to maximize or minimize the signal for a given heat
production rate~\cite{Fiorentini:2005ma}. We also remark that, in
comparison with the experimental error, the width of the allowed
band is so narrow that we can limit the discussion to the median
line in Fig.~\ref{fig:signHeatGeoConstraint}, which represents,
according to~\cite{Fiorentini:2005ma}, the best estimate for the
relationship between signal and power.

By considering only this median line in Fig.~\ref{fig:signHeatGeoConstraint}, the reported signal
$S(\U+\Th) = 57^{+33}_{-31}$~TNU
implies\footnote{By using   the result from the analysis including
the new values of $\nucleus[13]{C}(\alpha,n)\nucleus[16]{O}$, $S(U+Th)= 63^{+28}_{-25}$~TNU,
one finds  $H(\U+\Th)= 44^{+31}_{-27}$~TW.}
$H(U+Th) = 38^{+35}_{-33}$~TW (rate + spectrum) and the
99\% confidence limit on the signal (145 TNU) corresponds to 133 TW. By including the uncertainty
band of the theoretical models, one gets an upper bound of 162 TW, see point A in Fig.~\ref{fig:signHeatGeoConstraint}.
This point corresponds to a model with a total uranium mass $m(\U) = 8 \times 10^{17}$~kg,
an uranium poor crust, $m_{\mathrm{C}}(\U) = 0.3 \times 10^{17}$~kg, the rest of the uranium
being placed at the bottom of the mantle, and global chondritic thorium-to-uranium ratio.

This 162 TW upper bound is much higher than the 60 TW upper bound claimed in~\cite{Araki:2005qa},
which was obtained by using a family of geological models which are \emph{too narrow} and are also \emph{incompatible}
with well-known geochemical and geophysical data, see the discussion in~\cite{Fiorentini:2005ma}.

We remark that the bound $H(\U+\Th) < 162$~TW does not add any significant information on Earth's
interior, since anything exceeding $H(\U+\Th) = 37$~TW (\ie $H(\U+\Th+\K) = 44$~TW) is unrealistic.
The upper limit simply reflects the large uncertainty in this pioneering experiment.

In summary, KamLAND has shown that the technology for geo-neutrino detection is now available;
however, the determination of radiogenic heat power from geo-neutrino measurements is
still affected by a 70\% uncertainty.

An important quantity for deciding the potential of future experiments is the relationship between
geo-neutrino signal and heat production. The basic parameter is the slope, $dS/dH$, which
expresses how the experimental error translates into an uncertainty on the deduced heat production.
For our models we find from Fig.~\ref{fig:signHeatGeoConstraint}
$dS/dH \approx 1$~TNU/TW. Discrimination between BSE, $H(\U+\Th) \approx 16$~TW,
and fully radiogenic models $H(\U+\Th) \approx 37$~TW, requires a
precision $\Delta H \approx 7$~TW, and thus an experiment with an
accuracy $\Delta S \approx 7$~TNU.

\section{\label{sec:reactor}Background from reactor antineutrinos}

As first pointed out by Lagage~\cite{Lagage:1985tq}, antineutrinos form nuclear power plants can be a
significant background for geo-neutrino detection.

An order of magnitude estimate of the flux of antineutrinos from
reactor can be immediately found from the knowledge of the energy
produced per fission ($E_{\mathrm{fis}} \approx 200$~MeV) and the
number of antineutrinos resulting form each fission
($N_{\anu}\approx 6 $). The flux at a detector lying a distance $R$
from a reactor with thermal power $W$ is thus:
\begin{equation}
\Phi^{\mathrm{(arr)}} = \langle P_{ee} \rangle \frac{N_{\anu}
W}{4\pi R^2 E_{\mathrm{fis}}} \quad .
\end{equation}
For a typical value $W\approx 3$~GW at $R \approx 100$~km one has
$\Phi^{\mathrm{(arr)}}\approx 2.5\times 10^5$cm$^{-2}$s$^{-1}$. In a
region where there many reactors, as near Kamioka (21 nuclear
reactors already within 200 km) the signal of man-made antineutrinos
exceeds that from natural radioactivity in the Earth.

\begin{figure}[htbp]
\includegraphics[width=0.45\textwidth,angle=-90]{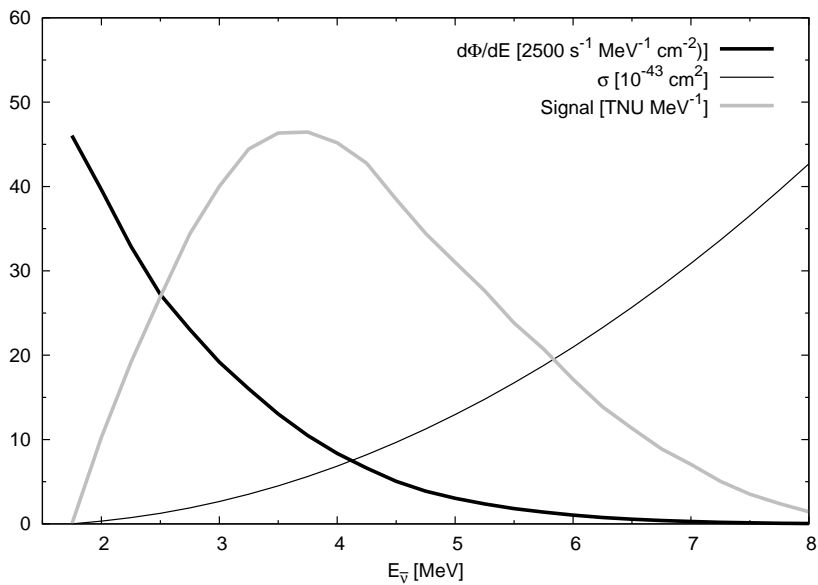}
\caption[avvv]{Antineutrinos from reactors. The differential
produced flux (no oscillations) in units of 2500 \anue\ s$^{-1}$
cm$^{-1}$ MeV$^{-1}$, inverse beta decay cross section in units of
$10^{-43}$cm$^{2}$, and the corresponding signal in TNU MeV$^{-1}$.
The flux corresponds to $10^{20}$ fissions per second (or about
3~GW) in a reactor 100~km from the detector.}
\label{fig:reactorSpectrum}
\end{figure}

In more detail, the four isotopes whose fission is the source of
virtually all the reactor power are \U[235], \U[238],
\nucleus[239]{Pu}, and \nucleus[241]{Pu}. Each isotope produces a
unique neutrino spectrum through the decay of its fission fragments
and their daughters. The instantaneous fission rates of the four
isotopes are used as an input for the evaluation of the antineutrino
spectrum. For all but \U[238], careful measurements of the
(electron) spectrum from fission by thermal neutrons have been
performed~\cite{Schreckenbach:1985ep,Hahn:1989zr}. In
Fig.~\ref{fig:reactorSpectrum} we show the differential antineutrino
spectrum calculated assuming $10^{20}$ fissions per second,
corresponding to about 3~GW, in a reactor 100~km from the detector.
The spectrum was calculated assuming a fuel composition
(0.568,0.297,0.078,0.057) for (\U[235], \nucleus[239]{Pu}, \U[238],
\nucleus[241]{Pu}); for \nucleus[239]{Pu} and \nucleus[241]{Pu} we
used spectra from~\cite{Hahn:1989zr},
 for \U[235] spectrum from~\cite{Schreckenbach:1985ep}, and for
 \U[238] the spectrum from~\cite{Vogel:1980bk} corresponding to 0.5~MeV neutrons.
For an extensive review of
reactor antineutrinos see~\cite{Bemporad:2001qy}.

About 450 reactors are operational all over the world. If they all
work at full power, this results in a total heat production of about
1 TW, just a factor 30 smaller than the natural heat flow from the
Earth. The man-made antineutrino luminosity of the Earth is thus
$\approx 10^{23}$~s$^{-1}$, a factor 10 below the natural luminosity
in geo-neutrinos from \U\ and \Th\ chains.

\begin{figure}[p]
\includegraphics[width=0.6\textwidth]{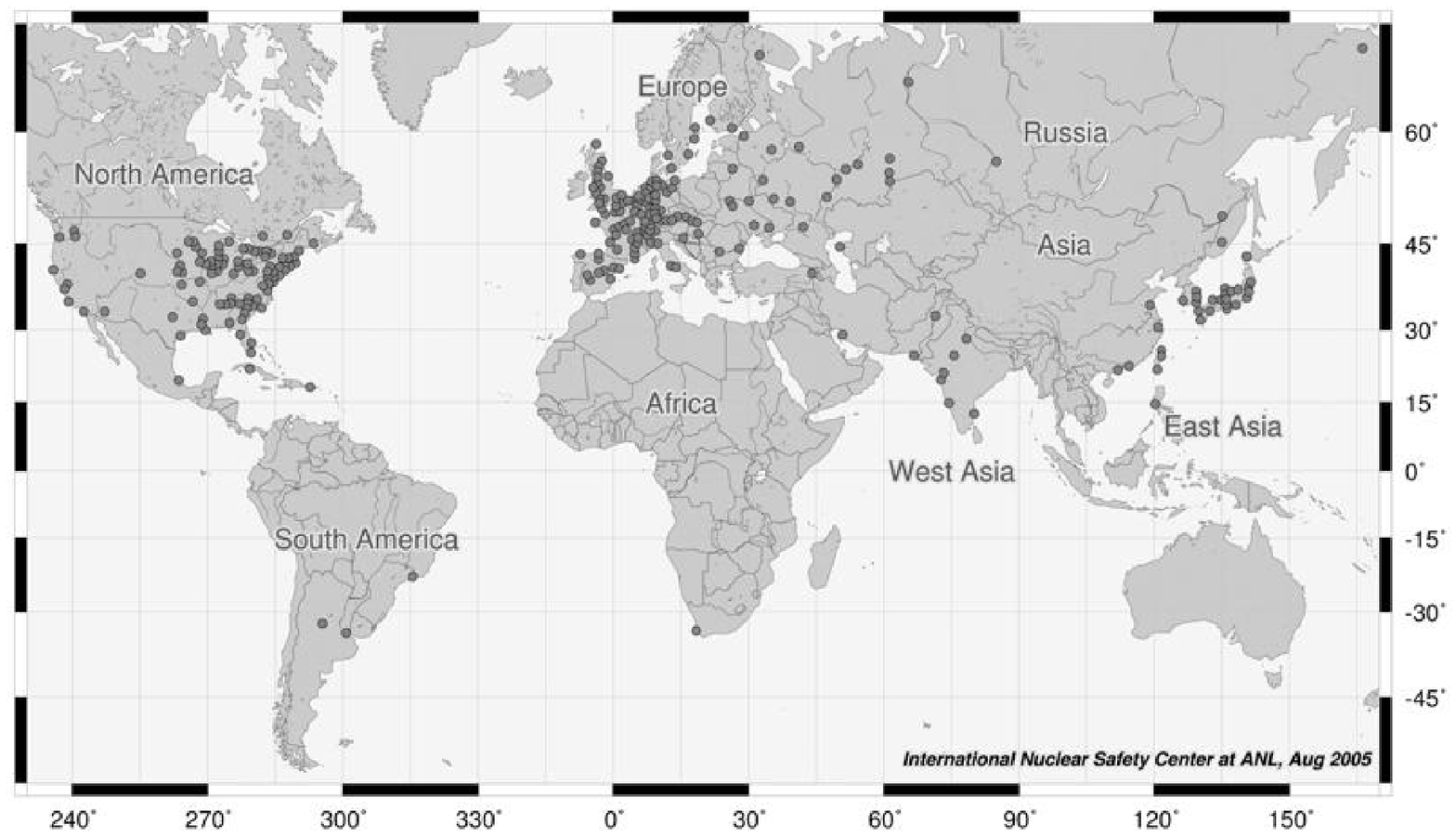}
\caption[azzzc]{Nuclear power plants in the world, from
\texttt{http://www.insc.anl.gov/pwrmaps/map/world\_map.php}.}
\label{fig:nuclearPlantsMap}
\vspace{4cm}
\includegraphics[width=0.7\textwidth]{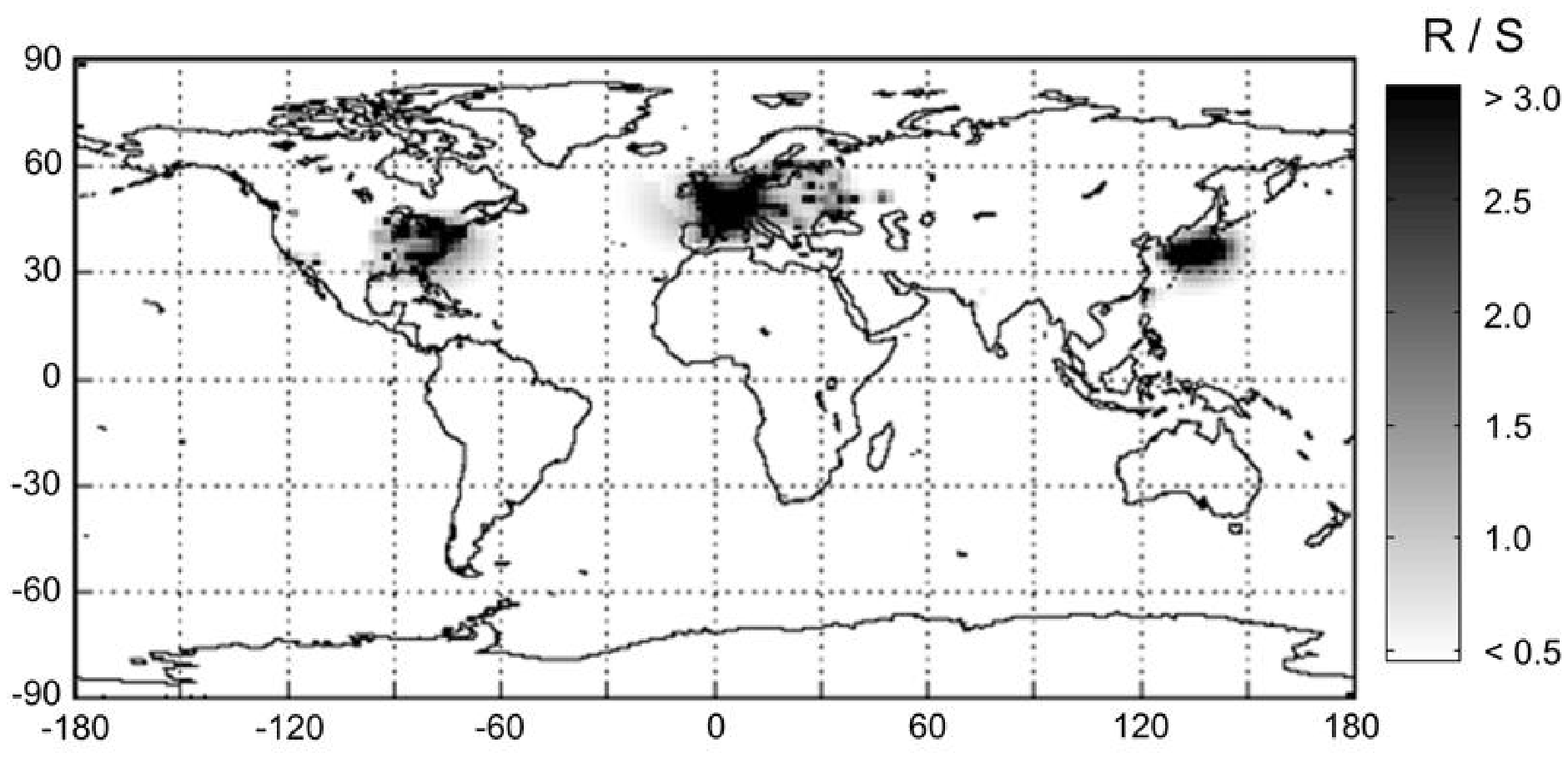}
\caption[bccc]{The ratio of reactor anti-neutrino events (in the geo-neutrino energy region)
to the expected geo-neutrino events all over the globe.} \label{fig:ratioReactorGeonu}
\end{figure}

Maps of the nuclear power plants in the world (see
Fig.~\ref{fig:nuclearPlantsMap}), together with information on power
and operational status are provided by several organizations, \eg
the International Nuclear Safety Center, United Nations Environment
Programme.

The ratio $r$ of reactor event rate in the geo-neutrino energy window to the geo-neutrino
signal $S(\U+\Th)$ predicted by the reference model is shown in Fig.~\ref{fig:ratioReactorGeonu}
all over the globe and its inverse $s=1/r$ in Table~\ref{tab:expGeoReaSignLocat} for eight locations.
Kamioka ($s \approx 0.14$) is clearly one of the worst locations over the globe.
At Gran Sasso and Sudbury the geo-neutrino event are comparable to reactor events,
whereas a place like Hawaii looks  much more favorable.

\section{\label{sec:future}Future prospects}
In summary, KamLAND has shown that the technique for exploiting
geo-neutrinos in the investigation of the Earth's interior is now
available. On the other hand, the determination of radiogenic heat
power from geo-neutrino measurements is still affected by a 70\%
uncertainty. The best fit of the KamLAND result implies a radiogenic
heat production close to the prediction of maximal and fully
radiogenic model; however, the BSE prediction is within $1\sigma$
from it. In order to discriminate among different models of heat
production in the Earth an accuracy of at least $\pm 7$~TW is
necessary. The relationship between geo-neutrino signal and
radiogenic heat, $dS/dH\approx 1$~TNU/TW, implies that the
experimental error has to be $\pm 7$~TNU, \ie a factor of four
improvement with respect to present.

It looks to us that the following questions are relevant for the future:
\begin{itemize}
  \item
  How shall we have definite (at least $3\sigma$) evidence of geo-neutrinos?
  \item
  How much uranium and thorium are in the Earth's crust?
  \item
  How much in the mantle?
  \item
   What can be said about the core?
\end{itemize}

A preliminary point for establishing suitable detector locations is the reactor background.
Figure~\ref{fig:ratioReactorGeonu} shows the ratio of reactor events (in the geo-neutrino energy
region) to the expected geo-neutrino events all over the globe.
KamLAND location is obviously one of the worst for the study of geo-neutrinos.

The potential of different locations is summarized in
Table~\ref{tab:expGeoReaSignLocat}, where we present the separate
contributions to the geo-neutrino signal rate from crust and mantle
according to our reference model, $S$, together with the reactor
event rate in the geo-neutrino energy window, $R$. In the same Table
we present two merit figures:
\begin{itemize}
  \item[(i)]
   $\Delta S_0=\sqrt{S+R}$ is the square root of the total counts expected in a detector with an effective exposure
   of $\alpha_0 = 10^{32}$ proton$\cdot$yr; it represents the limiting (\ie
  neglecting backgrounds other than reactors, uncertainties of oscillation parameters \ldots)
  statistical error on the geo-neutrino signal which might be achieved with such a detector.
  For an exposure $\alpha$  the statistical error is  $\Delta S = \Delta S_0 \times\sqrt{\alpha_0 / \alpha}$.
  \item[(ii)]
  The ratio of geo-neutrino events to reactors events in the geo-neutrino energy window
  $s = S / R$.
\end{itemize}

\begin{table}[htb]
\caption{The geo-neutrino (\U + \Th) signal rate expected from the
crust $S_{\mathrm{C}}$, from the mantle $S_{\mathrm{M}}$, and their
sum $S$, together with the reactor event rate $R$ in the
geo-neutrino energy window. $\Delta S_0$ represents the limiting
statistical error for an effective exposure $\alpha_0 = 10^{32}$
proton$\cdot$yr. All rates are in TNU. The $s$ factor is the ratio
between the geo-neutrino events and reactor events in the
geo-neutrinos energy window. } \label{tab:expGeoReaSignLocat}
\newcommand{\dg}{\hphantom{$0$}}
\newcommand{\cc}[1]{\multicolumn{1}{c}{#1}}
\renewcommand{\tabcolsep}{2pc} 
\begin{tabular}{lrrrrrc}
\hline
Location &  $S_{\mathrm{C}}$ & $S_{\mathrm{M}}$ & $S$ &  $R$  & $\Delta S_0$ & $s$ \\
\hline
Pyhasalmi  & 42.5 &   9.0 & 51.5 &  27.2 &  8.9 & 1.9  \\
Homestake  & 42.3 &   9.0 & 51.3 &  9.4  &  7.8 & 5.5  \\
Baksan     & 41.8 &   9.0 & 50.8 &  11.8 &  7.9 & 4.3  \\
Sudbury    & 41.8 &   9.0 & 50.8 &  53.7 &  10  &  0.9  \\
Gran Sasso & 31.7 &   9.0 & 40.7 &  35.1 &  8.7 & 1.2 \\
Kamioka    & 25.5 &   9.0 & 34.5 &  230  &  16  & 0.15 \\
Curacao    & 23.5 &   9.0 & 32.5 &  3.3  &  6.0 & 9.9 \\
Hawaii     & 3.5  &   9.0 & 12.5 &  1.4  &  3.7 & 9.2 \\
\hline
\end{tabular}\\[2pt]
\end{table}

With additional statistics KamLAND should be capable of providing
$3\sigma$ evidence of geo-neutrinos, but discrimination between BSE
and fully radiogenic models definitely requires new detectors, with
class and size similar to that of KamLAND, far away from nuclear
power plants. Borexino at Gran Sasso should reach the $3\sigma$
evidence, but cannot go much further due to its relatively small
size.

At Sudbury, SNO$^{+}$ with liquid scintillator will have excellent
opportunities to determine the uranium mass in the crust,
which accounts for about 80\% of the geo-neutrino signal.
This will provide an important test about models for the Earth's crust.

A detector at Hawaii, very far from the continental crust, will be
mainly sensitive to the mantle composition. We remind that the amount of radioactive
materials in this reservoir is the main uncertainty of geological models of the Earth.
Due to the absence of nearby reactors, the geo-neutrino signal can be measured with
a small error, such that different models for terrestrial heat generation can be
discriminated. On the other hand it is necessary that non-reactor backgrounds be kept at very small value.

For the very long term future, one can speculate about completely
new detectors, capable of providing (moderately) directional
information. These should allow the identification of the different
geo-neutrino sources (crust, mantle and possibly core) in the Earth;
in summary, \emph{se son rose fioriranno}\footnote{If they are
roses, they will blossom, \ie time will tell.}.

\section*{Acknowledgments}
We are grateful for enlightening discussions and valuable comments
to E.~Bellotti, C.~Broggini, A.~Bottino, L.~Carmignani, M.~Chen,
M.~Coltorti, S.~Enomoto, G.~Gratta, A.~Ianni, K.~Inoue, T.~Laserre,
E.~Lisi, W.~F.~McDonough, G.~Ottonello, R.~Raghavan, B.~Ricci,
C.~Rolfs, S.~Schoenert, A.~Suzuki, R.~Vannucci, and F.~Vissani.

This work was partially supported by MIUR (Ministero
dell'Istruzione, dell'Universit\`a e della Ricerca) under
MIUR-PRIN-2006 project ``Astroparticle physics''.

\bibliography{geonu}
\appendix
\section{Analytical estimates of the geo-neutrino flux}
\label{sec:appenAnalyticEst}

\subsection{The flux from a spherical shell}
\begin{figure}[hp]
\includegraphics[width=0.4\textwidth,angle=0]{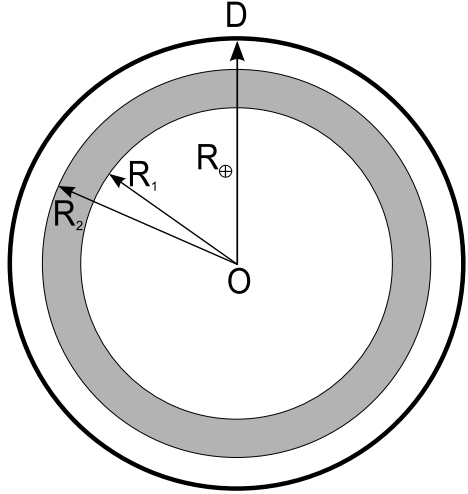}
\vspace{1cm}
 \caption[fff]{Relevant variables for parameterizing source positions in the Earth relative to the detector $D$ and
 spherical shell whose flux is given by Eq.~(\ref{eq:uniformShellFlux}).
\label{fig:geoVariable}}
\end{figure}

Assuming spherical symmetry, the (produced) antineutrino flux at  a
detector on the Earth surface originated from a spherical shell
centered at the Earth center and with radii $R_1$  and $R_2$ (see
Fig.~\ref{fig:geoVariable}) is
\begin{equation}\label{eq:uniformShellFlux}
    \Phi(X)=
     \frac{A_X \rE}{2} \left[
     \frac{R_2}{\rE}  -
     \frac{1}{2} \frac{\rE^2-R_2^2}{\rE^2}\log\left(\frac{\rE+R_2}{\rE-R_2}\right)
     -\frac{R_1}{\rE}  +
     \frac{1}{2} \frac{\rE^2-R_1^2}{\rE^2}\log\left(\frac{\rE+R_1}{\rE-R_1}\right)
     \right]
     \quad ,
\end{equation}
where $A$ is the specific geo-neutrino activity, \ie\ the number of
geo-neutrinos produced per unit time and volume, and $X$ stands for
\U\ or \Th.

From Eq.~(\ref{eq:uniformShellFlux}) one derives simple expressions
for the contributions of the crust and of the mantle, when treating
them as spherical shells of uniform density of heat generating
elements.

\subsection{Flux from the crust}
In this case $R_2=\rE$ and $R_1=\rE -\Delta$, where $\Delta\approx
30$~km is the thickness of the crust. Since $\Delta \ll \rE$,
Eq.~(\ref{eq:uniformShellFlux}) simplifies to:
\begin{equation}
   \label{eq:analyCrustTotFlux}
    \Phi_C(X)\approx
     \frac{A_X \Delta}{2} \left[ 1+\log\frac{2\rE}{\Delta} \right]
     \quad .
\end{equation}
In terms of the mass of the element $X$ contained in the crust
$m_C(X)$:
\begin{equation}
    A_X \approx \frac{n_X m_C(X)}{m_X \tau_X 4\pi \rE^2 \Delta} \quad ,
\end{equation}
where $m_X$ and $\tau_X$ are the mass and lifetime of the nucleus
$X$ and $n_X$ is the number of antineutrino produced in the decay
chain. This gives
\begin{equation}
    \Phi_C(X) \approx
     \frac{n_X m_C(X)}{m_X \tau_X 8\pi \rE^2 } \left[
     1+\log\frac{2\rE}{\Delta} \right]
     \quad .
\end{equation}
Note that this result is weakly dependent on $\Delta$.

By inserting the appropriate constants one finds
\begin{equation}
    \Phi_C(\U) =
     5.1\times 10^{6} \mathrm{\ cm}^{-2}\mathrm{s}^{-1}\times m_C(\U)
     \quad ,
\end{equation}
and
\begin{equation}
    \Phi_C(\Th) =
     1.1\times 10^{6} \mathrm{\ cm}^{-2}\mathrm{s}^{-1}\times m_C(\Th)
     \quad ,
\end{equation}
where \U\ and \Th\ masses are measured in units of $10^{17}$~kg.

For the values used in the reference model ($m_C(\U)=0.353$,
$m_C(\Th)=1.38$) one finds
\begin{equation}
    \Phi_C(\U) =
     1.8\times 10^{6} \mathrm{\ cm}^{-2}\mathrm{s}^{-1}
     \quad \quad\quad
  \Phi_C(\Th) =
     1.5\times 10^{6} \mathrm{\ cm}^{-2}\mathrm{s}^{-1}
     \quad .
\end{equation}
This provides an order of magnitude estimate of the flux originated
from the crust, however, as shown in Table~\ref{tab:fluxU}, there
can be substantial differences among different locations.

\subsection{Flux from the mantle}
In this case $R_2=\rE$ so that Eq.~(\ref{eq:uniformShellFlux})
simplifies to:
\begin{equation}
    \Phi_M(X) \approx
     \frac{A_X \rE}{2} \left[
     \frac{\rE-R_M}{\rE}  +
     \frac{1}{2} \frac{\rE^2-R_M^2}{\rE^2}\log\left(\frac{\rE+R_M}{\rE-R_M}\right)
     \right]
     \quad ,
\end{equation}
where $R_M\approx 3500$~km is the inner radius of the mantle. The
specific activity is in this case
\begin{equation}
    A_X = \frac{3 n_X m_M(X)}{m_X \tau_X 4\pi (\rE^3 - R_M^3 )} \quad
    .
\end{equation}
This gives
\begin{equation}
    \Phi_M(X) \approx
     \frac{n_X m_M(X)}{m_X \tau_X 8\pi \rE^2 }\frac{3\rE^2}{\rE^2+R_M \rE +R_M^2} \left[
     1+\frac{\rE+R_M}{2\rE}\log\frac{\rE+R_M}{\rE-R_M} \right]
     \quad .
\end{equation}
By inserting the appropriate constants one finds
\begin{equation}
    \Phi_M(\U) =
     2.30\times 10^{6} \mathrm{\ cm}^{-2}\mathrm{s}^{-1}\times m_M(\U)
\end{equation}
and
\begin{equation}
    \Phi_M(\Th) =
     0.50\times 10^{6} \mathrm{\ cm}^{-2}\mathrm{s}^{-1}\times m_M(\Th)
     \quad .
\end{equation}
For the values used in the reference model ($m_M(\U)=0.451$,
$m_M(\Th)=1.76$) one has
\begin{equation}
    \Phi_M(\U) =
     1.04\times 10^{6} \mathrm{\ cm}^{-2}\mathrm{s}^{-1}
     \quad \quad\quad
  \Phi_M(\Th) =
     0.88\times 10^{6} \mathrm{\ cm}^{-2}\mathrm{s}^{-1}
     \quad .
\end{equation}
These values are in agreement with the numerical calculation used
for the reference model, to the level of about 9\%.

\section{The contributed flux as function of the distance}
\begin{figure}[p]
\includegraphics[width=0.45\textwidth,angle=-90]{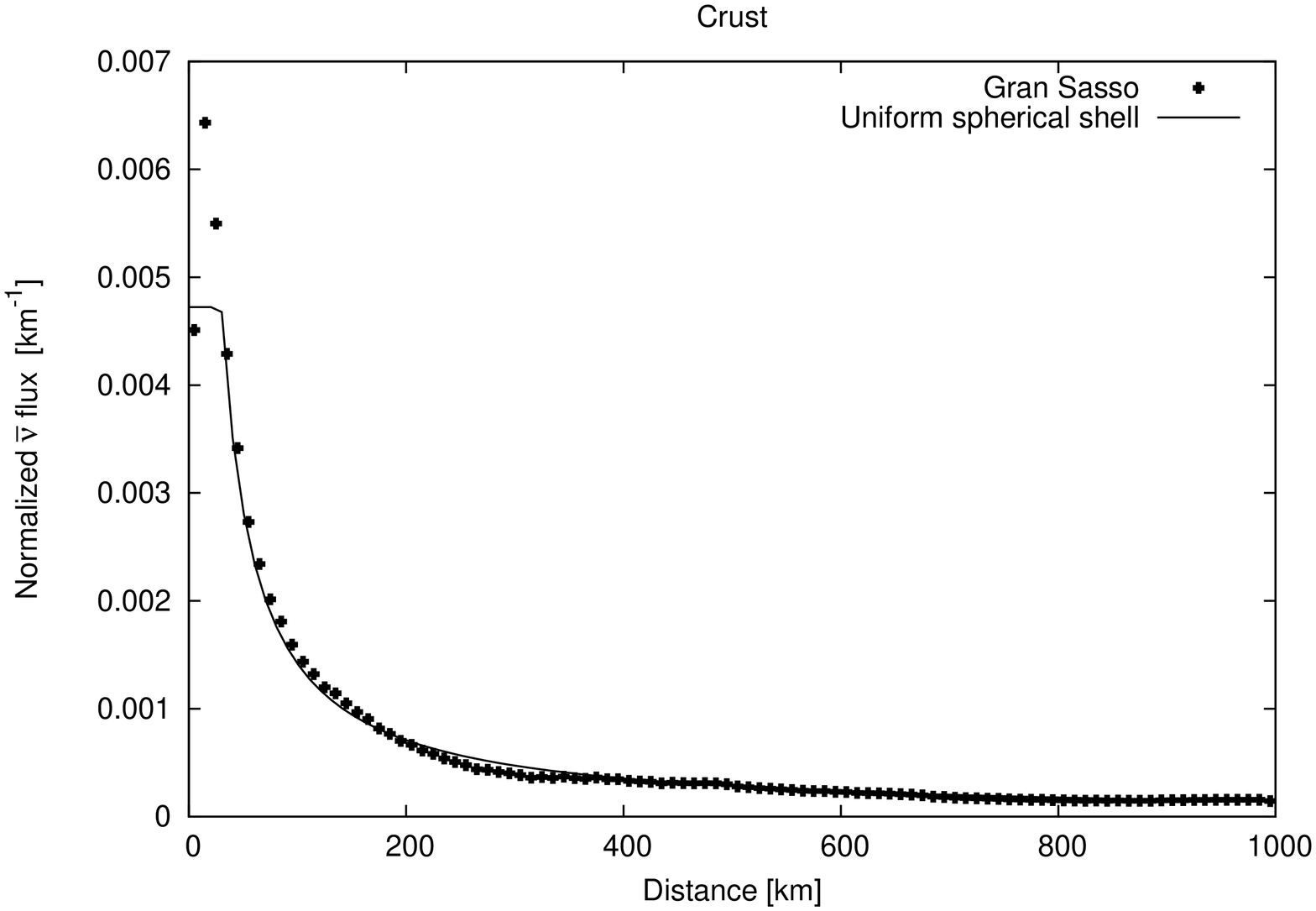}
 \caption[fff]{Antineutrino flux from the crust as function of the distance from a
 detector at the Gran Sasso underground laboratory, normalized to an unitary total  flux.
 Points corresponds to a detailed reference model~\cite{Mantovani:2003yd}. The thin line is the result of
 a crust model with
 uniform density spread over a spherical shell of thickness 30~km with the detector at
 its vertex, Eq.~(\ref{eq:analytCrustFluxDistance}).
\label{fig:crust}}
\vspace{1.5cm}
\includegraphics[width=0.45\textwidth,angle=-90]{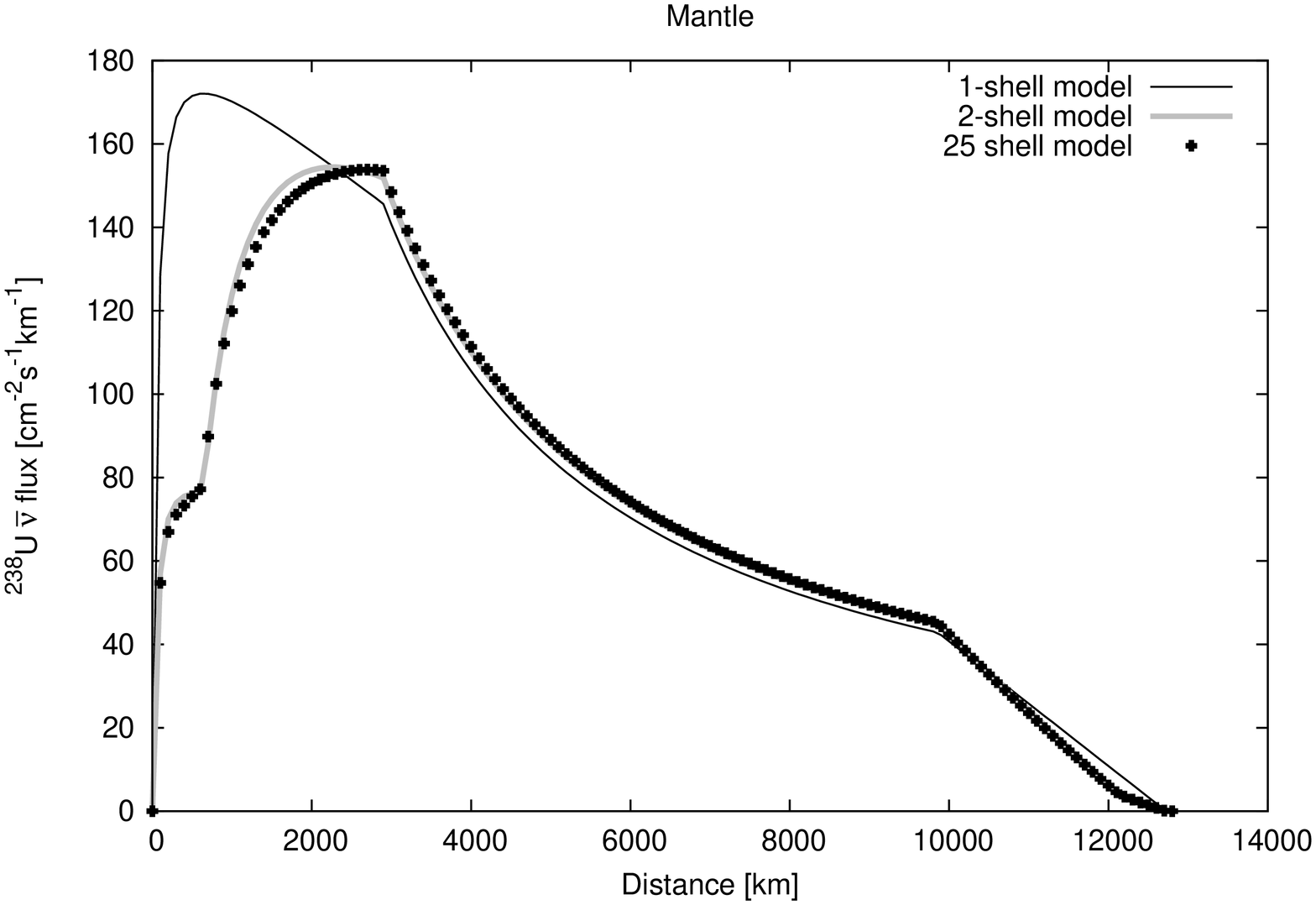}
 \caption[fff]{Uranium antineutrino flux from the mantle as function of the distance from the detector.
 Points corresponds the 25-shell reference model~\cite{Dziewonski:1981}. The thin line describes a uniform uranium density
 model with the same mass, with uniform uranium density $\rho_{U}=5.16\times 10^{-8}$~g~cm$^{-3}$
from a depth of 32~km up to one of $3478$~km: this can be obtained
with
 an average density $\rho=5$~g cm$^{-3}$ and an average abundance
$a_{\U}=10.32$~ppb.
  The thick gray line shows a two-shell model, with
 uranium density $2.29\times 10^{-8}$~g~cm$^{-3}$ in the upper mantle
 and $6.48\times 10^{-8}$~g~cm$^{-3}$ in the lower
 mantle: this can be obtained with
average densities $3.5$~g cm$^{-3}$ and $5$~g cm$^{-3}$ and average
abundances $a_{\U}=6.55$~ppb and $a_{\U}=12.96$~ppb, respectively.
\label{fig:mantle}}
\end{figure}

Again assuming spherical symmetry, the contribution to the flux from
the portion of the crust at a distance $x$ from the detector is
\begin{equation}
    \frac{d\Phi_C}{dx}=\left\{
\begin{array}{l}
  A/2  \\
  (A/2) \times (\Delta / x)
\end{array}
   \right.
\quad\quad
\begin{array}{l}
   x \le \Delta \\
   \Delta \le x \le 2\rE
\end{array}
\end{equation}
By using Eq.~(\ref{eq:analyCrustTotFlux}) the relative contribution
is
\begin{equation}
\label{eq:analytCrustFluxDistance}
    \frac{1}{\Phi_C} \frac{d\Phi_C}{dx}=\left\{
\begin{array}{l}
  (1+\log(2\rE/\Delta))^{-1}  \\
   (1+\log(2\rE/\Delta))^{-1} \times (\Delta / x )
\end{array}
   \right.
\quad\quad
\begin{array}{l}
   x \le \Delta \\
   \Delta \le x \le 2\rE
\end{array}
\end{equation}
This analytical estimate is shown in Fig.~\ref{fig:crust} together
with the corresponding numerical result for the calculation of the
reference model.

A similar calculation can be made for the mantle. The result for a
single- or double-shell mantle model is presented in
Fig.~\ref{fig:mantle} together with the numerical result of the
reference model, which considered 25 distinct shells.

While the uniform model overestimates the contribution from
distances smaller than about 2000~km, the two-shell model is quite
close to the reference
model\footnote{In fact the density of uranium in the mantle in our reference model
is not constant for two reason: the total density grows with depth
from $\rho=3.38$~g cm$^{-3}$ at $R=6346$~km (depth 32~km) to $\rho=5.41$~g cm$^{-3}$
at $R=2900$~km (we used 25 different layers~\cite{Dziewonski:1981})
and the abundance of uranium changes from $a=6.5$~ppb
in the upper mantle (above a depth of
about $632$~km) to $a=13.2$~ppb in the lower mantle.}.

\section{A comment on geological uncertainties}
\label{sec:appenErrors}
An assessment of uncertainties is most important for understanding
the significance of the theoretical predictions and the relevance of geo-neutrino experiments.

In the case of geological measurements, error determination is admittedly more complex than
for laboratory measurements since the quantity to be determined is often indirectly measured
or extrapolated from an incomplete set of samples, important examples being the elemental
abundances in the different Earth's reservoirs, and individual results are often published
without quoting an error. Nevertheless it is important to have an -- even if rough -- estimate
of  the uncertainties and to propagate it onto the predicted signals.

In the following we suggest an approach for estimating and combining
errors of the geological quantities relevant for geo-neutrino
calculations.

\subsection{Elemental abundances: selection and treatment of data.}
As for any experimental quantity, it is not possible to give completely objective
criteria for the choice of data. It is the somewhat
subjective judgment of experts in the field that selects the relevant data and
uses them to extract the ``best'' educated estimate and its error.

One has the choice of making his own compilation and selection of data,
as in~\cite{Mantovani:2003yd}  or of using some existing
compilation, as in~\cite{Fogli:2005qa}.

We believe that a robust procedure is to select all published
results after excluding measurements that have been superseded or
included in later results, or that are dependent on measurements
already included, measurements that are clearly inconsistent with
known more reliable information or that are based on questionable
assumptions. Then all the selected measurements are averaged and the
estimated standard deviation of the mean (the standard deviation of
the results divided by $\sqrt{n-1}$  where $n$ is the number of
independent results) is used as error. This procedure implies that
new independent measurements consistent with the existing ones
reduce the error and new information can exclude inconsistent
results.

In fact this is basically the procedure adopted by the Review of
Particle Physics~\cite{Yao:2006px}, with the addition that selected
data are weighted using their errors. However, geological abundances
are often published without standard errors: in fact it is not easy
to give a reliable estimate of the error, since the final number
depends not only on the measurements of the individual samples, but
also on how the samples are chosen to be representative of a much
larger portion of material.

A comprehensive critical compilation of data and estimation of errors is needed.

We remark that, when discussing uncertainties of crust and upper
mantle abundances, in the past our group  followed a different
procedure that should be upgraded: we have considered the spread of
published data as a full-range error, equivalent in same sense to a
$3\sigma$ error. The numerical difference between the two procedures
can be appreciated by looking at Table~\ref{tab:UabundCompare}.
Given the small number of independent published data, this procedure
underestimates the error compared to the standard deviation divided
by  $\sqrt{n-1}$: the case of only two data is paradigmatic.
Moreover, the errors can only become larger the more data become
available.

\subsection{Global and local source distributions: errors on theoretical hypotheses.}

In several instances, sought-after quantities depend on theoretical
hypotheses or unknown parameters, \eg  distribution of elements in
the mantle (see Sect.~\ref{subsec:mantleDataModels}) or regional
sources as a subducting slab under Japan (see
Sect.~\ref{subsec:subductingSlab}). In such cases, we suggest to
consider the minimal and maximal result as extremes of the $±\pm
3\sigma$  interval, so that the $±\pm 3\sigma$ interval cover the
complete range of theoretical hypotheses.

Our group has basically followed the above-described procedure using
full-range errors that included the spread of theoretical hypotheses.

\subsection{Combining errors: correlations.}

When building a reference model for geo-neutrinos, there are several
sources of uncertainties (abundances, source distributions,
oscillation parameters, cross sections, etc.) and one needs to
combine them to obtain the total error. When errors are independent,
one can combine them quadratically, but when they are correlated
(anti-correlated), this procedure underestimates (overestimates) the
total error. Correlation between \U\ and \Th\ abundances constitutes
an important example: the ratio of these abundances has a smaller
fluctuation than the individual abundances, indicating strong
correlation. In fact, the only other important (anti)correlation is
the one between abundances in the crust and in the mantle, if we
impose that the total element mass (in the mantle and crust) be
constrained to either the BSE value or to a given value used as free
parameter.

In principle it would be nice to have not only good determination of errors on the
abundances, but also of their correlations. Given the present situation,
we believe that a correct and robust approach is to select data and use
them in such a way as to reduce correlation.

As an example the $\U/\Th$ correlation can be tackled by performing the complete
calculation for uranium neutrinos and than scaling the result for $\U+\Th$ using a fixed ratio
of $\U/\Th$: this procedure is equivalent to a 100\% correlation
(correlation coefficient  $\rho=1$) and slightly overestimates the error.

This approach was used in Ref.~\cite{Mantovani:2003yd} and also in this paper.
In fact if one uses  $\rho=0.94$ as in Ref.~\cite{Fogli:2005qa}
 the part of the error on the total signal due to crust abundances changes by about
 $1\%$.
  For instance the
total error of our theoretical prediction at Kamioka is 5.9~TNU: if
this error were due only to  crust abundances, the use of
$\rho=0.94$ instead of $\rho=1$ would reduce the error to 5.85~TNU:
the error itself is not known with that precision.

The correlation between mantle and crust abundances is avoided not
using mantle abundances as variables, but instead total (crust plus
mantle) masses.
\end{document}